# A hybrid s-version isogeometric strategy for dynamic crack propagation in 2D and 3D problems

Tianyu He [a], Kosei Kurosaki [a], Naoki Morita [b], Naoto Mitsume [b], Kazuki Shibanuma [a,*]

[a] School of Engineering, The University of Tokyo, 7-3-1 Bunkyo-ku, Tokyo 113-8656, Japan
[b] Institute of Systems and Information Engineering, University of Tsukuba, 1-1-1 Tennodai, Tsukuba, Ibaraki 305-8573, Japan

*Corresponding author. E-mail: shibanuma@struct.t.u-tokyo.ac.jp

**Abstract**

A hybrid s-version of isogeometric analysis (hS-IGA) strategy is proposed for accurate and efficient evaluation of near-crack fracture quantities in dynamic crack propagation analysis. The strategy retains the global–local superposition framework of the conventional s-method, while introducing B-spline basis functions only into the global discretisation and preserving a Lagrange-based local mesh in the crack domain. This hybrid formulation is motivated by the continuity-related bottleneck in the global–local coupling integration of the conventional Lagrange-based s-method, and by the need to retain a Lagrange-based local mesh for crack representation and post-processing of the dynamic stress intensity factor (DSIF) and local stress. The resulting formulation removes discontinuities in the coupling integrands caused by the global approximation and enables accurate coupling integration by standard Gauss quadrature without recursive subdivision. The proposed strategy is verified using two-dimensional stationary and dynamic straight-crack problems against the standard finite element method and the conventional s-method, and is further assessed using three-dimensional stationary and dynamically propagating circular-crack problems against the conventional s-method. The results show that the proposed hS-IGA strategy accurately evaluates the DSIF and local stress while retaining the global–local modelling advantages of the s-method. It also substantially reduces the number of integration points required for coupling integration, by approximately 81% in the two-dimensional dynamic benchmark and 95.6% in the three-dimensional dynamic benchmark relative to the conventional s-method. These results demonstrate that the proposed hS-IGA framework provides an accurate and efficient global–local strategy for dynamic crack propagation analyses requiring reliable evaluation of near-crack fracture quantities.

## 1. Introduction

Dynamic crack propagation and arrest are commonly analysed using fracture parameters that characterise the driving force and the near-crack mechanical state. Among these parameters, the dynamic stress intensity factor (DSIF) has long been one of the most widely used fracture mechanics parameters and has therefore been extensively evaluated using numerical methods such as the standard finite element method (FEM) [1–4], the extended finite element method (XFEM) [5–7], and peridynamic approaches [8,9]. A recent study on ferritic steels has further suggested that, in addition to the conventional fracture mechanics parameters, the local stress ahead of the crack front can also serve as an informative quantity for describing brittle crack propagation and arrest [10]. In particular, the local stress has been reported to remain nearly constant during crack propagation and to take a characteristic value for a given type of steel under various loading and temperature conditions [10]. These findings provide an additional perspective for understanding brittle crack propagation and arrest, suggesting that, besides the DSIF, the local stress is also an important quantity in the numerical analysis of brittle fracture. Therefore, accurate evaluation of both the DSIF and the local stress is essential for the reliable simulation of dynamic crack propagation and arrest.

From a computational viewpoint, however, the accurate evaluation of these quantities remains challenging. Both the DSIF and the local stress are governed by the mechanical fields near the crack front and therefore require sufficiently fine spatial resolution in that region to ensure reliable numerical accuracy [11–13]. In particular, their accurate evaluation depends on whether the numerical model can reproduce the near-front fields with high fidelity, which in turn necessitates highly refined meshes around the crack front. However, the crack length and the structural dimensions of interest are typically several orders of magnitude larger than the refined region required for such evaluation, leading to a severe scale-gap problem in structural applications [14–16]. As a result, a straightforward discretisation of the entire domain using the standard FEM becomes computationally inefficient, because a sufficiently fine mesh is needed near the crack front while the structural-scale model must still be retained. This difficulty motivates the development of numerical frameworks that can resolve the crack-front fields locally while preserving structural-scale modelling efficiency.

To address this multiscale difficulty, our previous studies employed the s-version of the finite element method (s-method), in which a fine local mesh is superposed only in the vicinity of the crack front, while the overall structure is represented by a comparatively coarse global mesh [12,13]. By combining local refinement with structural-scale modelling within a unified framework, this global–local strategy enables accurate resolution of the crack-front fields without requiring uniform refinement of the entire domain. It is therefore well suited to dynamic crack analysis, in which reliable evaluation of the DSIF and the local stress depends on accurate reproduction of the near-front mechanical fields. Based on this framework, dynamic crack propagation simulations were successfully developed in our previous studies for both two-dimensional and three-dimensional problems [12,13,17,18].

Nevertheless, the conventional s-method still has an inherent computational bottleneck. In the overlapping domain, the global and local meshes contribute simultaneously to the discretised equations, and the corresponding coupling terms are integrated over the local domain [12,13]. In the conventional s-method, the global discretisation employs Lagrange basis functions, which have only $C^0$-continuity across element boundaries, and their derivatives are discontinuous there. When a local element overlaps multiple global elements, such discontinuities may occur within the local element, causing the coupling integrands to become discontinuous within that element. Consequently, a standard fixed-order Gauss quadrature rule becomes inefficient unless the integration domain is further partitioned along the global element boundaries to reduce the numerical integration error [17,19]. This leads to a significant increase in the number of sub-cells and integration points, and hence to a substantial additional computational cost.

From this perspective, isogeometric analysis (IGA), originally proposed by Hughes et al. [20], provides a promising basis for addressing this issue. By employing spline-based basis functions such as B-splines and NURBS, IGA offers higher-order continuity across element boundaries [21–26]. For the present problem, this feature is particularly attractive because it can improve the continuity of the global approximation across element boundaries and thereby reduce the need for element subdivision in coupling integration. As a natural extension in this direction, previous studies have reported an IGA-based formulation in which both the global and local domains in the s-method are discretised using IGA, namely, S-IGA [26].

However, for crack propagation analysis, it is not self-evident that such a full IGA discretisation is the most appropriate choice. In particular, while the higher continuity of IGA is beneficial for the global approximation, its application to the local crack domain may raise additional issues that must be considered, especially in relation to crack representation and the post-processing procedures required for the evaluation of fracture-mechanics parameters and local stress near the crack front. This suggests that IGA should be incorporated into the s-method for crack propagation analysis in a manner that carefully distinguishes the respective roles of the global and local discretisations.

Motivated by this perspective, the present study proposes a hybrid strategy, referred to as the hS-IGA, in which IGA is introduced only into the global mesh, while the local crack domain is retained as a standard Lagrange finite element discretisation. The proposed strategy is formulated and assessed through a series of two-dimensional straight-crack and three-dimensional circular-crack problems under pure tensile loading. These settings are adopted because they represent important fundamental configurations in dynamic crack propagation [10,27], and because such benchmark problems provide a clear and rigorous basis for examining how differences in the approximation scheme and the coupling-integration framework affect numerical accuracy and computational efficiency.

The proposed strategy is first verified using two-dimensional stationary and dynamic straight-crack benchmarks, where direct comparisons with the standard FEM and the conventional s-method are performed in terms of numerical accuracy, degrees of freedom, and the number of integration points. These benchmarks are used to systematically isolate, under controlled conditions, the effects of the approximation scheme and the coupling-integration framework on the evaluation of near-tip fracture quantities and on computational cost. They also clarify whether the proposed hS-IGA strategy retains the inherent advantages of the s-method framework, namely high local accuracy, reduced degrees of freedom, and a simple meshing procedure based on independent global and local discretisations, while substantially reducing the number of integration points required for global–local coupling. The strategy is then further assessed using three-dimensional stationary and dynamically propagating circular-crack benchmarks to examine whether the same hybrid rationale remains effective in crack-front analyses, where the coupling integration becomes more computationally demanding. In the following, the term “crack tip/front” is used, for brevity, to refer collectively to the crack tip in two-dimensional problems and the crack front in three-dimensional problems.

The remainder of this paper is organised as follows. Section 2 reviews the conventional s-method framework for crack propagation analysis, which serves as the basis of the present study. Section 3 presents the proposed hybrid s-version of isogeometric analysis (hS-IGA) strategy. Section 4 systematically verifies the proposed strategy through two-dimensional crack problems, with comparisons to the standard FEM and the conventional s-method in terms of numerical accuracy and computational efficiency. Section 5 further assesses its applicability to three-dimensional crack problems through comparison with the conventional s-method. Finally, Section 6 concludes the paper.

## 2. Conventional s-method framework for crack propagation analysis

This section introduces the conventional s-method as the baseline framework for crack propagation analysis in the present study. Section 2.1 briefly reviews the basic concept of the conventional s-method. Section 2.2 then presents the crack representation procedure, and Section 2.3 describes the methodology for evaluating two important fracture quantities considered in the present study, namely the stress intensity factor and the local stress. Finally, Section 2.4 discusses a key limitation of the conventional Lagrange-based formulation in numerical integration.

### *2.1. Overview of the conventional s-method*

The s-method is an advanced finite element strategy that superimposes multiple meshes within a single framework. First introduced by Fish [28], it facilitates multi-scale analysis by combining varying levels of refinement. Owing to this feature, it is particularly effective for problems involving strong stress gradients [29,30], including crack propagation [17,31]. In this study, the term “conventional (Lagrange-based) s-method” refers to the widely adopted implementation in which all domains are discretised using Lagrange basis functions [12,13,17–19,28–37]. This definition is introduced here because the choice of basis functions plays a key role in the computational cost of assembling the matrices, which will be discussed in Sections 2.4 and 3.

In the present study, to simulate dynamic crack propagation, we employ two types of meshes: a coarse global mesh, which discretises the target domain $\Omega^{\mathrm{G}}$, and a refined local mesh, which is superimposed over the local domain $\Omega^{\mathrm{L}}$ around the crack tip/front (see Fig. 1). This allows the computational effort to be concentrated in the most critical regions. Continuity of the displacement and acceleration fields between the global and local domains is ensured by the Dirichlet boundary conditions imposed on the boundary of the local domain, $\Gamma^{\mathrm{GL}}$, as illustrated in Fig. 1. This selective refinement considerably reduces computational cost without compromising accuracy, and the robustness of the approach has been confirmed in previous studies [12,13]. Detailed formulations of the conventional s-method and the associated numerical schemes are provided in Appendix A.

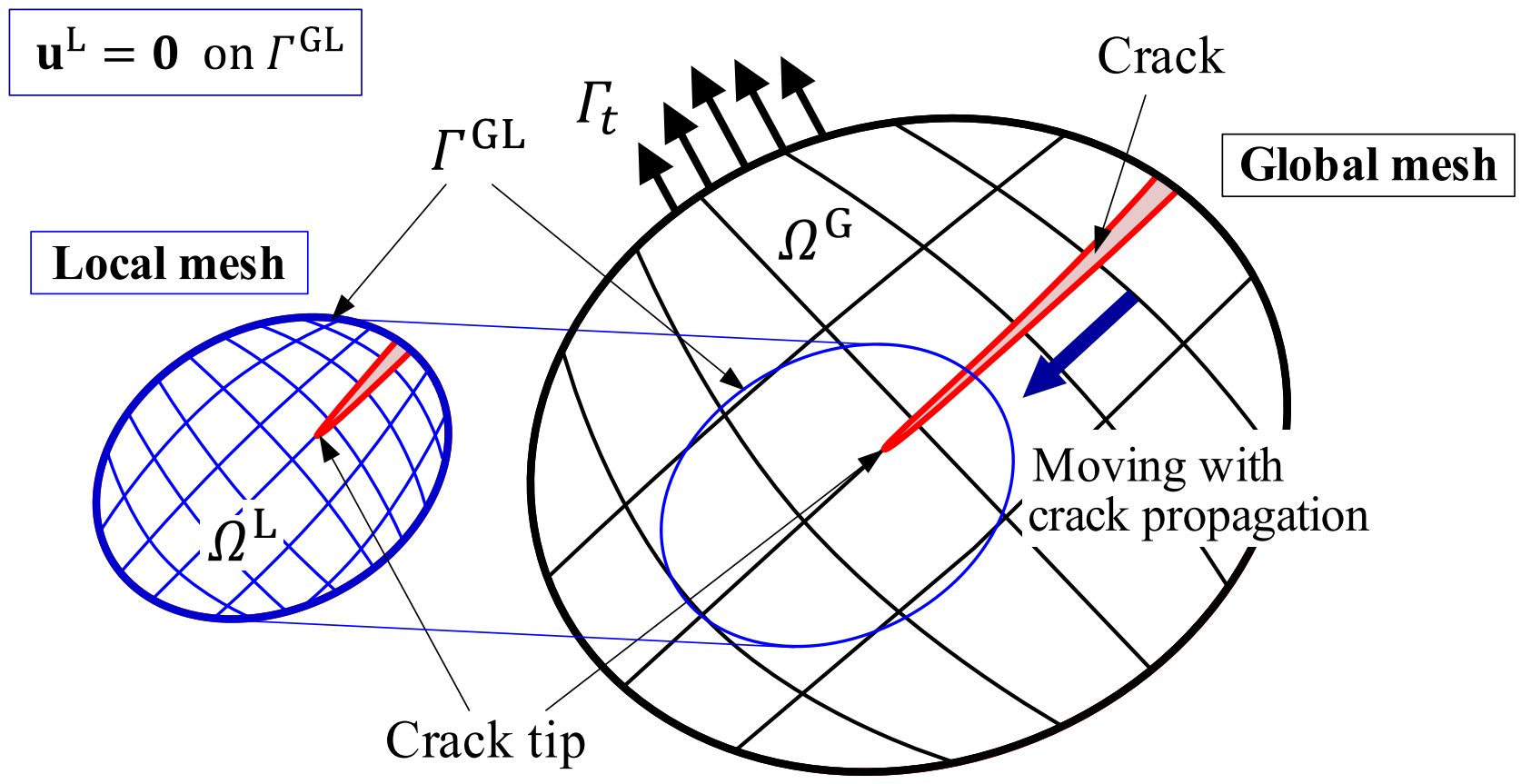

**Fig. 1.** Global and local meshes for a crack propagation analysis based on the s-method.

### *2.2. Crack representation*

This section introduces the conventional s-method framework for modelling dynamic crack propagation. The present study considers two-dimensional straight-crack problems and three-dimensional circular-crack problems subjected to remote tensile loading under Mode I conditions. These configurations represent fundamental settings for dynamic crack propagation analysis [10,27]. A schematic of the crack representation is shown in Fig. 2.

A structured local mesh is superimposed around the crack tip/front and defined in a local coordinate system. In the two-dimensional setting, the local coordinates are $(x', y')$, where $x'$ denotes the crack propagation direction and $y'$ is normal to the symmetry plane (see Fig. 2(a)). In the three-dimensional setting, the local coordinates are extended to $(x', y', z')$, where $z'$ denotes the crack-front direction (see Fig. 2(b2)). The crack-induced discontinuity is represented by imposing boundary conditions on the symmetry plane $y' = 0$. Specifically, the relevant portion of the symmetry plane ($y' = 0$) within the target domain is decomposed into the ligament part $\Gamma_{\mathrm{lg}}$ and the crack surface part $\Gamma_{\mathrm{cr}}$ as

$$\Gamma_{\mathrm{lg}} = \{\mathbf{x} | x' \geq 0, y' = 0\}, \tag{1}$$

$$\Gamma_{\mathrm{cr}} = \{\mathbf{x} | x' < 0, y' = 0\}. \tag{2}$$

Using these two parts of the symmetry plane, the set of nodes $\mathcal{S}_{\mathrm{D}}$, on which the Dirichlet boundary condition $u_{y'} = 0$ is imposed, and the set of nodes $\mathcal{S}_{\mathrm{N}}$, on which the Neumann boundary condition $t_{y'} = 0$ is imposed, are defined in both the global and local meshes in a unified manner, as follows:

$$\mathcal{S}_{\mathrm{D}} = \left\{ i \in \mathcal{N} \middle| \Omega_i \cap \Gamma_{\mathrm{lg}} \neq \emptyset,\ \mathbf{x}_i \notin \Gamma^{\mathrm{GL}} \right\}, \tag{3}$$

$$\mathcal{S}_{\mathrm{N}} = \{ i \in \mathcal{N} | \Omega_i \cap \Gamma_{\mathrm{cr}} \neq \emptyset,\ i \notin \mathcal{S}_{\mathrm{D}},\ \mathbf{x}_i \notin \Gamma^{\mathrm{GL}} \}, \tag{4}$$

where $\mathcal{N}$ is the set of all nodes, and $\Omega_i$ is the support associated with node $i$, expressed as

$$\Omega_i = \{\mathbf{x} | \phi_i(\mathbf{x}) \neq 0\}, \tag{5}$$

where $\phi_i$ is the Lagrange-basis shape function corresponding to node $i$. Note that the nodes immediately behind the crack tip/front in the global mesh (see Figs. 2(a) and 2(b2)) are included in $\mathcal{S}_{\mathrm{D}}$, even though they are located on the crack surface side [12,13].

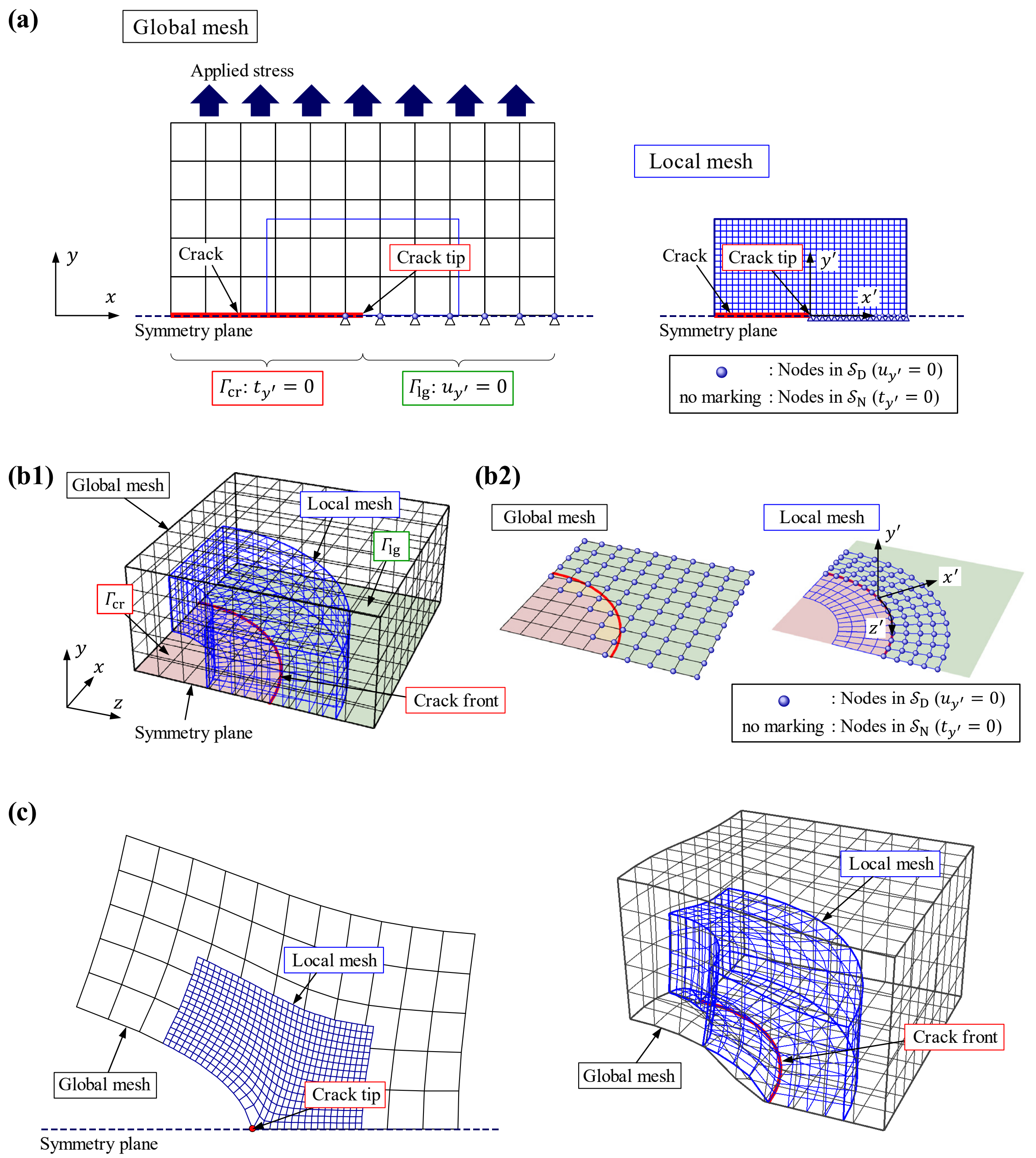

**Fig. 2.** Crack representation in the conventional s-method framework: (a) two-dimensional mesh configuration and node sets $\mathcal{S}_D$ and $\mathcal{S}_N$; (b1) three-dimensional mesh configuration; (b2) node sets $\mathcal{S}_D$ and $\mathcal{S}_N$ on the symmetry plane; and (c) deformation of the global and local meshes.

Dynamic crack propagation is modelled using the nodal force release technique [12,13,38,39] applied to the local mesh. Fig. 3(a) illustrates an example of this technique in a three-dimensional setting. At the beginning of a propagation step, the nodes in set $A$ correspond to the current crack front. The nodal forces at all nodes in set $A$ are released simultaneously, and the crack front advances to the nodes in set $B$ by the end of the step. Fig. 3(b) shows the corresponding release process in a representative two-dimensional section. In the two-dimensional straight-crack problem, the same procedure is directly applied to the crack-tip node, as illustrated in Fig. 3(b). Thus, in both two- and three-dimensional cases, the crack advances by one local element during each propagation step. Consequently, the crack velocity $V$ is prescribed by

$$V = \frac{h_{\mathrm{L}}}{\Delta t}, \tag{6}$$

where $h_{\mathrm{L}}$ is the local element size in the crack propagation direction ($x'$) and $\Delta t$ is the time increment.

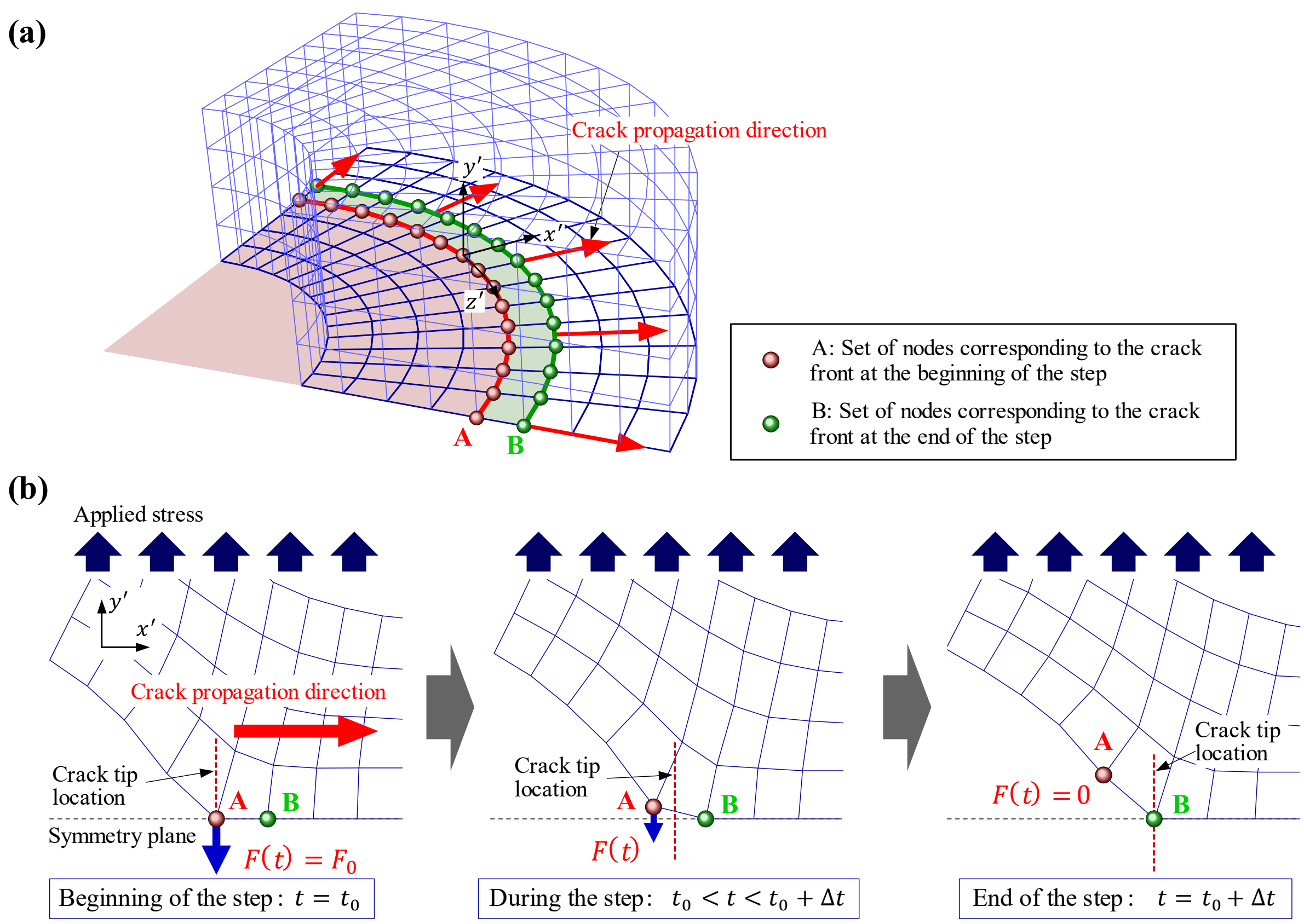

**Fig. 3.** Nodal force release technique applied to the local mesh: (a) crack-front node sets in a three-dimensional local mesh at the beginning and end of a propagation step; and (b) nodal force release process within one step, illustrated in a two-dimensional section.

Following the previous studies [12,13,38], a linear nodal force release assumption is adopted within each step. Specifically, the nodal force at the nodes corresponding to the crack front at the beginning of the step ($A$ in Fig. 3) is expressed as

$$F(t) = F_0\left(1 - \frac{t - t_0}{\Delta t}\right), \tag{7}$$

where $F_0$ and $t_0$ are the nodal force and the time at the beginning of the step, respectively.

The local mesh is updated to track the propagating crack tip/front, exploiting the flexibility of the s-method discretisation. Fig. 4 illustrates this update procedure for the two- and three-dimensional cases. During the transition from the end of the current step to the beginning of the next step, the rearmost nodes are removed, and a new layer of nodes is added ahead in the crack propagation direction. This operation updates the local mesh around the advancing crack tip in two dimensions (see Fig. 4(a)) and along the advancing crack front in three dimensions (see Fig. 4(b)). A smooth update is ensured by imposing homogeneous Dirichlet conditions on the boundary of the local domain, i.e., $\mathbf{u}^{\mathrm{L}}(\mathbf{x}) = \ddot{\mathbf{u}}^{\mathrm{L}}(\mathbf{x}) = \mathbf{0}$ on $\Gamma^{\mathrm{GL}}$. In the present implementation, all local node coordinates remain unchanged except for the newly added nodes; therefore, the nodal values $\mathbf{d}^{\mathrm{L}}$ and $\ddot{\mathbf{d}}^{\mathrm{L}}$ at the end of a step can be directly used as the initial values for the next step. For a more general update in which the local mesh changes size or topology, a projection of $\mathbf{d}^{\mathrm{L}}$ and $\ddot{\mathbf{d}}^{\mathrm{L}}$ between successive meshes can be employed [17,18].

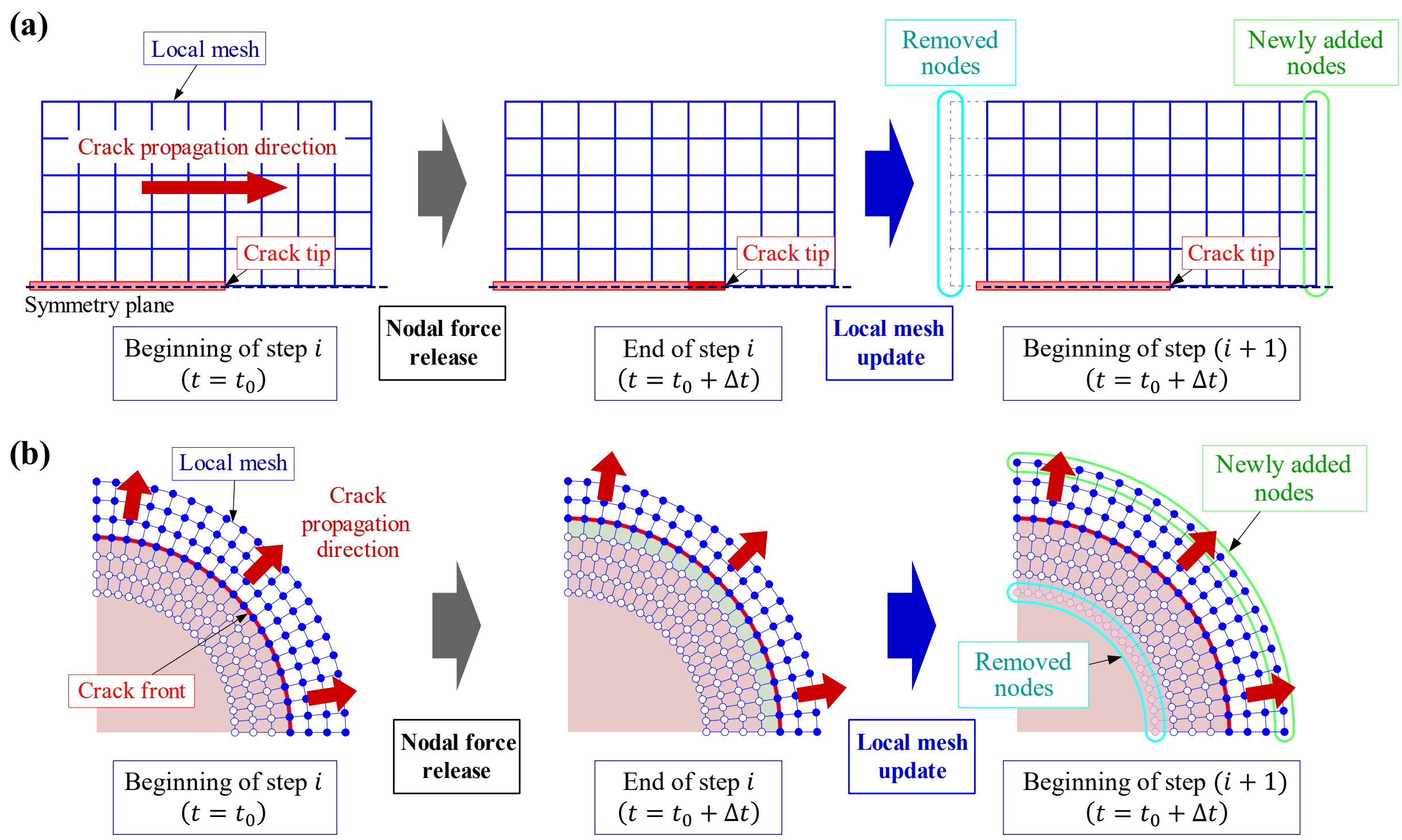

**Fig. 4.** Local mesh update during dynamic crack propagation: (a) two-dimensional straight-crack problem; and (b) three-dimensional circular-crack problem.

*2.3. Evaluation of stress intensity factor and local stress*

This section presents the methodology for evaluating the stress intensity factor and the local stress. In the following subsections, the local coordinate system introduced in Section 2.2 is employed. In the two-dimensional problem, the local coordinates are $(x', y')$, whereas in the three-dimensional problem they are $(x', y', z')$ with $x'$ aligned with the crack propagation direction, $y'$ normal to the symmetry plane, and $z'$ along the crack-front direction. For convenience, the coordinate origin is located at the evaluation point on the crack tip/front, i.e., $(x', y') = (0,0)$ in two dimensions and $(x', y', z') = (0,0,0)$ in three dimensions.

*2.3.1. Evaluation of stress intensity factor*

The stress intensity factor (SIF) is one of the most widely used fracture parameters for characterising crack propagation behaviour [10,14,40–47], and is adopted in the present study as one of the key evaluation quantities. In the SIF-based fracture criterion, the dynamic stress intensity factor (DSIF), $K_{\mathrm{I}}^{(\mathrm{d})}$, and the material resistance, $K_{\mathrm{D}}$, which is known as a function of crack velocity, $V$, must be in balance, expressed as:

$$K_{\mathrm{I}}^{(\mathrm{d})} = K_{\mathrm{D}}(V). \tag{8}$$

In the following, the static and dynamic SIFs are denoted by $K_{\mathrm{I}}^{(\mathrm{s})}$ and $K_{\mathrm{I}}^{(\mathrm{d})}$, respectively, and the corresponding J-integrals by $J^{(\mathrm{s})}$ and $J^{(\mathrm{d})}$. In the present framework, the stress intensity factor is computed from a domain integral formulation of the J-integral [48], as detailed below.

*(i) Common SIF/J-integral formulation*

In the pure Mode I crack problem considered here, $K_{\mathrm{I}}^{(\mathrm{d})}$ in dynamic crack propagation can be expressed in terms of the $J^{(\mathrm{d})}$ as [48]

$$K_{\mathrm{I}}^{(\mathrm{d})} = \sqrt{\frac{E \cdot J^{(\mathrm{d})}}{(1+\nu)A_{\mathrm{I}}}}, \tag{9}$$

and

$$A_{\mathrm{I}} = \frac{\beta_1(1-\beta_2{}^2)}{4\beta_1\beta_2 - (1+\beta_2{}^2)^2}, \tag{10}$$

$$\beta_1 = \sqrt{1-\left(\frac{V}{V_1}\right)^2}, \tag{11}$$

$$\beta_2 = \sqrt{1-\left(\frac{V}{V_2}\right)^2}, \tag{12}$$

where $\nu$ is Poisson's ratio, and $V_1$ and $V_2$ are the dilatational and shear wave velocities, respectively, expressed as

$$V_1 = \sqrt{\frac{(1-\nu)E}{(1+\nu)(1-2\nu)\rho}}, \tag{13}$$

$$V_2 = \sqrt{\frac{1}{2(1+\nu)}\frac{E}{\rho}}. \tag{14}$$

For static problems, the crack velocity is taken as $V = 0$. In this limit, $A_{\mathrm{I}}$ reduces to

$$\lim_{V\to 0} A_{\mathrm{I}} = 1-\nu, \tag{15}$$

and the SIF becomes the well-known static relation [49]

$$K_{\mathrm{I}}^{(\mathrm{s})} = \sqrt{\frac{E \cdot J^{(\mathrm{s})}}{1-\nu^2}}. \tag{16}$$

The domain-integral forms used to evaluate $J^{(\mathrm{d})}$ and $J^{(\mathrm{s})}$ are described separately for the two-dimensional and three-dimensional implementations below. In both cases, the function $q(\mathbf{x})$ is used as a weighting function associated with virtual crack extension. It defines the effective integration domain in the domain-integral formulation, rather than representing the physical crack extension amount itself.

*(ii) Two-dimensional implementation*

For two-dimensional problems, the crack-front direction is not explicitly modelled, and the J-integral formulation is interpreted per unit thickness in the out-of-plane direction. In this case, the dynamic J-integral is evaluated directly using the two-dimensional domain-integral form

$$J^{(\mathrm{d})} = \int_{\Omega}\left(\left(\sigma_{ij}\frac{\partial u_j}{\partial x_1'} - (W+K)\delta_{1i}\right)\frac{\partial q}{\partial x_i'} + \rho\left(\ddot{u}_j\frac{\partial u_j}{\partial x_1'} - \dot{u}_j\frac{\partial \dot{u}_j}{\partial x_1'}\right)q\right)d\Omega, \tag{17}$$

where $u_j$ is the displacement, $\sigma_{ij}$ and $\varepsilon_{ij}$ are the stress and strain tensors, $\delta_{1i}$ is the Kronecker delta, and $x_1' = x'$ denotes the crack propagation direction. The strain energy density $W$ and the kinetic energy density $K$ are defined as

$$W = \frac{1}{2}\sigma_{ij}\varepsilon_{ij}, \tag{18}$$

$$K = \frac{1}{2}\rho\dot{u}_j\dot{u}_j, \tag{19}$$

Correspondingly, the static J-integral is written as

$$J^{(\mathrm{s})} = \int_{\Omega}\left(\sigma_{ij}\frac{\partial u_j}{\partial x_1'} - W\delta_{1i}\right)\frac{\partial q}{\partial x_i'}\,d\Omega, \tag{20}$$

In the two-dimensional implementation, the weighting function $q(\mathbf{x})$ is defined using the local mesh basis functions $\phi_i^{\mathrm{L}}(\mathbf{x})$ as

$$q(\mathbf{x}) = \sum_i \phi_i^{\mathrm{L}}(\mathbf{x})q_i^{\mathrm{L}}, \tag{21}$$

The nodal value $q_i^{\mathrm{L}}$ is given by the radial selection function in the $x'y'$-plane as

$$q_i^{\mathrm{L}} = q_i^R, \tag{22}$$

where

$$q_i^R = \begin{cases} 1, & \text{if } \sqrt{(x_i')^2 + (y_i')^2} \le R_J, \\ 0, & \text{if } \sqrt{(x_i')^2 + (y_i')^2} > R_J. \end{cases} \tag{23}$$

Here, $R_J$ defines the size of the effective integration domain in the $x'y'$-plane. Based on preliminary investigations [13], $R_J$ is set to $1.5h_{\mathrm{L}}$, where $h_{\mathrm{L}}$ is the local element size in the $x'$-direction. Because only the regions where $q \neq 0$ or $\partial q/\partial x_i' \neq 0$ contribute to the domain integral, this definition restricts the effective integration domain to the local elements around the crack tip, as illustrated by the $x'y'$-section in Fig. 5. This localisation reduces the cost of SIF evaluation while maintaining sufficient numerical accuracy.

*(iii) Three-dimensional implementation*

For three-dimensional problems, the crack-front direction is explicitly considered, and the J-integral is evaluated near selected points on the crack front. In contrast to the two-dimensional formulation, the three-dimensional domain integral accumulates contributions over a finite segment in the crack-front direction. Therefore, a crack-front normalisation factor, $\delta A$, is introduced to obtain the J-integral value associated with a selected crack-front evaluation point.

For each selected crack-front evaluation point, the dynamic J-integral is evaluated as

$$J^{(\mathrm{d})} = \frac{1}{\delta A}\int_{\Omega}\left(\left(\sigma_{ij}\frac{\partial u_j}{\partial x_1'} - (W+K)\delta_{1i}\right)\frac{\partial q}{\partial x_i'} + \rho\left(\ddot{u}_j\frac{\partial u_j}{\partial x_1'} - \dot{u}_j\frac{\partial \dot{u}_j}{\partial x_1'}\right)q\right)d\Omega, \tag{24}$$

Here, $\delta A$ is the normalisation factor associated with the three-dimensional virtual crack-extension function and is defined as

$$\delta A = \int_S q(\mathbf{x})\,dz', \tag{25}$$

where $S$ denotes the portion of the crack front selected by the three-dimensional virtual crack-extension function, $z'$ is the local coordinate along the crack-front direction, and $\mathbf{x}$ denotes a point on the crack front. This normalisation is necessary because the three-dimensional domain integral is evaluated over a finite width in the crack-front direction; division by $\delta A$ gives the J-integral value associated with the selected crack-front point, rather than the integrated value over the selected crack-front segment.

The corresponding static J-integral is written as

$$J^{(s)} = \frac{1}{\delta A}\int_{\Omega}\left(\sigma_{ij}\frac{\partial u_j}{\partial x_1'} - W\delta_{1i}\right)\frac{\partial q}{\partial x_i'}\,d\Omega, \tag{26}$$

The three-dimensional weighting function $q(\mathbf{x})$ is defined using the same interpolation form as Eq. (21). In the three-dimensional implementation, however, the nodal value $q_i^{\mathrm{L}}$ is defined by combining the radial selection function in the $x'y'$-plane and the window function in the crack-front direction as

$$q_i^{\mathrm{L}} = q_i^R \cdot q_i^W. \tag{27}$$

Here, $q_i^R$ is the same radial selection function as that used in the two-dimensional implementation, defined by Eq. (23), and $q_i^W$ is defined as

$$q_i^W = \begin{cases} 1 & \text{if } |z_i' - z_0'| \leq \dfrac{1}{2}W_J, \\ 0 & \text{if } |z_i' - z_0'| > \dfrac{1}{2}W_J. \end{cases} \tag{28}$$

Here, $W_J$ defines the width of the integration domain in the crack-front direction. The coordinate $z_0'$ denotes the local coordinate of the evaluation point on the crack front, whereas $z_i'$ denotes the local coordinate of node $i$ in the crack-front direction. Thus, $q_i^W$ selects the nodal values within a width $W_J$ centred at the evaluation point on the crack front.

As in the two-dimensional implementation, $R_J$ is set to $1.5h_{\mathrm{L}}$. In addition, based on preliminary investigations [13], $W_J$ is set to $h_{\mathrm{L}(z')}$, where $h_{\mathrm{L}(z')}$ is the local element size in the crack-front direction. Thus, the three-dimensional effective integration domain is obtained by extending the $x'y'$-plane domain defined by $q_i^R$ over a width $W_J$ centred at the evaluation point on the crack front, as illustrated in Fig. 5.

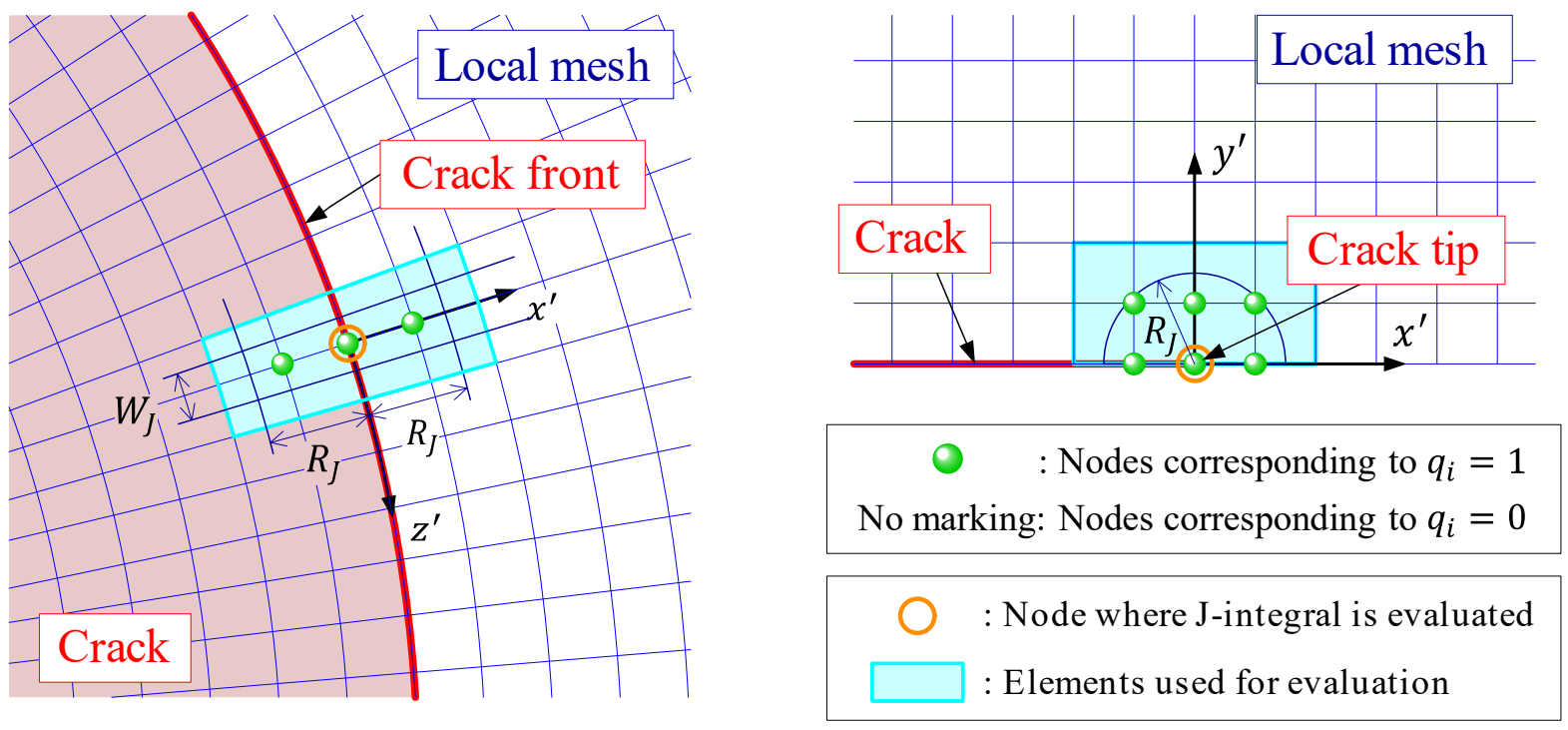

**Fig. 5.** Integration domain for J-integral-based evaluation of the stress intensity factor.

### *2.3.2. Evaluation of local stress*

The local stress is another fracture quantity considered in the present study and is associated with the local fracture stress criterion for brittle crack propagation and arrest in steel structures [10,27,50–52]. As discussed in Section 1, its accurate numerical evaluation is challenging within conventional finite-element frameworks because of the severe scale-gap problem between the highly refined discretisation required near the crack tip/front and the overall structural domain. In the s-method framework, this difficulty is alleviated by superposing a fine local mesh only in the vicinity of the crack tip/front while retaining a comparatively coarse global mesh for the structural-scale domain.

According to the local fracture stress criterion, the tensile stress at a characteristic distance $x_c$ ahead of the crack tip/front in the propagation direction ($x'$) remains approximately constant during crack propagation. The fracture condition is written as

$$\sigma_{yy}\big|_{x'=x_{\rm c}} = \sigma_{\rm f}, \tag{29}$$

where $\sigma_{\rm f}$ is the local fracture stress, which can be interpreted as an intrinsic material resistance against cleavage crack propagation in steel [52,53].

In the proposed strategy, the local stress $\sigma_{yy}\big|_{x'=x_{\rm c}}$ is evaluated from the local mesh quantities. In two-dimensional problems, the evaluation point is the crack tip, and the characteristic point is defined as $(x', y') = (x_{\rm c}, 0)$ in the local coordinate system. In three-dimensional problems, the same evaluation is performed at each selected point on the crack front; for each crack-front point, the local coordinate origin is placed at that point, and the characteristic point is defined as $(x', y', z') = (x_{\rm c}, 0,0)$. For both cases, the characteristic point is denoted here by $\mathbf{x}_{\rm c}$, and the tensile stress is interpolated from nodal values using the local Lagrange basis functions $\phi_i^{\rm L}(\mathbf{x})$ as

$$\sigma_{yy}\big|_{x'=x_{\rm c}} = \sum_i \phi_i^{\rm L}(\mathbf{x}_{\rm c})\sigma_{yy}(\mathbf{x}_i), \tag{30}$$

where $\mathbf{x}_{\rm c} = (x_{\rm c}, 0)$ for two-dimensional problems and $\mathbf{x}_{\rm c} = (x_{\rm c}, 0,0)$ for three-dimensional problems. The nodal tensile stress $\sigma_{yy}(\mathbf{x}_i)$ is calculated from the $y'$-component of the nodal external force $\left(f_y^{\rm L}\right)_i$ as

$$\sigma_{yy}(\mathbf{x}_i) = \frac{\left(f_y^{\rm L}\right)_i}{\int_{\Gamma_{\rm lg}\cup\Gamma_{\rm cr}} \phi_i^{\rm L}(\mathbf{x})\, d\Gamma}. \tag{31}$$

Here, $d\Gamma$ denotes the line measure in two-dimensional problems and the surface measure in three-dimensional problems. Note that smoothing techniques for displacement or stress distributions along the crack front are often employed in three-dimensional analyses [54]. In the present study, however, $\sigma_{yy}(\mathbf{x}_i)$ is evaluated directly using Eq. (31), without introducing additional smoothing, so that the accuracy of the proposed discretisation itself can be assessed directly.

*2.4. Limitation of the conventional Lagrange-based s-method*

As introduced in Section 2.1, the conventional s-method employs Lagrange basis functions for the discretisation of both the global and local domains. Although this choice is straightforward, it leads to a practical difficulty in the numerical integration required for assembling the global–local coupling terms (Eqs. (A15)–(A16) and (A20)–(A21)), thereby significantly reducing the efficiency of matrix generation. This difficulty is caused by the inter-element continuity characteristics of the basis functions used in the global approximation across element boundaries. Accordingly, Section 2.4.1 briefly summarises the relevant continuity characteristics, and Section 2.4.2 explains how they affect the numerical integration of the coupling matrices in the conventional s-method.

*2.4.1. Lagrange basis functions*

To clarify the source of the integration bottleneck, we briefly review the inter-element continuity of Lagrange basis functions.

In the standard finite element formulations, Lagrange basis functions are typically constructed on a parent element and transferred to the physical element through the isoparametric mapping.

Consider a one-dimensional parent element with $\hat{\xi} \in [-1,1]$. The $p$-th order Lagrange basis function associated with node $i$ is given by

$$\phi_{i,p}(\hat{\xi}) = \prod_{j=1, j\neq i}^{p+1} \frac{\hat{\xi} - \hat{\xi}_j}{\hat{\xi}_i - \hat{\xi}_j}, \tag{32}$$

where $i = 1, 2, \cdots p+1$, and $\hat{\xi}_i$ is the parent coordinate of the node $i$ (with $\hat{\xi}_1 = -1$, $\hat{\xi}_{p+1} = 1$). The basis function at a physical point $x$ is evaluated by composition as $\phi_{i,p}(x) = \phi_{i,p}\left(\hat{\xi}(x)\right)$, where $\hat{\xi}(x)$ is obtained from the inverse isoparametric mapping.

Fig. 6 illustrates representative linear and quadratic Lagrange basis functions assembled over adjacent elements, together with their first derivatives. As shown in Fig. 6(a), the basis functions remain polynomial and smooth within each individual element. However, Lagrange discretisation generally provides only $C^0$-continuity across element boundaries, irrespective of the polynomial order. Consequently, although the basis functions themselves are continuous, their derivatives are discontinuous across the element boundaries, as shown in Fig. 6(b). This derivative discontinuity is the key reason why the coupling-term integrands in the conventional s-method may become discontinuous over a local element when that local element overlaps multiple global elements, thereby motivating the specialised integration treatment described in Section 2.4.2.

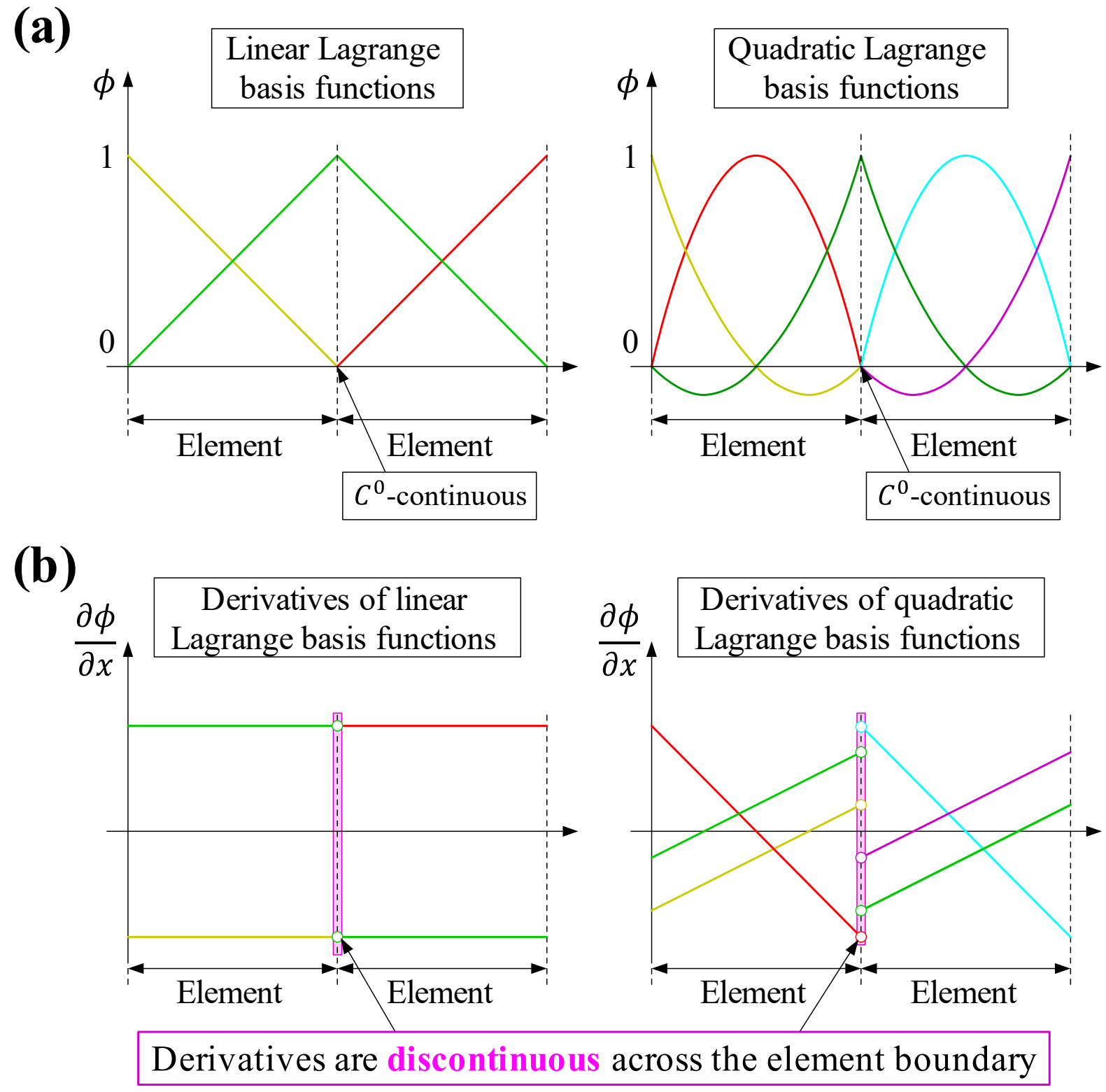

**Fig. 6.** Representative linear and quadratic Lagrange basis functions assembled over adjacent elements: (a) basis functions and (b) their first derivatives.

### *2.4.2. Integration bottleneck in the coupling matrices*

The fact that the Lagrange basis functions provide only $C^0$-continuity across element boundaries becomes problematic when assembling the coupling stiffness and mass matrices. For the standard terms $\mathbf{K}^{\mathrm{G}}$, $\mathbf{K}^{\mathrm{L}}$, $\mathbf{M}^{\mathrm{G}}$, and $\mathbf{M}^{\mathrm{L}}$ (Eqs. (A13)–(A14) and (A18)–(A19)), the integrands have the same structure as in the standard FEM; therefore, the standard Gauss quadrature can be applied efficiently. In contrast, the coupling matrices $\mathbf{K}^{\mathrm{GL}}$, $\mathbf{K}^{\mathrm{LG}}$, $\mathbf{M}^{\mathrm{GL}}$, and $\mathbf{M}^{\mathrm{LG}}$ (Eqs. (A15)–(A16) and (A20)–(A21)) are defined over the local domain $\Omega^{\mathrm{L}}$, and it is natural to perform the integration elementwise on the local mesh. Their integrands involve not only the local basis functions or their derivatives related to the local mesh (e.g., $\boldsymbol{\phi}^{\mathrm{L}}$, $\mathbf{B}_{\phi}^{\mathrm{L}}$) but also the corresponding global quantities (e.g., $\boldsymbol{\phi}^{\mathrm{G}}$, $\mathbf{B}_{\phi}^{\mathrm{G}}$). Therefore, when a local element overlaps multiple global elements, the global quantities appearing in the integrand are evaluated across a global element boundary within that local element.

Fig. 7 schematically illustrates this situation using a representative coupling integrand involving basis-function derivatives. As shown in the figure, the derivatives of the local basis functions remain continuous within the local element, whereas the derivatives of the global basis functions are discontinuous at the global element boundary because Lagrange basis functions provide only $C^0$-continuity across element boundaries. Consequently, the discontinuity of the coupling integrand within the local element is caused by the discontinuity of the global basis-function derivatives, rather than by the local ones. Owing to this discontinuity, directly applying a standard fixed-order Gauss quadrature over the local element may not provide sufficient accuracy.

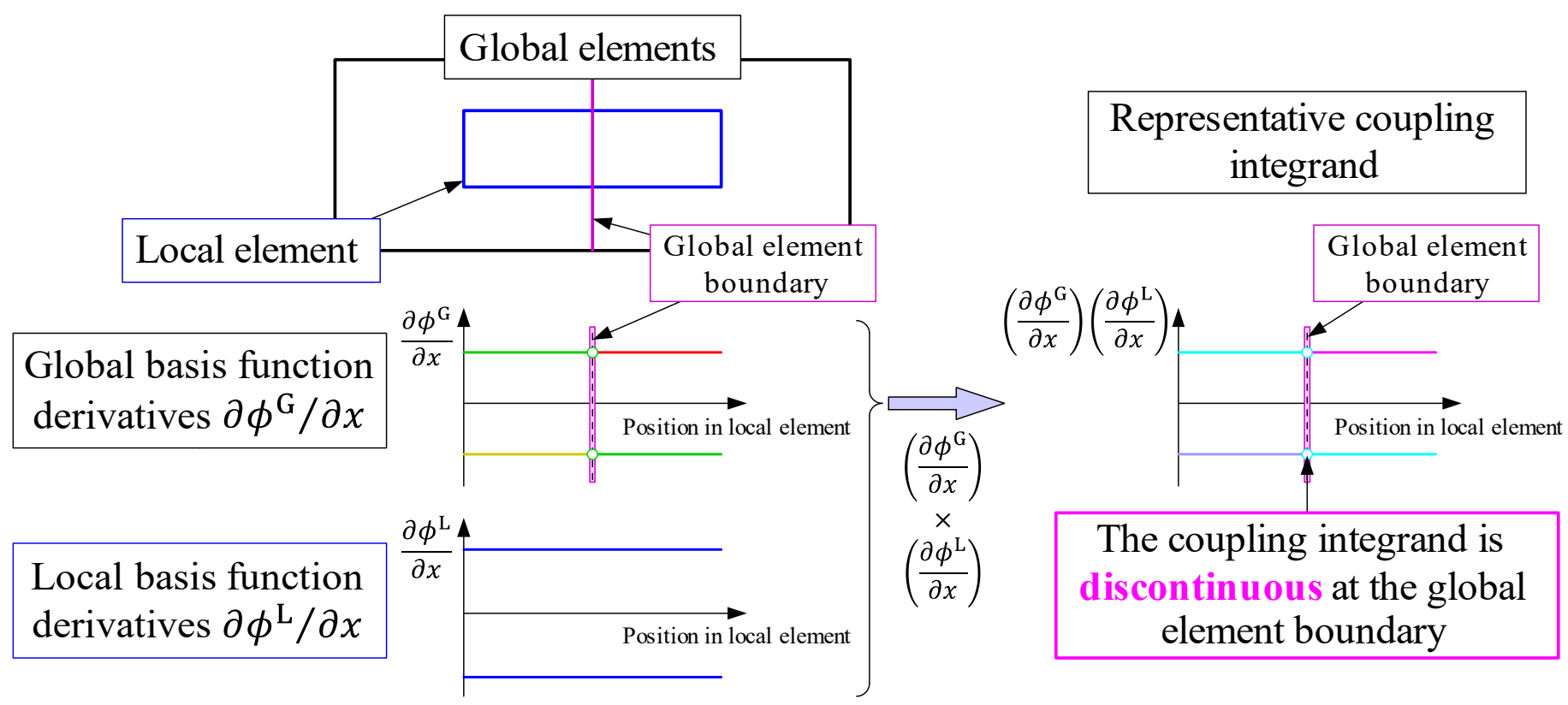

**Fig. 7.** Schematic illustration of a discontinuous coupling integrand within a local element in the conventional Lagrange-based s-method.

To address this issue, previous implementations typically employ recursive subdivision [13,19], as schematically shown in Fig. 8. Local elements intersected by global element boundaries are recursively subdivided into smaller integration sub-domains. In the sub-domains that do not contain a global element boundary, standard Gauss quadrature can be applied accurately. In the remaining sub-domains containing the boundary, further subdivision is performed so that their size becomes sufficiently small and the resulting integration error is reduced to an acceptable level. Although the conventional s-method reduces the total degrees of freedom by restricting mesh refinement to $\Omega^{L}$, the recursive subdivision required for accurate evaluation of the coupling terms can greatly increase the number of integration points. As a result, the associated matrix-assembly cost may offset this advantage and even become the dominant computational expense.

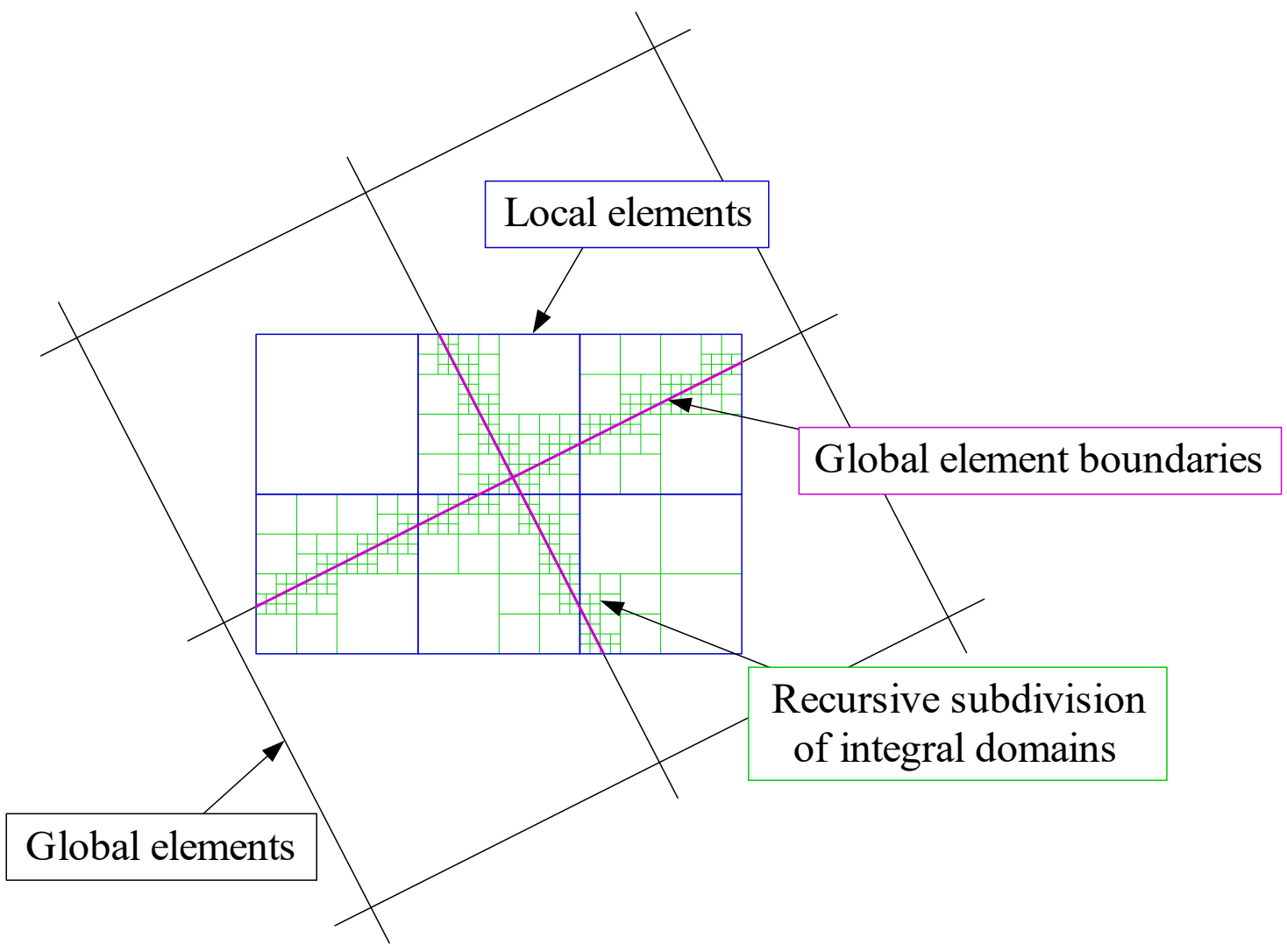

**Fig. 8.** Schematic illustration of recursive subdivision for coupling integration in the conventional s-method.

## 3. Hybrid s-version of isogeometric analysis for crack propagation analysis

As discussed in Section 2.4, in the conventional Lagrange-based s-method, the primary computational bottleneck arises from the numerical integration required for assembling the global–local coupling matrices, owing to the limited inter-element continuity of the global approximation across element boundaries. To address this issue [12,13], the present study proposes a hybrid s-version of isogeometric analysis (hS-IGA), in which the global and local meshes are assigned different approximation spaces. Rather than introducing IGA into both domains as in S-IGA [26], the proposed formulation employs B-spline basis functions only in the global mesh, while retaining Lagrange basis functions in the local mesh. Section 3.1 presents the rationale for this hybrid formulation. Section 3.2 then describes the formulation of the proposed hybrid s-version of isogeometric analysis (hS-IGA), with emphasis on the different basis functions adopted in the global and local meshes. Finally, Section 3.3 introduces the crack representation employed in the present framework.

### *3.1. Rationale for the hybrid formulation*

This section presents the rationale for the proposed hybrid formulation. Rather than discretising both the global and local domains using B-spline basis functions, the present study adopts a hybrid discretisation in which B-spline basis functions are introduced only in the global mesh, while the local mesh is discretised using Lagrange basis functions. In this paper, knot spans in the IGA discretisation are treated as elements and are referred to as IGA (B-spline) elements. The rationale for the hybrid formulation is discussed in terms of (i) the continuity issue in the conventional s-method, (ii) crack representation in the local mesh, and (iii) post-processing procedures for fracture quantity evaluation.

*(i) Continuity issue in the conventional s-method*

In the conventional s-method, the essential continuity issue arises on the global-mesh side. As discussed in Section 2.4, the global-local coupling matrices are integrated over the local domain, but their integrands include not only the local basis functions and their derivatives but also the corresponding global quantities. When a local element overlaps multiple global elements, the global basis functions are therefore evaluated across global element boundaries within that local element. Because standard Lagrange basis functions provide only $C^0$-continuity across element boundaries, their derivatives are discontinuous there, which causes the coupling integrands to become discontinuous and gives rise to the coupling-integration bottleneck. By contrast, the local basis functions and their derivatives remain smooth within each local element, and thus the local mesh itself is not the source of this difficulty. This point indicates that the bottleneck of the conventional formulation originates from the continuity characteristics of the global approximation, and suggests that the improvement should be introduced on the global-mesh side.

*(ii) Crack representation in the local mesh*

As described in Section 2.2, a dynamically propagating crack is represented in the present study by boundary conditions imposed on the symmetry plane $y' = 0$. The rationale for using a standard Lagrange discretisation in the local mesh is explained below using a representative two-dimensional straight-crack example. This example also corresponds to the local $x'y'$-section at an arbitrary evaluation point on the crack front in the three-dimensional problem. Therefore, the following discussion applies to both the two-dimensional and three-dimensional problems.

For a straight Mode I crack propagating dynamically in a linear elastic solid, the $y'$-direction displacement $u_y$ on $y' = 0$ in the vicinity of the crack tip is known to be expressed as follows [48,55,56]:

$$u_y(x', y' = 0) = \begin{cases} 0 & (x' \geq 0), \\ C(V)\, K_{\mathrm{I}}^{(\mathrm{d})}(V)\sqrt{-x'} & (x' < 0), \end{cases} \tag{33}$$

where $C(V)$ is a coefficient dependent on the crack velocity $V$, and $K_{\mathrm{I}}^{(\mathrm{d})}(V)$ denotes the dynamic stress intensity factor. As indicated by the above expression and schematically illustrated in Fig. 9(a), the exact displacement field on $y' = 0$ is $C^0$-continuous but not $C^1$-continuous at the crack tip ($x' = 0$). This indicates that the approximation adopted in the local mesh must be able to represent the crack-tip location naturally as a point at which the displacement field is $C^0$-continuous but not $C^1$-continuous, while separating the crack surface side ($x' < 0$) from the ligament side ($x' \geq 0$).

In the present study, as described in Section 2.2, dynamic crack propagation is modelled on the basis of the nodal force release technique. In this framework, the crack-tip location moves successively along the local element boundaries as the analysis proceeds from one step to the next. Therefore, the approximation used in the local mesh must be able to represent such a $C^0$-continuous but not $C^1$-continuous displacement field naturally at each updated crack-tip location.

From this viewpoint, a standard finite element approximation based on Lagrange basis functions is well suited to the local crack domain, because Lagrange basis functions provide $C^0$-continuity across element boundaries, but not $C^1$-continuity. As illustrated in Fig. 9(b), this enables the crack tip to be represented naturally at an element boundary as a point that is $C^0$-continuous but not $C^1$-continuous. By contrast, B-spline basis functions used in IGA generally satisfy $C^1$- or higher-order continuity across element boundaries and therefore do not straightforwardly represent the crack tip in the local mesh unless special treatment is introduced, as illustrated in Fig. 9(c). These considerations suggest that, for the purpose of representing a dynamically propagating crack tip in the local mesh, a standard Lagrange discretisation is more appropriate than a B-spline-based IGA discretisation.

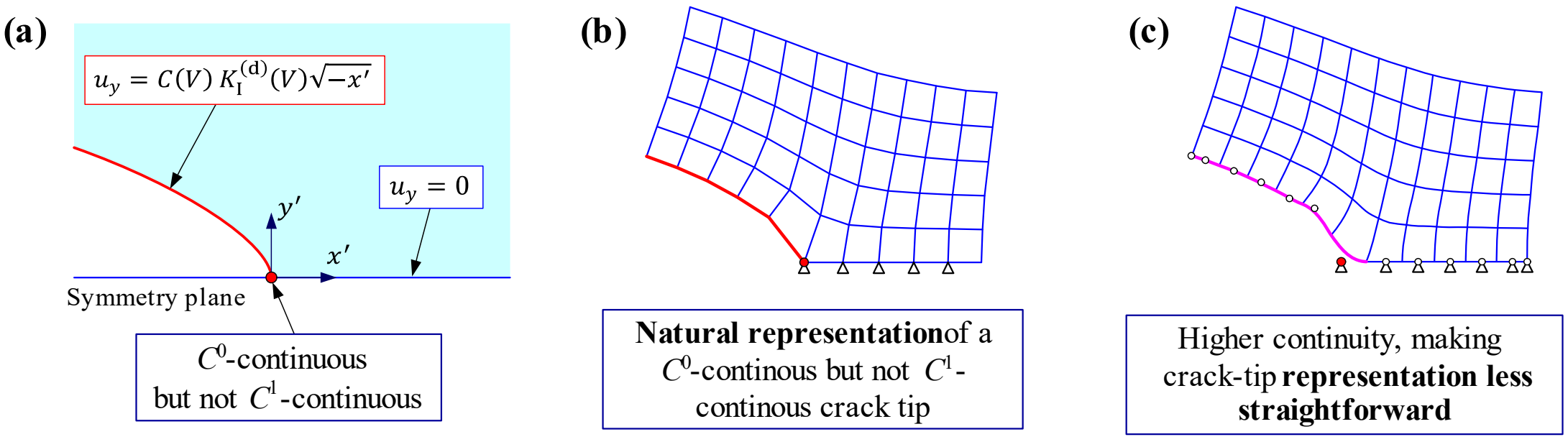

**Fig. 9.** Schematic illustration of crack-tip representation in the local mesh: (a) exact displacement field $u_y$ on $y' = 0$; (b) natural crack-tip representation by Lagrange basis functions; and (c) non-straightforward crack-tip representation by B-spline basis functions due to their higher inter-element continuity.

*(iii) Post-processing procedures for fracture quantity evaluation*

In addition to crack representation, the use of IGA in the local mesh would also complicate the post-processing procedures used to evaluate fracture quantities. In the present study, as described in Section 2.3, the stress intensity factor is evaluated by a domain-integral formulation, whereas the local stress is evaluated using a nodal-force-based procedure. These procedures are naturally implemented when the local mesh is discretised using standard Lagrange finite elements, because the crack tip/front and the crack-propagation direction can be represented straightforwardly by a structured element layout. If IGA is adopted in the local mesh, however, additional treatment becomes necessary in defining and updating the crack tip/front region and related quantities used for post-processing, particularly as the crack advances. This added complexity does not provide a clear advantage for the present purpose. Therefore, from the viewpoint of fracture quantity evaluation as well, standard Lagrange discretisation is more appropriate for the local mesh.

Based on the above considerations, the proposed formulation introduces IGA only into the global discretisation, while retaining Lagrange discretisation in the local domain. The following sections first present the formulation of the proposed hS-IGA, with emphasis on the different basis functions adopted in the global and local meshes (Section 3.2), and then describe the crack representation procedure in the hS-IGA (Section 3.3).

*3.2. Formulation*

This section presents the formulation of the proposed hS-IGA strategy. While the proposed method retains the superposition framework of the conventional s-method, it adopts different approximation spaces in the global and local meshes. Section 3.2.1 introduces the B-spline basis functions employed in the global discretisation and summarises the continuity properties relevant to the present formulation. Section 3.2.2 then presents the corresponding hS-IGA formulation, including the resulting coupling integrals and the associated coordinate-mapping procedure.

*3.2.1. B-spline basis functions and continuity properties*

IGA commonly employs B-spline basis functions, whose continuity across element boundaries is controlled by the knot multiplicity. In one dimension, a knot vector is a non-decreasing sequence in the parametric domain,

$$\Xi = \{\xi_1, \xi_2, \cdots, \xi_{n+p+1}\}, \tag{34}$$

where $\xi_i \in \mathbb{R}$ is the $i$-th knot, $p$ is the polynomial degree, and $n$ is the number of B-spline basis functions. The knots partition the parameter space into elements (knot spans).

Given $\Xi$, the B-spline basis functions are defined recursively starting with piecewise constants $(p = 0)$

$$N_{i,0}(\xi) = \begin{cases} 1 & \text{if } \xi_i \le \xi < \xi_{i+1}, \\ 0 & \text{otherwise.} \end{cases} \tag{35}$$

For $p = 1,2,3,\cdots$, these functions are defined by

$$N_{i,p}(\xi) = \frac{\xi - \xi_i}{\xi_{i+p} - \xi_i} N_{i,p-1}(\xi) + \frac{\xi_{i+p+1} - \xi}{\xi_{i+p+1} - \xi_{i+1}} N_{i+1,p-1}(\xi), \tag{36}$$

which is the Cox-de Boor recursion formula.

Fig. 10 illustrates the quadratic ($p = 2$) B-spline basis functions over multiple elements together with their first derivatives. As shown in Fig. 10(a), B-spline basis functions possess higher inter-element continuity across element boundaries than standard Lagrange basis functions. In general, a B-spline basis function of degree $p$ is $C^{p-m_i}$-continuous at a knot $\xi_i$ with multiplicity $m_i$. In this study, the interior knots are assigned multiplicity $m_i = 1$, and thus the basis functions are $C^{p-1}$-continuous across element boundaries. Consequently, their first derivatives, shown in Fig. 10(b), are also continuous across the element boundaries.

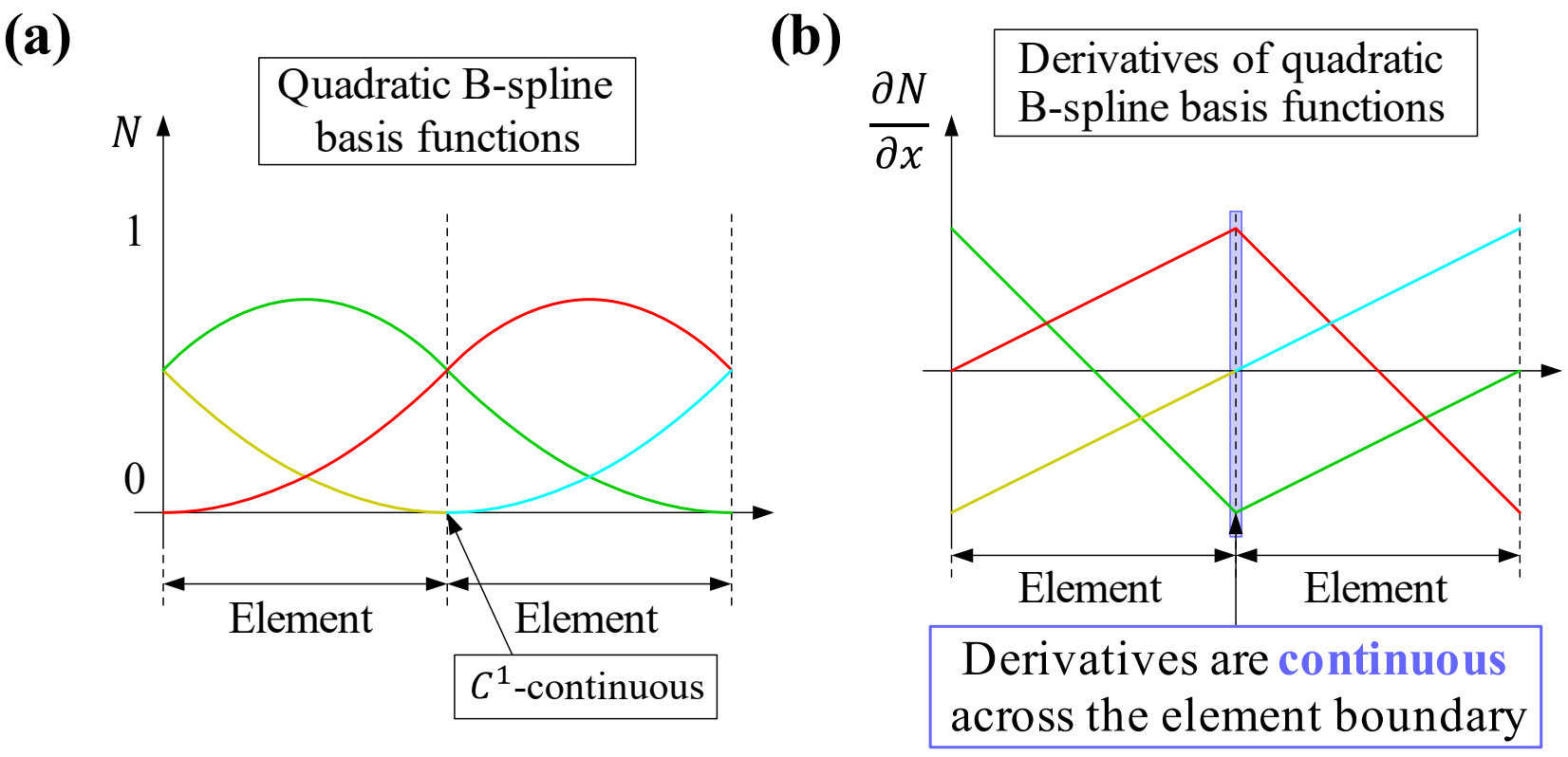

**Fig. 10.** Representative quadratic ($p = 2$) B-spline basis functions and their first derivatives.

In the present study, quadratic B-spline basis functions ($p = 2$) are adopted for the global discretisation. This choice is sufficient for the present purpose, because the stiffness-type quantities involved in the coupling matrices require only first derivatives of the global basis functions. With quadratic B-splines, the basis functions are $C^1$-continuous across element boundaries, and their first derivatives therefore remain $C^0$-continuous.

*3.2.2. Formulation of the proposed hS-IGA strategy*

Because the B-spline basis functions introduced in Section 3.2.1 possess higher inter-element continuity, they can remove the discontinuities in the global–local coupling integrands that arise from the conventional global approximation based on Lagrange basis functions within local elements. Specifically, as illustrated in Fig. 11, when B-spline basis functions are employed in the global mesh, the coupling integrand remains continuous even at a global element boundary located inside a local element, unlike in the conventional s-method shown in Fig. 7. As a result, the coupling terms can be evaluated with sufficient accuracy by standard Gauss quadrature without recursive subdivision, as discussed in Section 3.1(i).

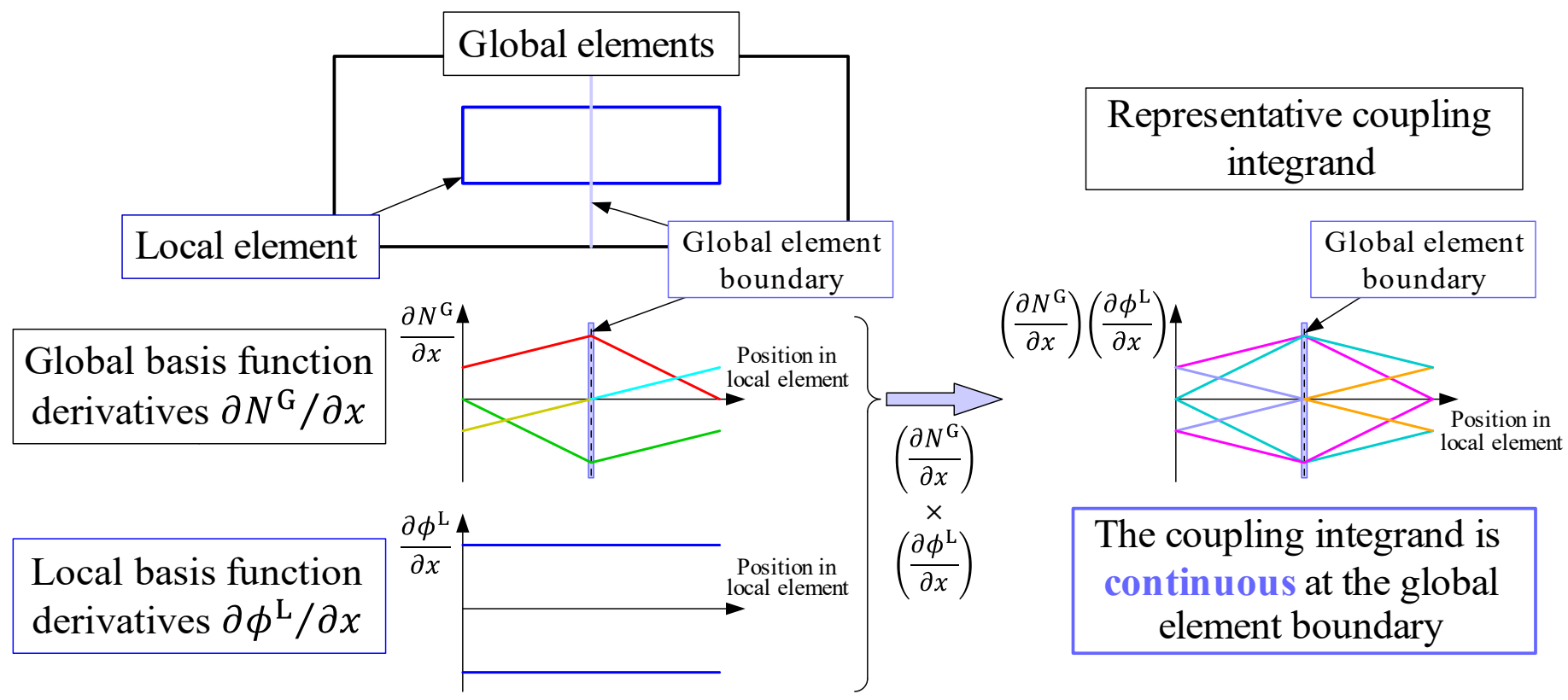

**Fig. 11.** Schematic illustration of continuous coupling integration in the proposed hS-IGA strategy.

By contrast, for the basis functions adopted in the local mesh, the continuity-related issue discussed in Section 3.1(i) does not arise in the same manner. Moreover, as discussed in Sections 3.1(ii) and (iii), a Lagrange-based discretisation is more suitable for the local crack domain from the viewpoints of representing a dynamically propagating crack tip/front and performing the post-processing procedures required for the evaluation of stress intensity factor and local stress.

Based on the above considerations, the proposed hS-IGA strategy retains the superposition framework of the conventional s-method, while adopting different basis functions for discretising the global and local domains. Specifically, B-spline basis functions are introduced in the global mesh, whereas standard Lagrange basis functions are retained in the local mesh.

In the present study, the vector of global B-spline basis functions is denoted by $\mathbf{N}^{\mathrm{G}}$, whereas the vector of local Lagrange basis functions is denoted by $\boldsymbol{\phi}^{\mathrm{L}}$. For the global IGA mesh, the individual components of $\mathbf{N}^{\mathrm{G}}$ are constructed in tensor-product form. In the two-dimensional setting, they are written as

$$N_{i,j}^{\mathrm{G}}(\xi,\eta) = N_i^{\mathrm{G}}(\xi)N_j^{\mathrm{G}}(\eta), \tag{37}$$

whereas in the three-dimensional setting they are written as

$$N_{i,j,k}^{\mathrm{G}}(\xi,\eta,\zeta) = N_i^{\mathrm{G}}(\xi)N_j^{\mathrm{G}}(\eta)N_k^{\mathrm{G}}(\zeta). \tag{38}$$

Similarly, for the local Lagrange mesh, the individual components of $\boldsymbol{\phi}^{\mathrm{L}}$ are constructed in tensor-product form in the parent coordinate system of each local element. In the two-dimensional setting, they are expressed as

$$\phi_{a,b}^{\mathrm{L}}(\hat{\xi},\hat{\eta}) = \phi_a^{\mathrm{L}}(\hat{\xi})\phi_b^{\mathrm{L}}(\hat{\eta}), \tag{39}$$

whereas in the three-dimensional setting they are expressed as

$$\phi_{a,b,c}^{\mathrm{L}}(\hat{\xi},\hat{\eta},\hat{\zeta}) = \phi_a^{\mathrm{L}}(\hat{\xi})\phi_b^{\mathrm{L}}(\hat{\eta})\phi_c^{\mathrm{L}}(\hat{\zeta}). \tag{40}$$

Here, $(\xi,\eta)$ and $(\xi,\eta,\zeta)$, which are collectively denoted by $\boldsymbol{\xi}$, are the parametric coordinates of the global IGA mesh in two and three dimensions, respectively. Similarly, $(\hat{\xi},\hat{\eta})$ and $(\hat{\xi},\hat{\eta},\hat{\zeta})$ are the parent coordinates of the local Lagrange element. The functions $N_i^{\mathrm{G}}$, $N_j^{\mathrm{G}}$ and $N_k^{\mathrm{G}}$ denote the one-dimensional global B-spline basis functions introduced in Section 3.2.1, whereas $\phi_a^{\mathrm{L}}$, $\phi_b^{\mathrm{L}}$ and $\phi_c^{\mathrm{L}}$denote the corresponding one-dimensional linear Lagrange basis functions introduced in Section 2.4.1.

Using these basis functions, the displacement and acceleration fields in the global and local meshes are approximated as

$$\mathbf{u}^{\mathrm{G}}(\mathbf{x}) = \mathbf{N}^{\mathrm{G}}(\mathbf{x})\mathbf{d}^{\mathrm{G}}, \tag{41}$$

$$\ddot{\mathbf{u}}^{\mathrm{G}}(\mathbf{x}) = \mathbf{N}^{\mathrm{G}}(\mathbf{x})\ddot{\mathbf{d}}^{\mathrm{G}}, \tag{42}$$

$$\mathbf{u}^{\mathrm{L}}(\mathbf{x}) = \boldsymbol{\phi}^{\mathrm{L}}(\mathbf{x})\mathbf{d}^{\mathrm{L}}, \tag{43}$$

$$\ddot{\mathbf{u}}^{\mathrm{L}}(\mathbf{x}) = \boldsymbol{\phi}^{\mathrm{L}}(\mathbf{x})\ddot{\mathbf{d}}^{\mathrm{L}}. \tag{44}$$

Although these basis functions are defined in different coordinate systems, the physical coordinate $\mathbf{x}$ is used here for brevity to denote the composed evaluations of the basis functions. The local approximations remain the same as those in the conventional s-method, whereas the global approximations are reformulated using the B-spline basis functions. Accordingly, the corresponding stiffness and mass matrix expressions (Eqs. (A13), (A15)–(A16), (A18), and (A20)–(A21)) take the following forms

$$\mathbf{K}^{\mathrm{G}} = \int_{\Omega^{\mathrm{G}}} \mathbf{B}_{\mathrm{N}}^{\mathrm{G}}(\mathbf{x})^{\mathrm{T}}\mathbf{D}\mathbf{B}_{\mathrm{N}}^{\mathrm{G}}(\mathbf{x})d\Omega, \tag{45}$$

$$\mathbf{K}^{\mathrm{GL}} = \int_{\Omega^{\mathrm{L}}} \mathbf{B}_{\mathrm{N}}^{\mathrm{G}}(\mathbf{x})^{\mathrm{T}}\mathbf{D}\mathbf{B}_{\phi}^{\mathrm{L}}(\mathbf{x})d\Omega, \tag{46}$$

$$\mathbf{K}^{\mathrm{LG}} = \int_{\Omega^{\mathrm{L}}} \mathbf{B}_{\phi}^{\mathrm{L}}(\mathbf{x})^{\mathrm{T}}\mathbf{D}\mathbf{B}_{\mathrm{N}}^{\mathrm{G}}(\mathbf{x})d\Omega \ \left(= \mathbf{K}^{\mathrm{GL}^{\mathrm{T}}}\right), \tag{47}$$

$$\mathbf{M}^{\mathrm{G}} = \int_{\Omega^{\mathrm{G}}} \rho\mathbf{N}^{\mathrm{G}}(\mathbf{x})^{\mathrm{T}}\mathbf{N}^{\mathrm{G}}(\mathbf{x})d\Omega, \tag{48}$$

$$\mathbf{M}^{\mathrm{GL}} = \int_{\Omega^{\mathrm{L}}} \rho\mathbf{N}^{\mathrm{G}}(\mathbf{x})^{\mathrm{T}}\boldsymbol{\phi}^{\mathrm{L}}(\mathbf{x})d\Omega, \tag{49}$$

$$\mathbf{M}^{\mathrm{LG}} = \int_{\Omega^{\mathrm{L}}} \rho\boldsymbol{\phi}^{\mathrm{L}}(\mathbf{x})^{\mathrm{T}}\mathbf{N}^{\mathrm{G}}(\mathbf{x})d\Omega \ \left(= \mathbf{M}^{\mathrm{GL}^{\mathrm{T}}}\right). \tag{50}$$

Here, $\mathbf{B}_{\mathrm{N}}^{\mathrm{G}}$ and $\mathbf{B}_{\phi}^{\mathrm{L}}$ are the strain–displacement matrices formed from the derivatives of the global B-spline basis functions and local Lagrange basis functions, respectively. In the present study, standard Gauss quadrature with three points in each parametric direction is employed for both the global and local elements in the evaluation of the stiffness and mass matrices; that is, $3 \times 3$ quadrature is used in two dimensions and $3 \times 3 \times 3$ quadrature in three dimensions. This choice is adopted consistently throughout the verification analyses and corresponds to the use of $p + 1$ quadrature points in each parametric direction for the quadratic global B-spline basis [21,57].

Since the coupling integrations in Eqs. (46), (47), (49) and (50) are performed on the local elements, the global basis functions must be evaluated at the physical locations of the local integration points. This requires a coordinate-mapping procedure to determine the corresponding parametric coordinates within the identified global IGA element. A schematic is shown in Fig. 12, and the procedure is summarised as follows:

(i). Specify the integration points in the parent coordinates of each local element, and evaluate the local quantities $\boldsymbol{\phi}^{\mathrm{L}}$ and $\mathbf{B}_{\phi}^{\mathrm{L}}$ accordingly.

(ii). Map these local integration points to physical coordinates, and identify the global IGA element containing each point.

(iii). For each mapped point, compute the corresponding parametric coordinates $\boldsymbol{\xi}$ in the identified global IGA element using an iterative algorithm, such as the Newton–Raphson method, and evaluate $\mathbf{N}^{\mathrm{G}}$ and $\mathbf{B}_{\mathrm{N}}^{\mathrm{G}}$ at $\boldsymbol{\xi}$.

(iv). Substitute the quantities obtained in steps (i)–(iii) into Eqs. (46), (47), (49) and (50) to assemble the coupling matrices.

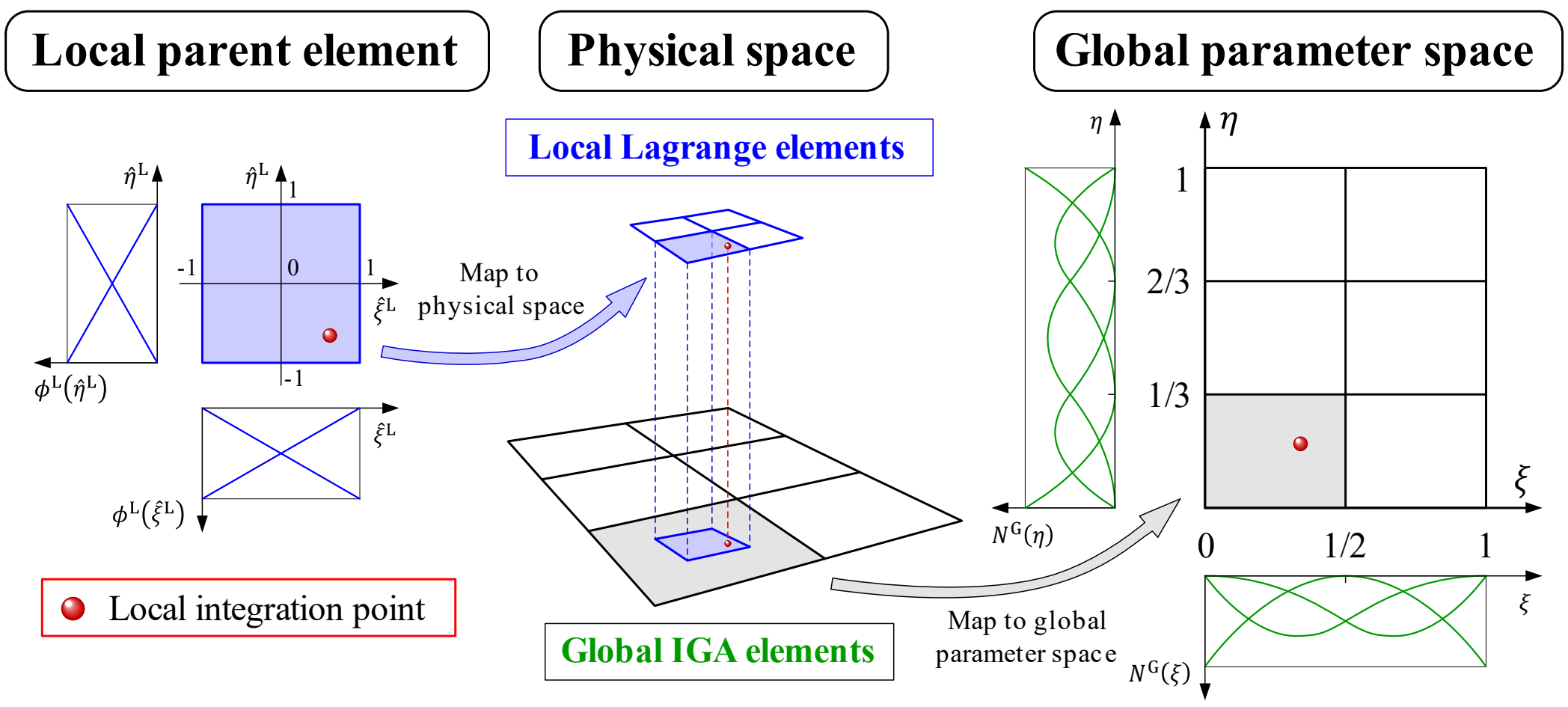

**Fig. 12.** Schematic illustration of the mapping from the physical coordinates of a local integration point to the parametric coordinates of the corresponding global IGA element.

### *3.3. Crack representation*

The crack representation procedure introduced in Section 2.2 is retained in the proposed hS-IGA strategy. That is, the crack is represented by imposing Dirichlet and Neumann boundary conditions along the symmetry plane ($y' = 0$). Because the local domain remains discretised using standard Lagrange basis functions, the crack-tip/front representation, nodal force release, and local mesh update can be applied in the same manner as in the conventional s-method.

Fig. 13 schematically illustrates the crack representation in the proposed hS-IGA strategy using a representative two-dimensional setting. The global mesh is discretised by B-spline basis functions, whereas the local crack domain is discretised by Lagrange basis functions so that the crack surface and ligament can be represented explicitly by the boundary conditions introduced in Section 2.2. The final displacement field in the local domain is obtained by superposing the global and local displacement components. Although Fig. 13 is shown for a two-dimensional problem for clarity, the same representation concept applies to the three-dimensional problems by considering the corresponding local $x'y'$-section at each point along the crack front.

Although the local treatment is unchanged, the boundary-condition assignment on the global side must be reinterpreted because the global mesh is discretised using B-spline basis functions. Specifically, the sets $\mathcal{S}_{\mathrm{D}}$ and $\mathcal{S}_{\mathrm{N}}$, originally defined by Eqs. (3) and (4) in the conventional s-method as the sets of nodes on which the Dirichlet and Neumann conditions are imposed, are generalised so that they can also be interpreted as sets of control points in the IGA mesh. Accordingly, the support $\Omega_i$ is redefined as

$$\Omega_i = \begin{cases} \{\mathbf{x} \mid N_i^{\mathrm{G}}(\mathbf{x}) \neq 0\} & \text{for global control points,} \\ \{\mathbf{x} \mid \phi_i^{\mathrm{L}}(\mathbf{x}) \neq 0\} & \text{for local nodes,} \end{cases} \tag{51}$$

where $N_i^{\mathrm{G}}$ and $\phi_i^{\mathrm{L}}$ denote the global B-spline basis function and the local Lagrange basis function corresponding to the $i$-th global control point and the $i$-th local node, respectively.

It is worth emphasising that, for the global IGA elements, boundary enforcement should be interpreted in terms of the support of control points rather than the pointwise location of the control point itself. As illustrated in Fig. 13, although control point $A$ lies in the crack surface part, its support still intersects the ligament part; therefore, the Dirichlet condition is imposed on the corresponding degree of freedom.

The dynamic crack propagation procedure described in Section 2.2 is directly retained in the proposed hS-IGA strategy. Because the local crack domain remains discretised using Lagrange elements, the nodal force release technique and the associated local mesh update can be applied in the same manner as in the conventional s-method (see Figs. 3 and 4).

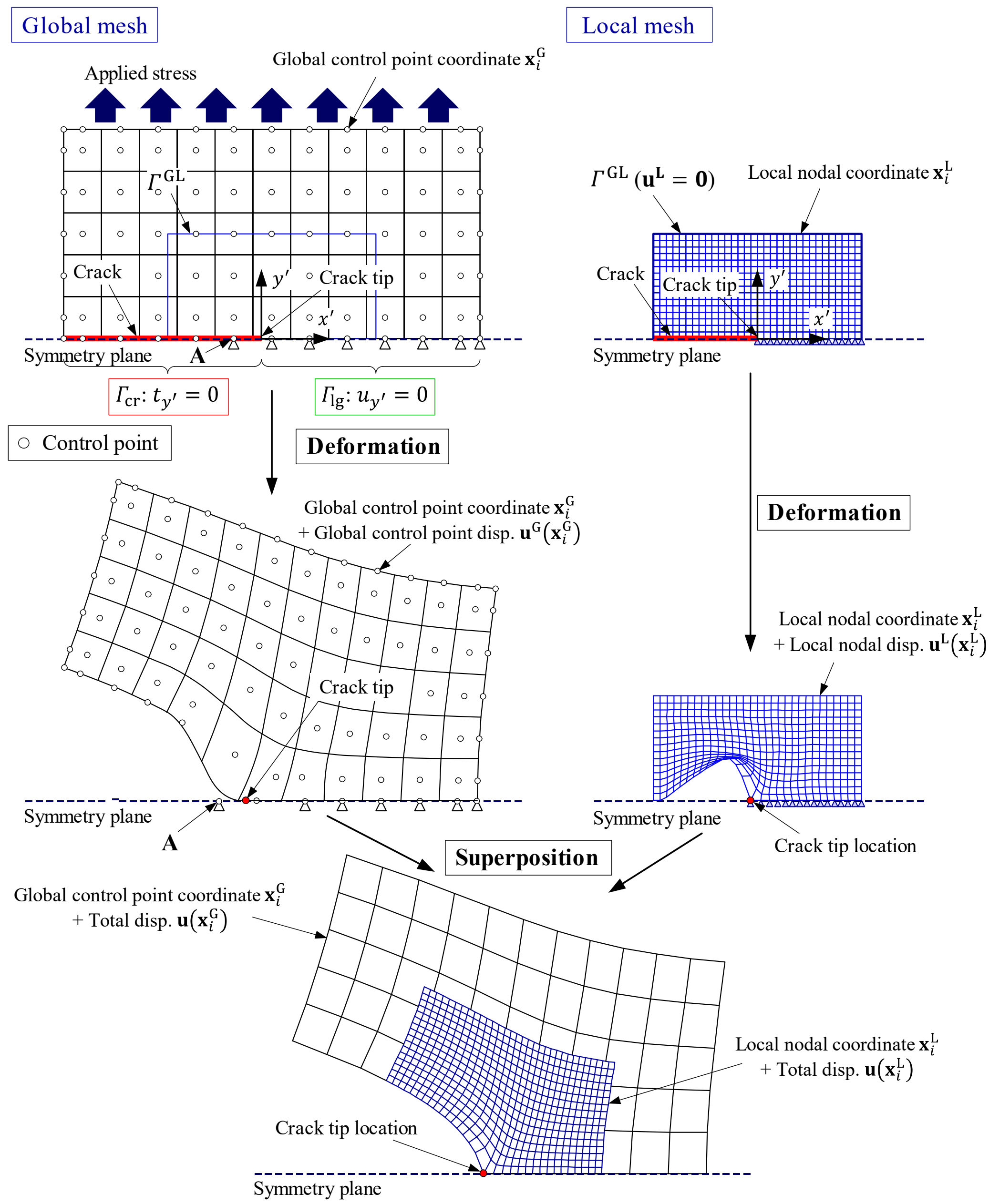

**Fig. 13.** Schematic illustration of crack representation in the proposed hS-IGA strategy using a representative two-dimensional setting.

## 4. Verification and discussion: two-dimensional crack problems

This section systematically verifies the proposed hS-IGA strategy through two-dimensional straight-crack problems. Two benchmark problems are considered: a stationary crack problem and a dynamic crack propagation problem. The stationary crack problem is used to examine the fundamental accuracy of near-tip field evaluation, including the overall displacement field and the local stress ahead of the crack tip. The dynamic crack propagation problem is then used to assess the applicability of the proposed strategy to crack propagation analyses with different crack velocities and to evaluate the dynamic stress intensity factor (DSIF) and the local stress. For these two-dimensional problems, the proposed hS-IGA strategy is directly compared with the standard FEM and the conventional s-method in terms of numerical accuracy, degrees of freedom, and the number of integration points. Through these comparisons, this section clarifies whether the proposed strategy can retain the inherent advantages of the s-method framework while substantially reducing the coupling-integration cost of the conventional s-method. Material properties representative of ferritic steel are adopted: Young's modulus $E = 206$ GPa, Poisson's ratio $\nu = 0.3$, and density $\rho = 7{,}800\ \mathrm{kg/m^3}$.

*4.1. Two-dimensional stationary crack problem*

The first two-dimensional benchmark considers a stationary straight crack. This problem is used to examine the fundamental accuracy of the proposed hS-IGA strategy in near-tip field evaluation under conditions where the exact solutions are available.

*4.1.1. Problem description*

The stationary benchmark considered here is a straight Mode I crack in an infinite plate (Fig. 14(a)). An idealised asymptotic field near a crack tip is adopted, with the static stress intensity factor prescribed as $K_{\mathrm{I}}^{\mathrm{ex(s)}} = 1$. The corresponding Mode I asymptotic displacement field is adopted as the exact solution for the stationary benchmark and is denoted by $\mathbf{u}^{\mathrm{ex(s)}}$ [49]:

$$\mathbf{u}^{\mathrm{ex(s)}}(\mathbf{x}) = \frac{1}{\mu}\sqrt{\frac{r}{2\pi}}\begin{Bmatrix} \cos\frac{\theta}{2}\left(1 - 2\nu + \sin^2\frac{\theta}{2}\right) \\ \sin\frac{\theta}{2}\left(2 - 2\nu - \cos^2\frac{\theta}{2}\right) \end{Bmatrix}, \tag{52}$$

where $\mu$ is the shear modulus and $(r, \theta)$ are the local polar coordinates centred at the crack tip.

The target domain consists of a small region extracted from the infinite plate (Fig. 14(a)). Exploiting symmetry, only the half-domain on one side of the crack plane is analysed (shown in Fig. 14(b)). The width and height of the target domain are set to $W_{\mathrm{G}} = 2$ and $H_{\mathrm{G}} = 1$, respectively, and crack length in the target domain is $a_{\mathrm{G}} = 1$ (i.e., half of $W_{\mathrm{G}}$). To approximate the infinite plate benchmark using this finite target domain, the asymptotic displacement field is prescribed as Dirichlet boundary conditions on the outer boundary of target domain $\Gamma^{\mathrm{G}}$ [12,13]. This treatment enables direct comparison between the numerical and exact solutions.

In addition to the displacement field, the exact tensile stress distribution ahead of the crack tip for the stationary asymptotic field, denoted by $\sigma_{yy}^{\mathrm{ex(s)}}$, is given by [49]

$$\sigma_{yy}^{\mathrm{ex(s)}}(\mathbf{x}) = \frac{1}{\sqrt{2\pi r}}\cos\frac{\theta}{2}\left(1 + \sin\frac{\theta}{2}\sin\frac{3\theta}{2}\right), \tag{53}$$

and is used to verify the accuracy of the local stress evaluation $\sigma_{yy}$ obtained by the proposed strategy.

Since the exact solution is available, the relative $L_2$ error norms for the stationary crack problem can be evaluated by

$$\begin{aligned} e_{L_2} &= \left|\frac{\int_{\Omega^{\mathrm{G}}}\left(\mathbf{u}(\mathbf{x}) - \mathbf{u}^{\mathrm{ex(s)}}(\mathbf{x})\right)^2 d\Omega}{\int_{\Omega^{\mathrm{G}}}\mathbf{u}^{\mathrm{ex(s)}}(\mathbf{x})^2 d\Omega}\right| \\ &= \left|\frac{\int_{\Omega^{\mathrm{G}}\setminus\Omega^{\mathrm{L}}}\left(\mathbf{u}^{\mathrm{G}}(\mathbf{x}) - \mathbf{u}^{\mathrm{ex(s)}}(\mathbf{x})\right)^2 d\Omega + \int_{\Omega^{\mathrm{L}}}\left(\mathbf{u}^{\mathrm{G}}(\mathbf{x}) + \mathbf{u}^{\mathrm{L}}(\mathbf{x}) - \mathbf{u}^{\mathrm{ex(s)}}(\mathbf{x})\right)^2 d\Omega}{\int_{\Omega^{\mathrm{G}}}\mathbf{u}^{\mathrm{ex(s)}}(\mathbf{x})^2 d\Omega}\right|, \end{aligned} \tag{54}$$

where $\mathbf{u}$ and $\mathbf{u}^{\mathrm{ex(s)}}$ denote the numerical and exact displacement fields defined by Eq. (A1) and Eq. (52), respectively.

The integrations over $\Omega^{\mathrm{G}} \setminus \Omega^{\mathrm{L}}$ and $\Omega^{\mathrm{G}}$ are evaluated using the global mesh, whereas the integral over $\Omega^{\mathrm{L}}$ is evaluated using the local mesh. An eighth-order Gauss quadrature rule is employed to approximate these integrals. It is known that the optimal slope of $e_{L_2}$ evaluated by the standard finite element approximation is 1/2 because of the singularity at the crack tip [12,13].

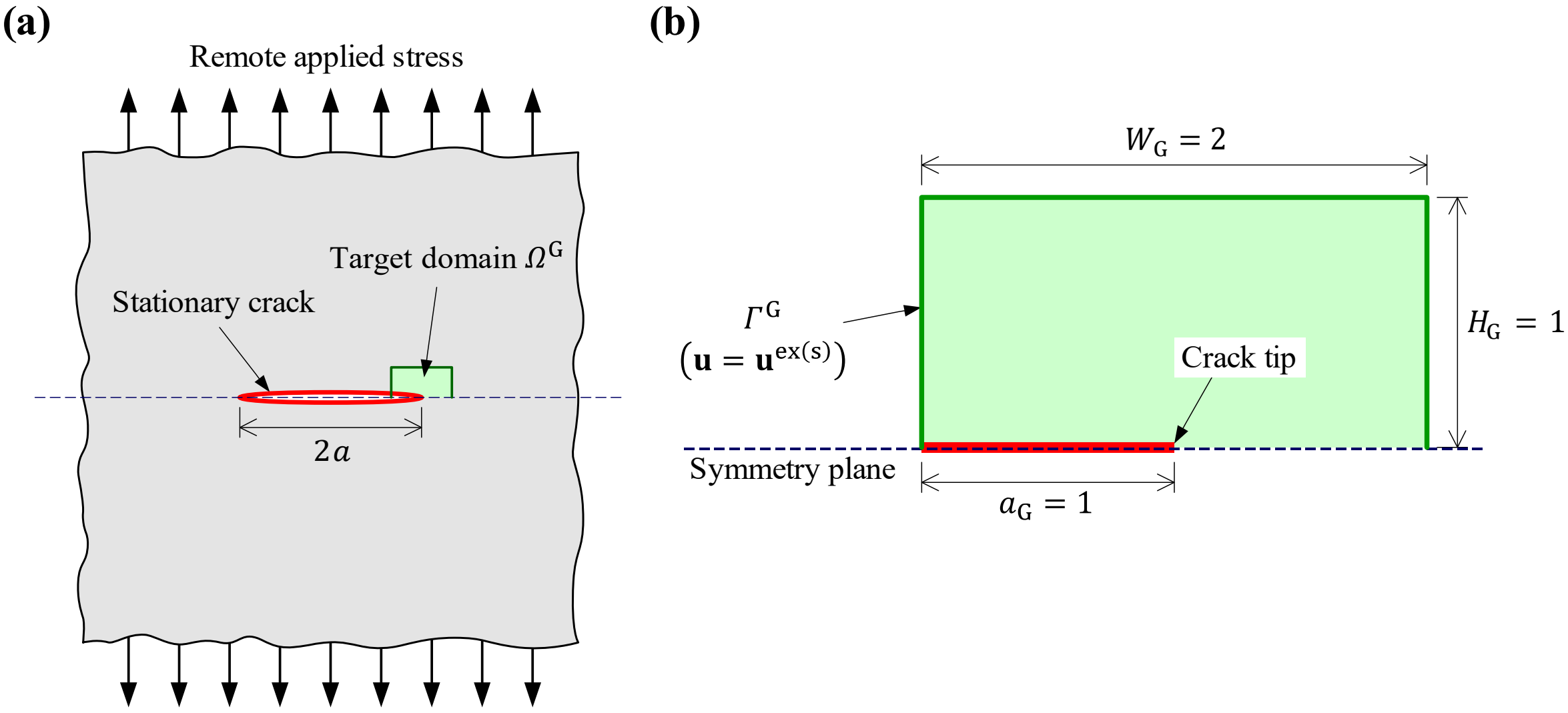

**Fig. 14.** Two-dimensional stationary crack benchmark problem: (a) problem schematic; (b) target domain for the numerical analysis.

*4.1.2. Verification of the two-dimensional stationary crack problem*

The global and local meshes used in the verification, together with the corresponding boundary conditions, are shown in Fig. 15. The global mesh is generated over the target global domain $\Omega^{G}$, and the global element size is denoted by $h_G$. As described in Section 4.1.1, the exact displacement field (Eq. (52)) is imposed as the Dirichlet boundary condition on the outer boundary of the target domain $\Gamma^{G}$.

For the local mesh, the width of the local domain $\Omega^{L}$ in the $x'$-direction (i.e., the crack propagation direction) is defined as $W_L = 1$, and is divided into two parts: the crack surface part ($x' < 0$) with length $a_L = 0.5$, and the crack ligament part ($x' \geq 0$) with length $l_L = 0.5$. The height of the local domain $\Omega^{L}$ is set to $H_L = 0.5$. The size of the local element is denoted by $h_L$.

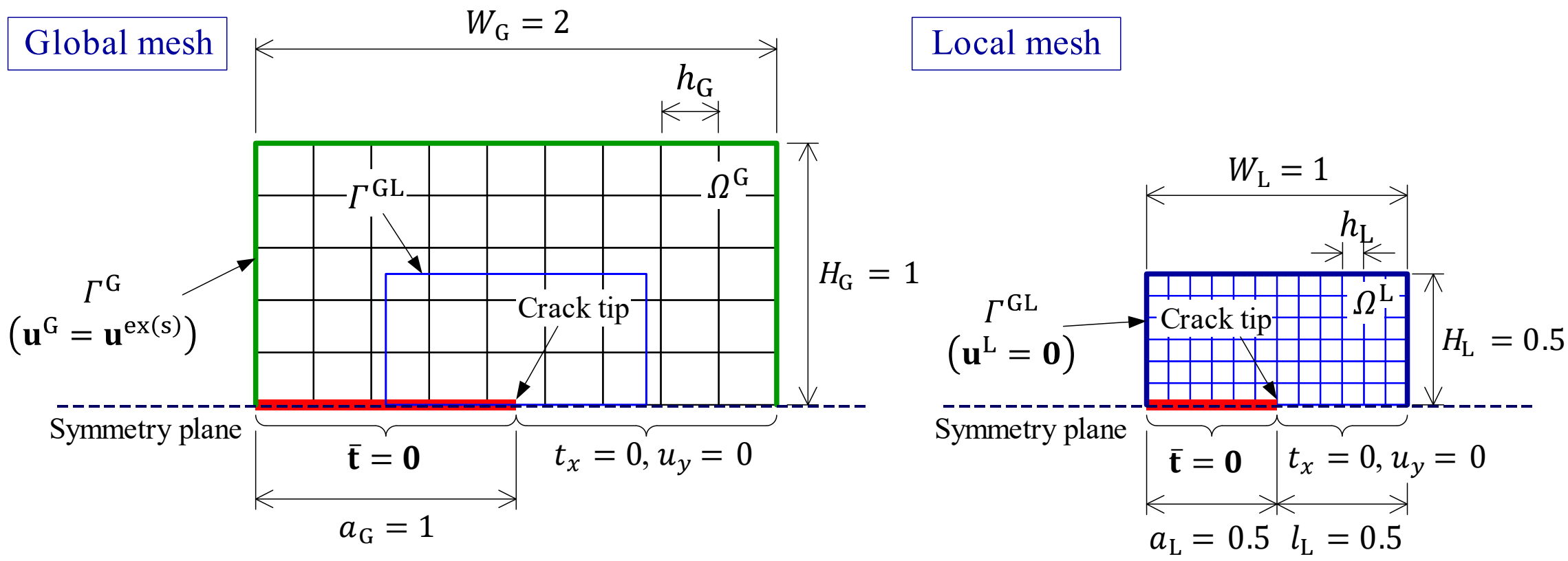

**Fig. 15.** Global and local mesh configurations and associated boundary conditions used in the verification analyses for the two-dimensional stationary crack problem.

Convergence studies of the relative $L_2$ error norm, $e_{L_2}$, defined in Eq. (54), were conducted for the global-to-local element-size ratios $r_{GL} = h_G/h_L = 2,\ 4,\ 6$, and $8$. The convergence results of $e_{L_2}$ versus the degrees of freedom (DOF), together with those of the standard FEM, are summarised in Fig. 16(a). For all examined $r_{GL}$, the proposed hS-IGA strategy yields lower $e_{L_2}$ values than the standard FEM over the examined range of discretisations. Equivalently, to attain a comparable level of accuracy, the proposed hS-IGA strategy requires fewer degrees of freedom than the standard FEM, while maintaining the optimal convergence behaviour.

The proposed hS-IGA strategy is further compared with the conventional s-method for the representative cases $r_{\mathrm{GL}} = 4$ and 8, as shown in Figs. 16(b1) and 16(b2), respectively. In both cases, the hS-IGA results remain comparable to those of the conventional s-method in terms of the relative $L_2$ error norm and the convergence trend. These results indicate that the proposed strategy preserves the fundamental approximation accuracy of the conventional s-method while requiring a comparable number of degrees of freedom to achieve a similar level of accuracy.

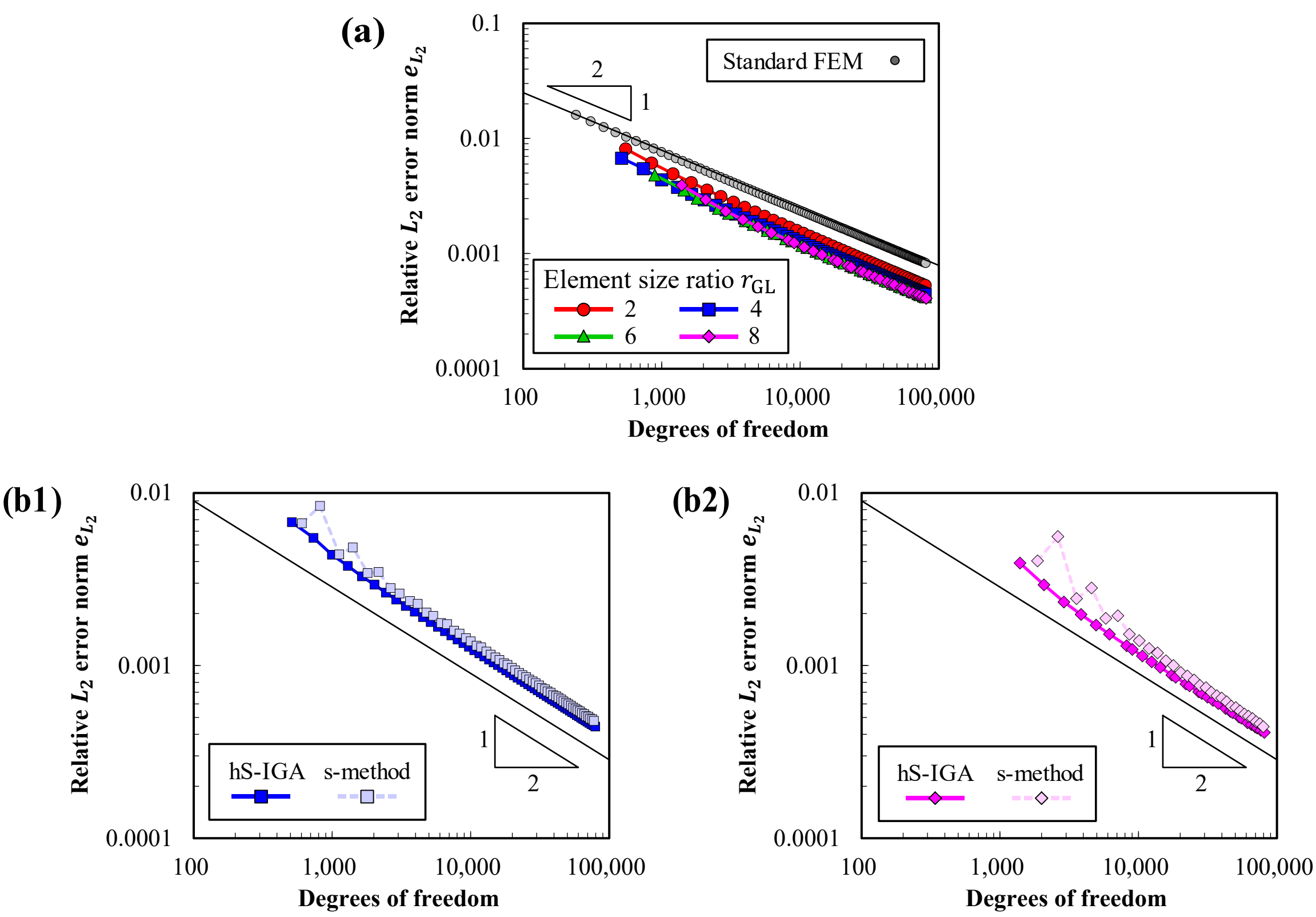

**Fig. 16.** Convergence results of the $L_2$ error norm under fixed $r_{\mathrm{GL}}$: (a) comparison with the standard FEM; and comparison with the conventional Lagrange-based s-method for (b1) $r_{\mathrm{GL}} = 4$, and (b2) $r_{\mathrm{GL}} = 8$.

To clarify the principal computational advantage of the proposed strategy, the total numbers of integration points required by the hS-IGA strategy and the conventional s-method are compared for the representative cases $r_{\mathrm{GL}} = 4$ and 8, corresponding to the results shown in Figs. 16(b1) and 16(b2). The relationship between the relative $L_2$ error norm and the number of integration points is summarised in Fig. 17, with both axes plotted on logarithmic scales. In both cases, the data points for the two methods are distributed approximately along straight lines with similar slopes, indicating a similar scaling relationship between the attained accuracy and the required integration cost. However, the hS-IGA curves are consistently shifted downward relative to those of the conventional s-method. This indicates that, for a comparable level of accuracy, the proposed hS-IGA strategy requires substantially fewer integration points than the conventional s-method.

These results highlight the main bottleneck of the conventional s-method, namely, the extremely large number of integration points required in the coupling integration. To quantify the advantage of the proposed strategy more rigorously, a shared-slope linear regression is performed for the results obtained by the two methods at each $r_{\mathrm{GL}}$. Since the fitted lines exhibit nearly identical slopes, the difference between the two methods can be characterised by the vertical offset between the fitted lines in the log-log plots. This offset corresponds to an approximately constant reduction factor in the number of integration points required to achieve a matched error level. Based on the fitted relations, the required number of integration points is reduced by approximately 98.1% for $r_{\mathrm{GL}} = 4$ and 98.0% for $r_{\mathrm{GL}} = 8$. These results confirm that the proposed hS-IGA strategy effectively resolves the principal integration-cost bottleneck of the conventional s-method while preserving comparable accuracy.

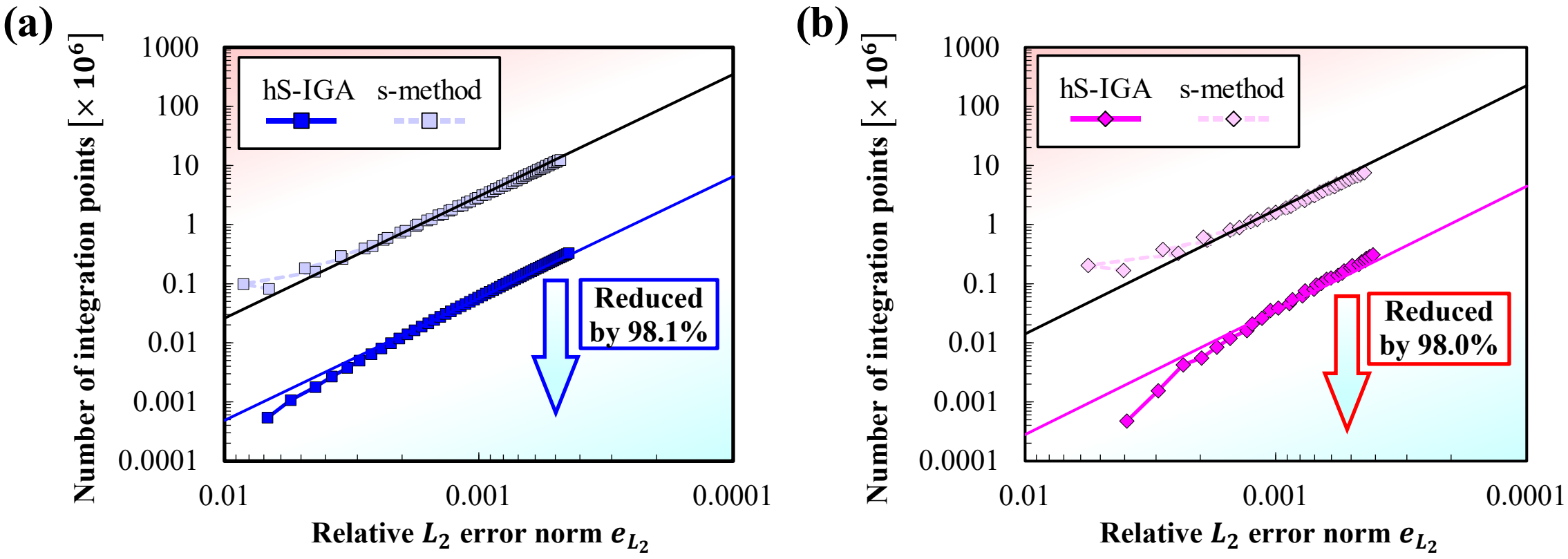

**Fig. 17.** Relationship between the relative $L_2$ error norm and the number of integration points for the hS-IGA and the conventional Lagrange-based s-method under: (a) $r_{\mathrm{GL}} = 4$, (b) $r_{\mathrm{GL}} = 8$.

To investigate the respective influences of $h_{\mathrm{G}}$ and $h_{\mathrm{L}}$ for the proposed hS-IGA strategy, $e_{L_2}$ is evaluated under fixed-$h_{\mathrm{G}}$ and fixed-$h_{\mathrm{L}}$ settings. The resulting relationships between $e_{L_2}$ and the DOF are summarised in Fig. 18. When $h_{\mathrm{G}}$ is fixed, the convergence curves are downward-convex in the low-DOF range, and their slopes gradually stabilise as the DOF increases. In contrast, under fixed-$h_{\mathrm{L}}$settings, the convergence curves exhibit an L-shaped trend. These results indicate that the local-mesh approximation governs the error reduction behaviour, whereas, once the local mesh becomes sufficiently fine relative to the global mesh, further improvement may be limited by the approximation accuracy of the global mesh.

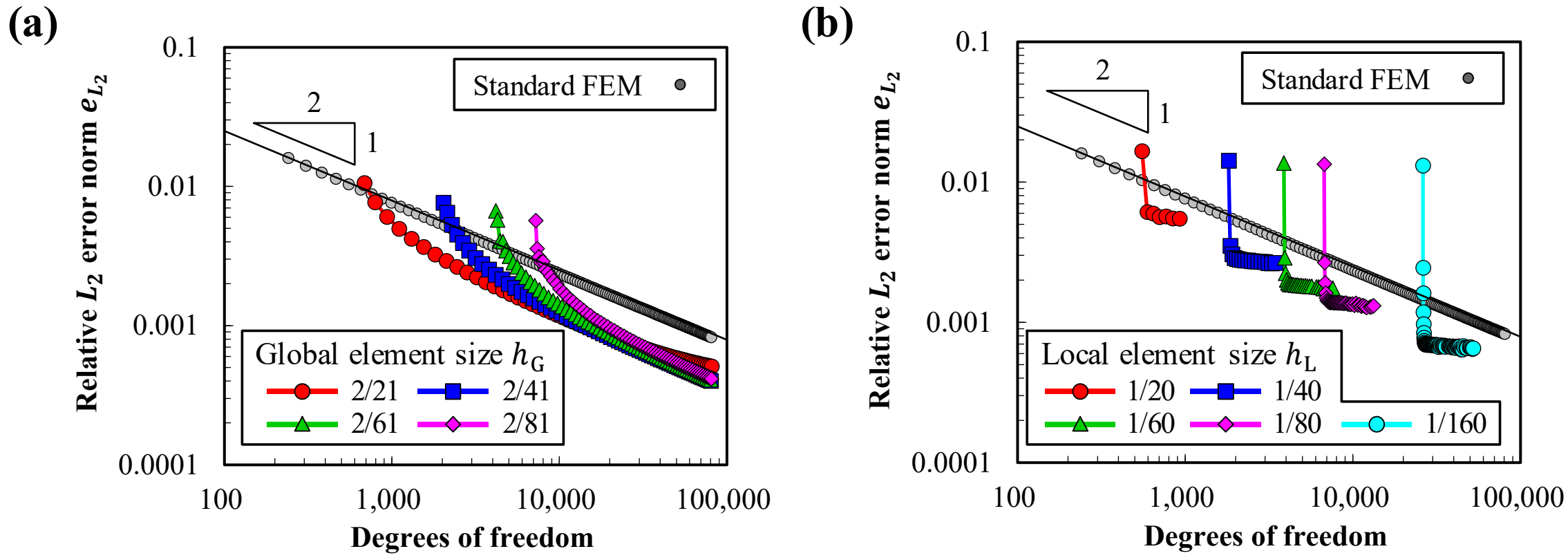

**Fig. 18.** Convergence results of the $L_2$ error norm under fixed global and local element sizes: (a) fixed $h_{\mathrm{G}}$, (b) fixed $h_{\mathrm{L}}$.

Fig. 19 presents four representative meshing conditions together with the evaluated tensile stress fields, $\sigma_{yy}$, as well as their differences from the exact solution $\sigma_{yy}^{\mathrm{ex(s)}}$given in Eq. (53). The results indicate that the accuracy of the simulated near-tip tensile stress field depends more strongly on $h_{\mathrm{L}}$ than on $h_{\mathrm{G}}$. In addition, some disturbances can be observed in the tensile stress field near $\Gamma^{\mathrm{GL}}$, possibly owing to weak discontinuities in the approximated displacement field $\mathbf{u}(\mathbf{x})$ across $\Gamma^{\mathrm{GL}}$. However, these disturbances are effectively reduced by mesh refinement.

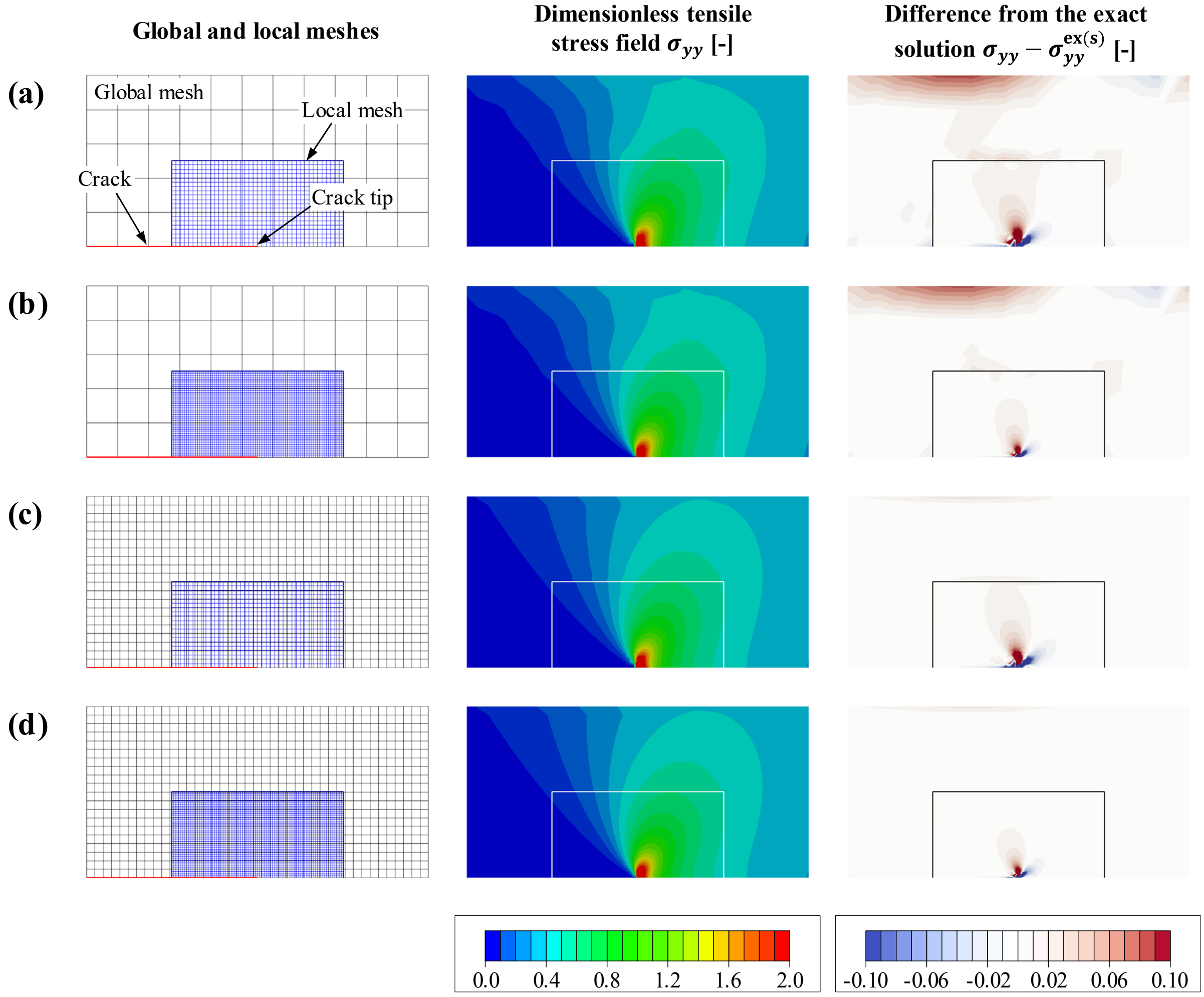

**Fig. 19.** Evaluation of the dimensionless tensile stress field and its difference from the exact solution for four mesh conditions: (a) $h_{\mathrm{G}} = 2/11,\ h_{\mathrm{L}} = 1/40$, (b) $h_{\mathrm{G}} = 2/11,\ h_{\mathrm{L}} = 1/80$, (c) $h_{\mathrm{G}} = 2/41,\ h_{\mathrm{L}} = 1/40$, (d) $h_{\mathrm{G}} = 2/41,\ h_{\mathrm{L}} = 1/80$.

Based on the methodology presented in Section 2.3.2, the local stress $\sigma_{yy}$ in front of the crack tip is evaluated using the nodal forces on the symmetry plane. To verify the accuracy of the proposed strategy, the results are compared with those obtained by the standard FEM and the conventional s-method under the same mesh density near the crack tip. Specifically, two crack-tip mesh resolutions are examined: (a) $1/40$ and (b) $1/80$. For the hS-IGA strategy and the conventional s-method, the local element size is set to $h_{\mathrm{L}} = 1/40$ and $1/80$, with the corresponding global element sizes set to $h_{\mathrm{G}} = 2/21$ and $2/41$, respectively. For the standard FEM, the element size is set to $h_{\mathrm{FE}} = 1/40$ and $1/80$, respectively.

The normalised local stress in front of the crack tip, $\sigma_{yy}/\sigma_{yy}^{\mathrm{ex(s)}}$, obtained by the three methods is shown in Fig. 20. The results indicate that all three methods can accurately evaluate the local stress. The proposed hS-IGA strategy, however, provides slightly higher accuracy than the standard FEM and the conventional s-method when the evaluation point is located close to the crack tip. In addition, Fig. 20(a) shows that, in the conventional s-method, the accuracy of $\sigma_{yy}$ decreases when the evaluation point approaches $\Gamma^{\mathrm{GL}}$. This issue is caused by the dependence of the accuracy of $\sigma_{yy}$ on $h_{\mathrm{G}}$, as reported in our previous work [12]. In contrast, the results obtained using the proposed strategy show no such $h_{\mathrm{G}}$-dependence and instead exhibit a tendency similar to that of the standard FEM. These results confirm that the proposed hS-IGA strategy can accurately reproduce the local stress ahead of the crack tip while avoiding the mesh-interface sensitivity observed in the conventional s-method.

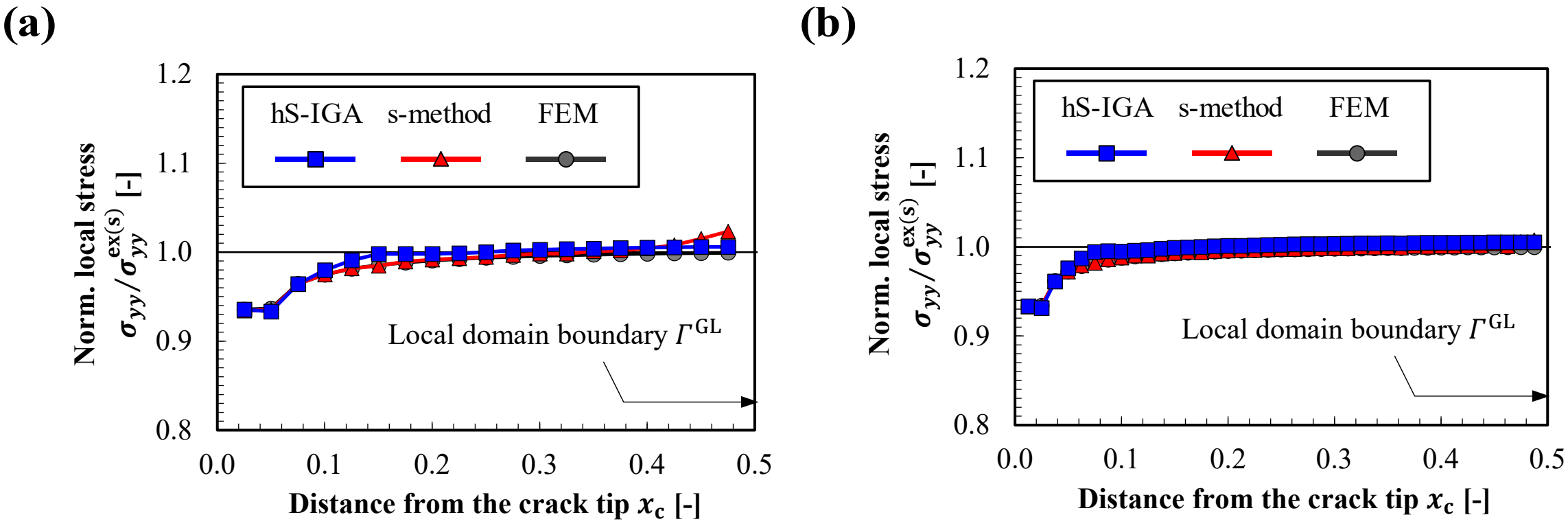

**Fig. 20.** Normalised local stress ahead of the crack tip under two crack-tip mesh resolutions: (a) $1/40$, (b) $1/80$.

To further examine the applicability of the proposed hS-IGA strategy to local stress evaluation under different mesh settings, analyses are conducted under three conditions: (a) fixed $h_{\mathrm{G}}/h_{\mathrm{L}}$ $(= 4)$, (b) fixed $h_{\mathrm{G}}$ $(= 2/21)$, and (c) fixed $h_{\mathrm{L}}$ $(= 1/80)$. The normalised local stress in front of the crack tip, $\sigma_{yy}/\sigma_{yy}^{\mathrm{ex(s)}}$, is shown in Fig. 21. The results indicate that errors in $\sigma_{yy}$ of less than 2.5% can be obtained when the evaluation point satisfies $x' \geq 4h_{\mathrm{L}}$ near the crack tip. In other words, the accuracy of $\sigma_{yy}$ near the crack tip strongly depends on $h_{\mathrm{L}}$, but is insensitive to $h_{\mathrm{G}}$ (Fig. 21(a) and 21(c)). For the present two-dimensional stationary benchmark, the condition for accurate local stress evaluation is identified as $4h_{\mathrm{L}} \leq x' < l_{\mathrm{L}}$.

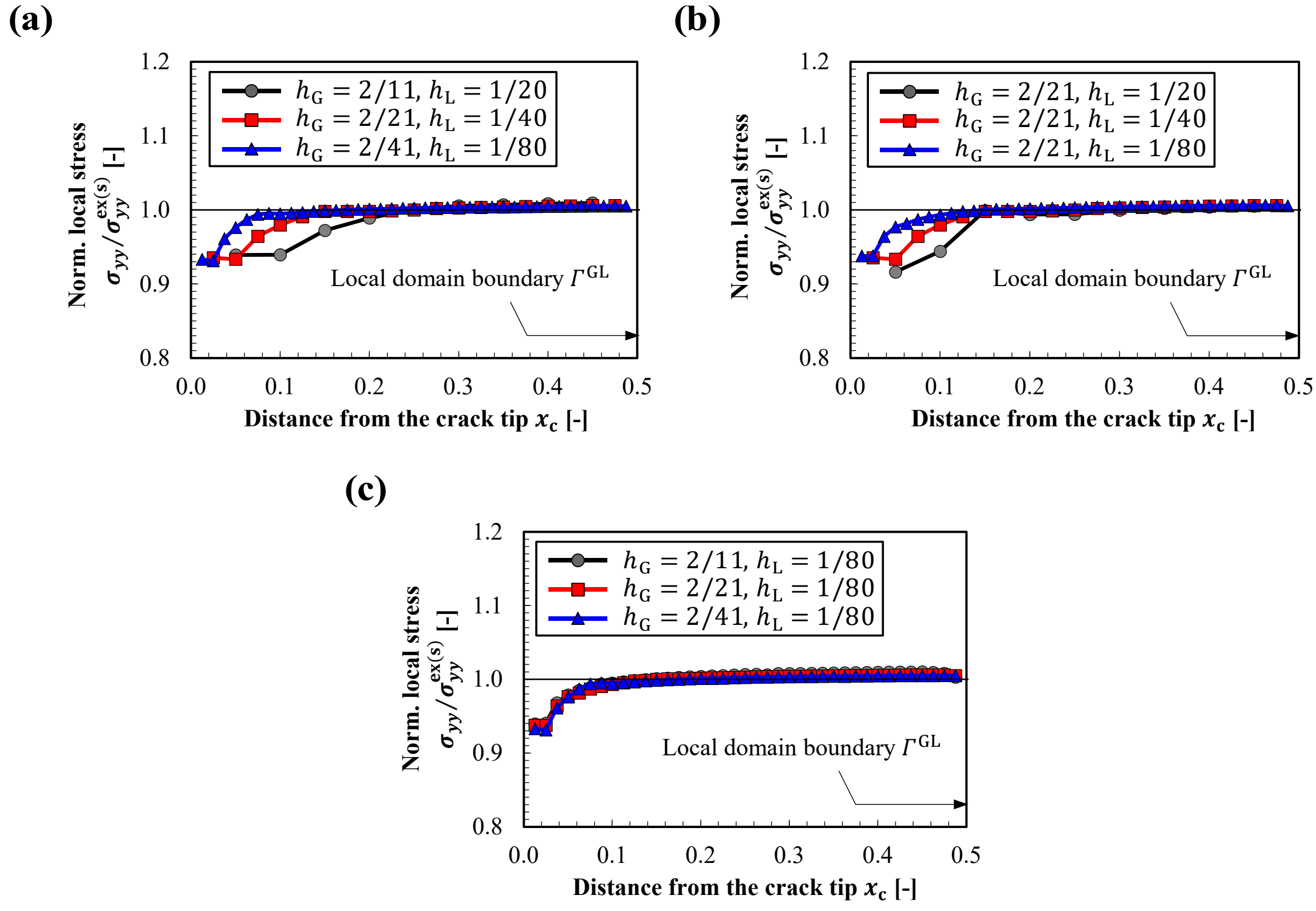

**Fig. 21.** Normalised local stress in front of the crack tip: (a) fixed $r_{\mathrm{GL}} = 4$, (b) fixed $h_{\mathrm{G}} = 2/21$, (c) fixed $h_{\mathrm{L}} = 1/80$.

*4.2. Two-dimensional dynamic crack problem*

The second two-dimensional benchmark considers a dynamically propagating straight crack. This problem is used to assess the accuracy and efficiency of the proposed hS-IGA strategy for evaluating the DSIF and the local stress under different crack lengths/velocities and mesh conditions.

*4.2.1. Problem description*

The dynamic benchmark considered in this section is a straight Mode I dynamically propagating crack at a prescribed constant velocity $V$ in an infinite plate subjected to a remote applied stress. A schematic of this benchmark is shown in Fig. 22(a). Crack propagation from an initially uncracked state ($a = 0$) to a maximum crack length $a_{\max}$ (which depends on the analysis conditions and is specified in Section 4.2.2) is analysed. As in the stationary benchmark in Section 4.1, a symmetric model is employed (Fig. 22(b)). The target domain has width $W_{\mathrm{G}} = (9/5)a_{\max}$ and height $H_{\mathrm{G}} = 6.0$ mm, such that the domain fully covers the maximum crack extension considered in the analysis. The crack velocities used for verification are specified in Section 4.2.2.

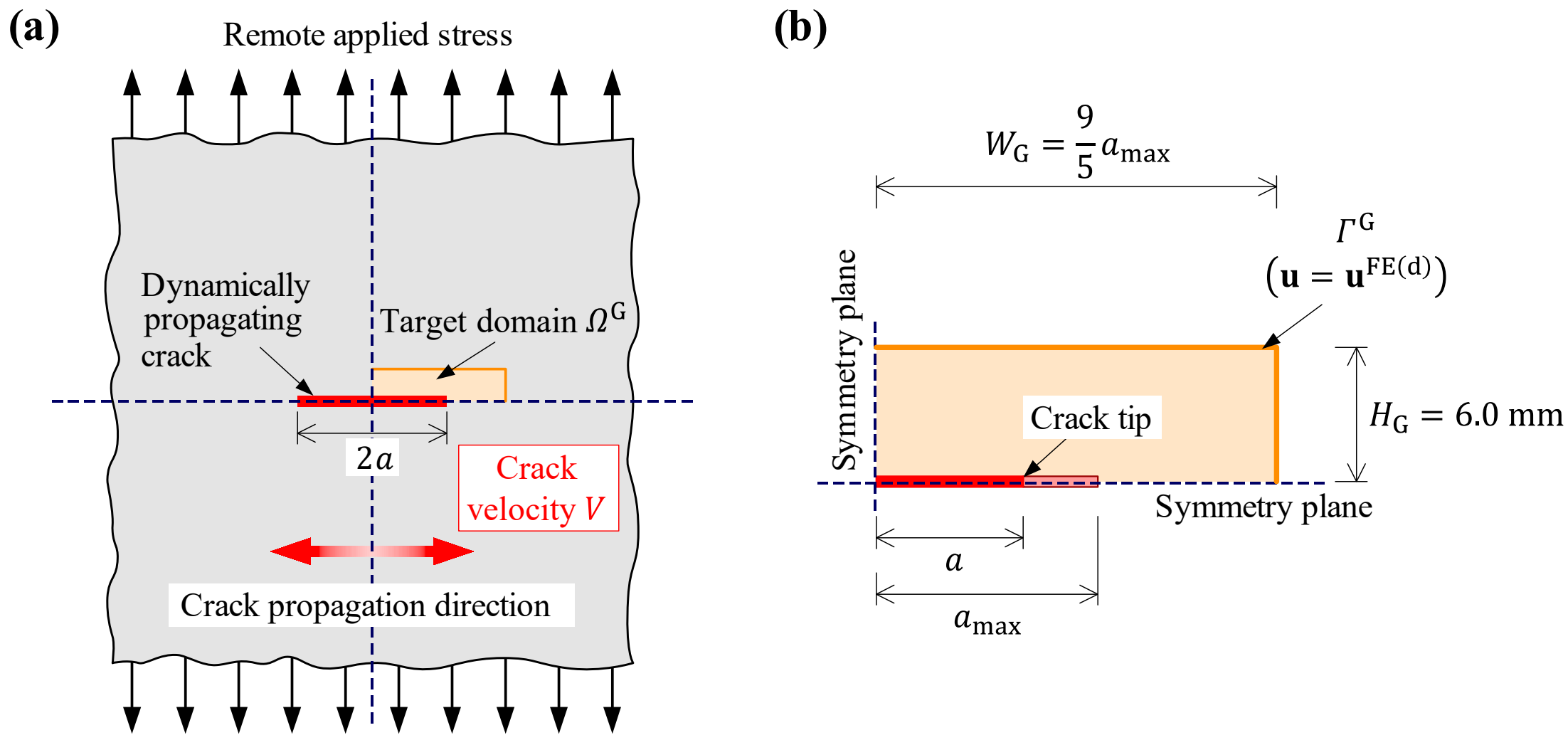

**Fig. 22.** Two-dimensional dynamic crack benchmark problem: (a) problem schematic; (b) target domain for the numerical analysis.

As in the stationary benchmark in Section 4.1, the ideal dynamic crack propagation in an infinite plate is reproduced by prescribing Dirichlet boundary conditions on the outer boundary of the global domain. However, because a corresponding closed-form exact solution for the displacement field is not available for the present setting, sufficiently accurate histories of the displacement field, $\mathbf{u}^{\mathrm{FE(d)}}$, are obtained by finite element analysis (FEA) using a highly refined mesh in the vicinity of the crack tip.

In the FEA used to obtain $\mathbf{u}^{\mathrm{FE(d)}}$, a sufficiently large domain is employed so that reflected Rayleigh waves do not affect the crack-opening response within the time window of interest. Accordingly, the width and height of the FEA domain, $W_{\mathrm{FE}}$ and $H_{\mathrm{FE}}$, are chosen to satisfy

$$W_{\mathrm{FE}} > \left(1 + \frac{V_{\mathrm{R}}}{V}\right) a_{\max}, \tag{55}$$

$$H_{\mathrm{FE}} > \frac{V_{\mathrm{R}}}{V} a_{\max}, \tag{56}$$

where $V_{\mathrm{R}}$ is the Rayleigh wave velocity. The minimum element size in the FEA is set to $h_{\min} = h_{\mathrm{L}}$, which is consistent with the local element size used in the subsequent hS-IGA simulations. This setting ensures consistency in the crack advance per time increment between the FEA and the subsequent verification analyses. An example of the mesh used in the FEA is shown in Fig. 23.

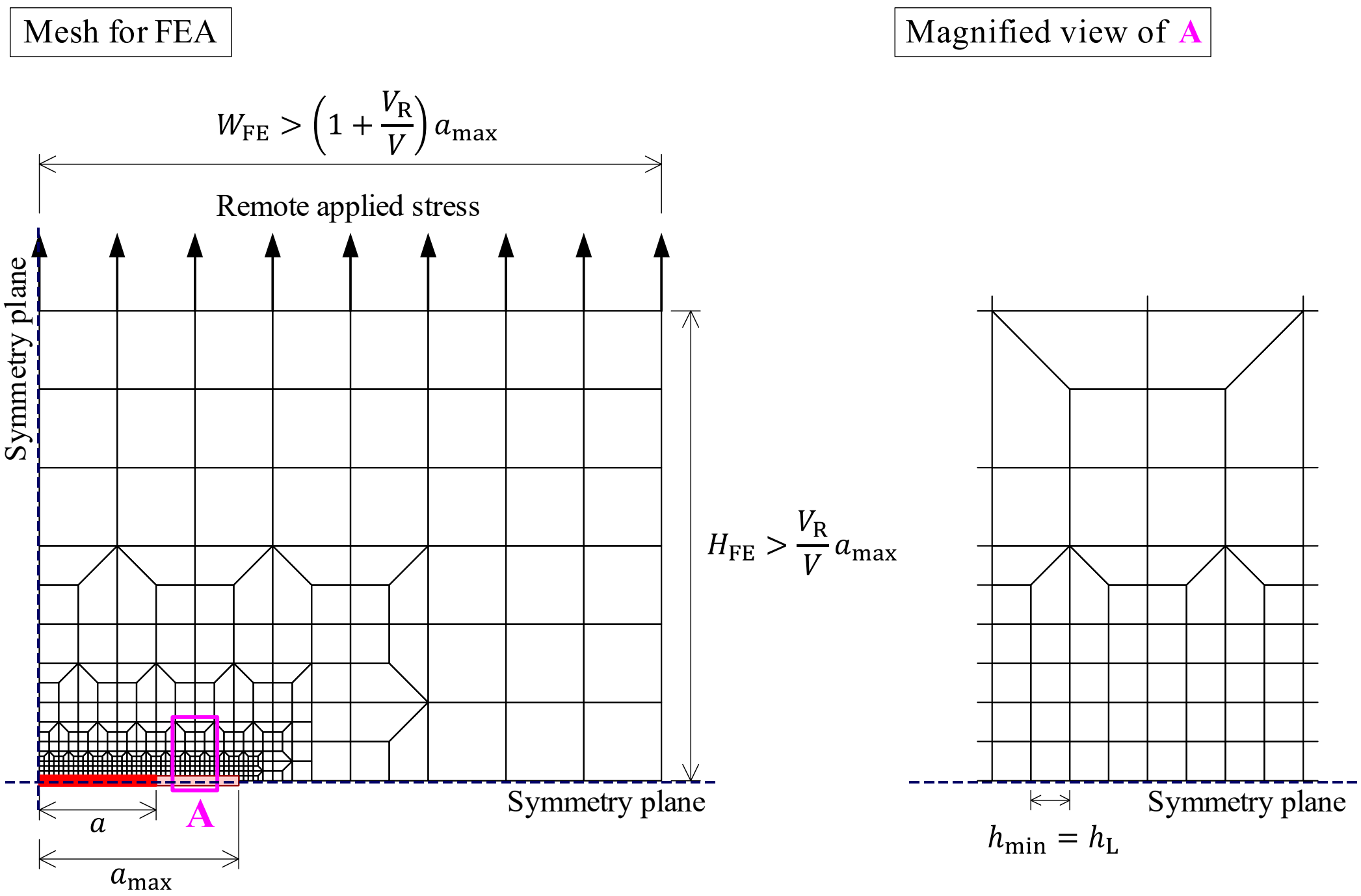

**Fig. 23.** Schematic of the finite element model for generating the reference histories of the displacement field.

The accuracy of the FEA model used to generate the boundary displacement histories is assessed by comparing the dynamic stress intensity factor, $K_I^{FE(d)}$, and the local stress, $\sigma_{yy}^{FE}$, obtained from the FEA with Broberg's analytical solutions, $K_I^{B(d)}$ and $\sigma_{yy}^{B(d)}$, respectively [48]. The detailed expressions of $K_I^{B(d)}$ and $\sigma_{yy}^{B(d)}$ are presented in Appendix B. Fig. 24 shows the normalised dynamic stress intensity factor, $K_I^{FE(d)}/K_I^{B(d)}$, and the normalised local stress, $\sigma_{yy}^{FE}/\sigma_{yy}^{B(d)}$, evaluated at $x' = 0.2$ mm ahead of the crack tip for $V = 500$ m/s and $V = 1000$ m/s. The maximum crack lengths are $a_{max} = 2.5,\ 6.25,\ 12.5,\ 25$ and $50$ mm, corresponding to 50 to 1000 nodal-force-release steps. This evaluation position is consistent with the characteristic distance adopted for local-stress evaluation in studies of brittle crack propagation in steels [10,27,50,51]. The results demonstrate stable agreement with Broberg's analytical solutions, with the maximum errors remaining below 0.9% for the dynamic stress intensity factor and 2.6% for the local stress. These results confirm that the FEA model provides sufficiently accurate displacement histories for prescribing the time-dependent boundary conditions in the subsequent verification analyses.

Accordingly, the histories of the displacement field, $\mathbf{u}^{FE(d)}$, are used to prescribe time-dependent Dirichlet boundary conditions on the outer boundary of the target domain, $\Gamma^G$ (see Fig. 22(b)). In the present study, the same target domain and the same boundary-condition treatment are applied consistently to the standard FEM, the conventional s-method, and the proposed hS-IGA strategy, so that all three methods are examined under identical domain and boundary conditions. This unified setting enables a direct comparison of the three approximation frameworks while excluding differences arising from the computational domain or the boundary conditions. The accuracy of each framework, including the proposed strategy, is then assessed by comparing the evaluated dynamic stress intensity factor, $K_I^{(d)}$, and local stress, $\sigma_{yy}$, with Broberg's analytical solutions, $K_I^{B(d)}$ and $\sigma_{yy}^{B(d)}$, respectively.

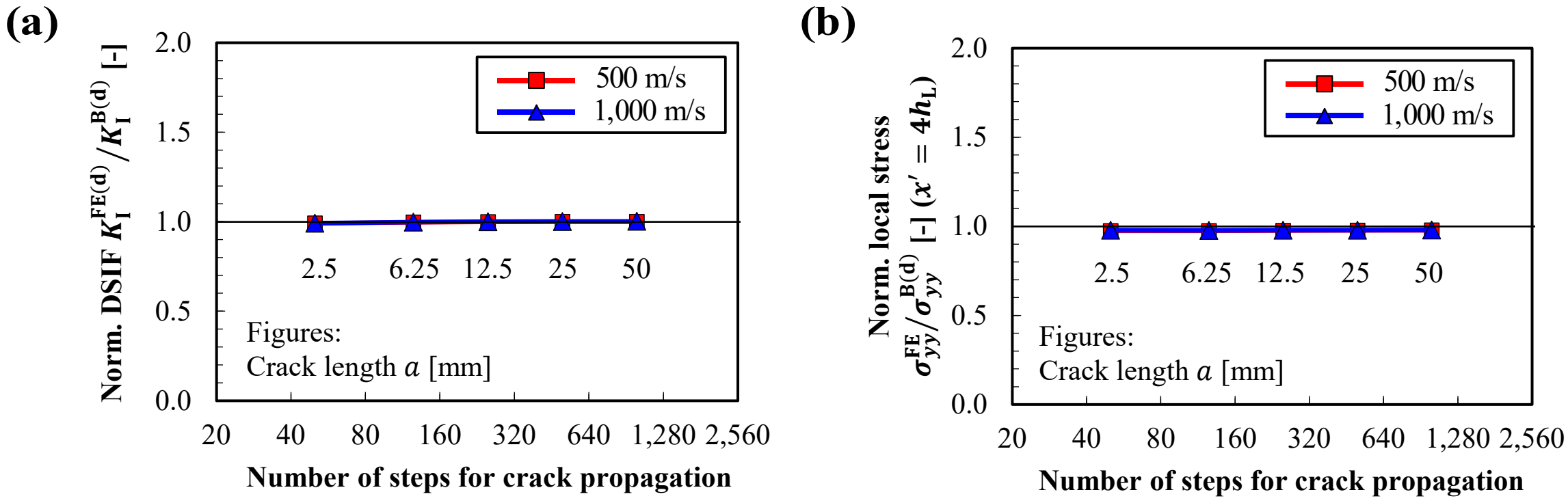

**Fig. 24.** Comparison of the normalised DSIF and local stress obtained by the FEA with Broberg's analytical solutions for the two-dimensional dynamic crack problem.

*4.2.2. Verification of the two-dimensional dynamic crack problem*

The global and local meshes, together with the associated boundary conditions used in the verification analyses, are shown in Fig. 25. As described in Section 4.2.1, the time-dependent Dirichlet boundary conditions imposed on the outer boundary of the target domain, $\Gamma^{\mathrm{G}}$, are prescribed from the displacement histories obtained by the FEA, $\mathbf{u}^{\mathrm{FE(d)}}$. The global mesh is defined over the target domain, while the local mesh is superposed in the vicinity of the propagating crack tip. The height of the global domain is set to $H_{\mathrm{G}} = 6.0$ mm, whereas the width is determined by the maximum crack length $a_{\mathrm{max}}$ as $W_{\mathrm{G}} = (9/5)a_{\mathrm{max}}$. The local element size in the $x'$-direction is fixed at $h_{\mathrm{L}} = 0.05$ mm. This choice is motivated by two considerations: (i) the stationary crack results in Section 4.1.2, where the local stress $\sigma_{yy}$ was evaluated with sufficient accuracy (errors below 2.5%) at $x' = 4h_{\mathrm{L}}$; and (ii) the characteristic distance reported for brittle crack propagation in steels, $x' = 0.2$ mm [10,27,50,51].

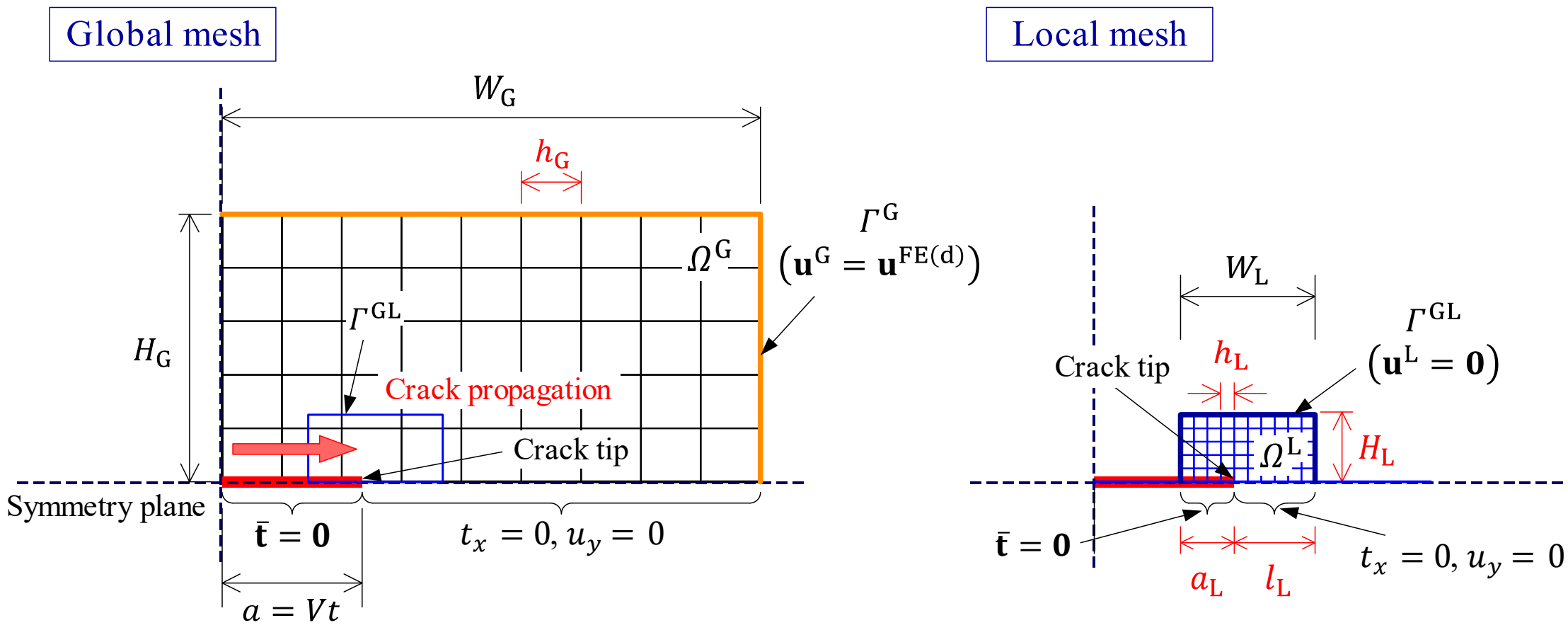

**Fig. 25.** Global and local mesh configurations and associated boundary conditions used in the verification analyses for the two-dimensional dynamic crack problem.

With the analysis framework defined above, the next step is to identify the local domain and mesh conditions that enable accurate and computationally efficient evaluation of the near-tip fracture quantities during dynamic crack propagation. The primary parameters investigated in this section are: (a) the global-to-local element-size ratio, $r_{\mathrm{GL}} = h_{\mathrm{G}}/h_{\mathrm{L}}$; (b) the crack length in the local domain, $a_{\mathrm{L}}$; (c) the ligament length in the local domain, $l_{\mathrm{L}}$; and (d) the height of the local domain, $H_{\mathrm{L}}$. The effects of these parameters on the dynamic stress intensity factor, $K_{\mathrm{I}}^{(\mathrm{d})}$, and the local stress, $\sigma_{yy}$, are assessed at the maximum crack length, $a_{\mathrm{max}} = 10$mm, which corresponds to $200h_{\mathrm{L}}$. Accordingly, crack propagation from $a = 0$ to $a = a_{\mathrm{max}}$ is simulated using 200 nodal-force-release steps.

Systematic analyses are performed for crack velocities of $V = 500$ m/s and $V = 1000$ m/s. The resulting accuracy requirements can be expressed in a unified form for both $K_{\mathrm{I}}^{(\mathrm{d})}$ and $\sigma_{yy}$ as follows: $r_{\mathrm{GL}} \geq 4$, $a_{\mathrm{L}} \geq 2.5h_{\mathrm{G}}$, $l_\mathrm{L} \geq 1.2h_\mathrm{G}$, and $H_\mathrm{L} \geq 1.8h_\mathrm{G}$. Under these conditions, the deviations from the corresponding Broberg's analytical solutions are within 1.8% for $K_\mathrm{I}^{(\mathrm{d})}$ and 4.0% for $\sigma_{yy}$. Although the parameter settings are explored systematically, only representative results are presented below to show the influence of each parameter on the accuracies of $K_\mathrm{I}^{(\mathrm{d})}$ and $\sigma_{yy}$. Unless otherwise stated, the parameter values satisfying the above criteria are used for all parameters except the one being varied.

*(a) Global-to-local element-size ratio $r_{GL}$ (Fig. 26)*

Instabilities in the evaluated $K_\mathrm{I}^{(\mathrm{d})}$ and $\sigma_{yy}$ are observed when $r_\mathrm{GL} < 4$, suggesting that the appropriate condition is $r_\mathrm{GL} \geq 4$. These results also suggest that the accuracies of $K_\mathrm{I}^{(\mathrm{d})}$ and $\sigma_{yy}$ depend only weakly on the global element size $h_\mathrm{G}$, and are governed mainly by the local element size $h_\mathrm{L}$. It should be noted that, if the ligament-length condition $l_\mathrm{L} = 1.2h_\mathrm{G}$ identified below is applied directly, then for $r_\mathrm{GL} = 2$ and $3$ the resulting ligament length becomes smaller than $4h_\mathrm{L}$, making the prescribed local-stress evaluation no longer possible. Therefore, in these two cases, $l_\mathrm{L}$ is instead set to $5h_\mathrm{L}$.

*(b) Crack length in the local domain $a_L$ (Fig. 27)*

The crack length in the local domain, $a_\mathrm{L}$, has a strong influence on the accuracy. The results indicate that the accuracies of $K_\mathrm{I}^{(\mathrm{d})}$ and $\sigma_{yy}$ are governed by the ratio $a_\mathrm{L}/h_\mathrm{G}$, and remain stable when $a_\mathrm{L} \geq 2.5h_\mathrm{G}$.

*(c) Ligament length in the local domain $l_L$ (Fig. 28)*

The ligament length in the local domain, $l_\mathrm{L}$, has a moderate influence on the accuracy. As discussed in Section 4.1.2, unlike in the conventional s-method, the accuracy of $\sigma_{yy}$ near $\Gamma^\mathrm{GL}$ is not sensitive to $h_\mathrm{G}$. The present dynamic results show the same tendency. However, instabilities are observed when $l_\mathrm{L} \leq 1.0h_\mathrm{G}$, and the appropriate condition is therefore identified as $l_\mathrm{L} \geq 1.2h_\mathrm{G}$.

*(d) Height of the local domain $H_L$ (Fig. 29)*

From the results, the appropriate condition for $H_\mathrm{L}$ to ensure accurate evaluation of both $K_\mathrm{I}^{(\mathrm{d})}$ and $\sigma_{yy}$ is identified as $H_\mathrm{L} \geq 1.8h_\mathrm{G}$.

In addition to the four primary parameters discussed above, the stability of the proposed strategy is further examined over a wide range of crack velocities. The parameters are set to $r_\mathrm{GL} = 8$, $a_\mathrm{L} = 2.5h_\mathrm{G}$, $l_\mathrm{L} = 1.2h_\mathrm{G}$, and $H_\mathrm{L} = 1.8h_\mathrm{G}$, which satisfy the identified conditions. Under these settings, the accuracies of $K_\mathrm{I}^{(\mathrm{d})}$ and $\sigma_{yy}$ are evaluated for crack velocities ranging from $V = 200\ \mathrm{m/s}$ to $1500\ \mathrm{m/s}$ in increments of $100\ \mathrm{m/s}$, with a maximum crack length of $a_\mathrm{max} = 10\ \mathrm{mm}$. The results, summarised in Fig. 30, indicate that the proposed hS-IGA strategy shows good agreement with Broberg's analytical solutions over the entire velocity range considered. For the normalised dynamic stress intensity factor (Fig. 30(a)), the maximum deviation remains below 1.4%, whereas the corresponding value for the normalised local stress (Fig. 30(b)) remains below 3.7%.

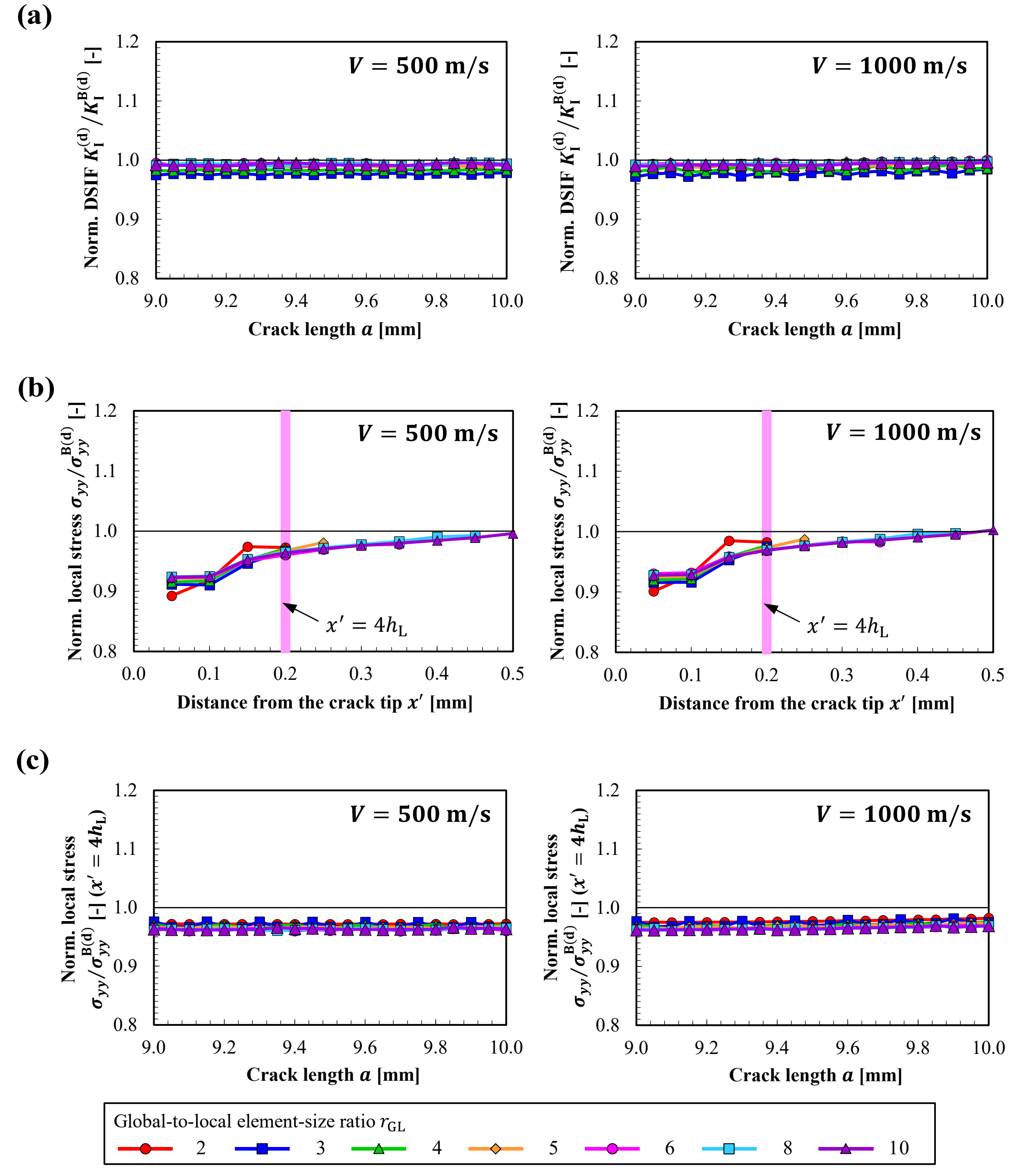

**Fig. 26.** Influence of the global-to-local element-size ratio $r_{GL}$ ($a = 10$ mm, $a_L = 2.5h_G$, $l_L = 1.2h_G$, $H_L = 1.8h_G$): (a) Histories of dynamic stress intensity factor; (b) local stress distributions in crack propagation direction; (c) histories of local stress at $x' = 0.2$ mm $(= 4h_L)$.

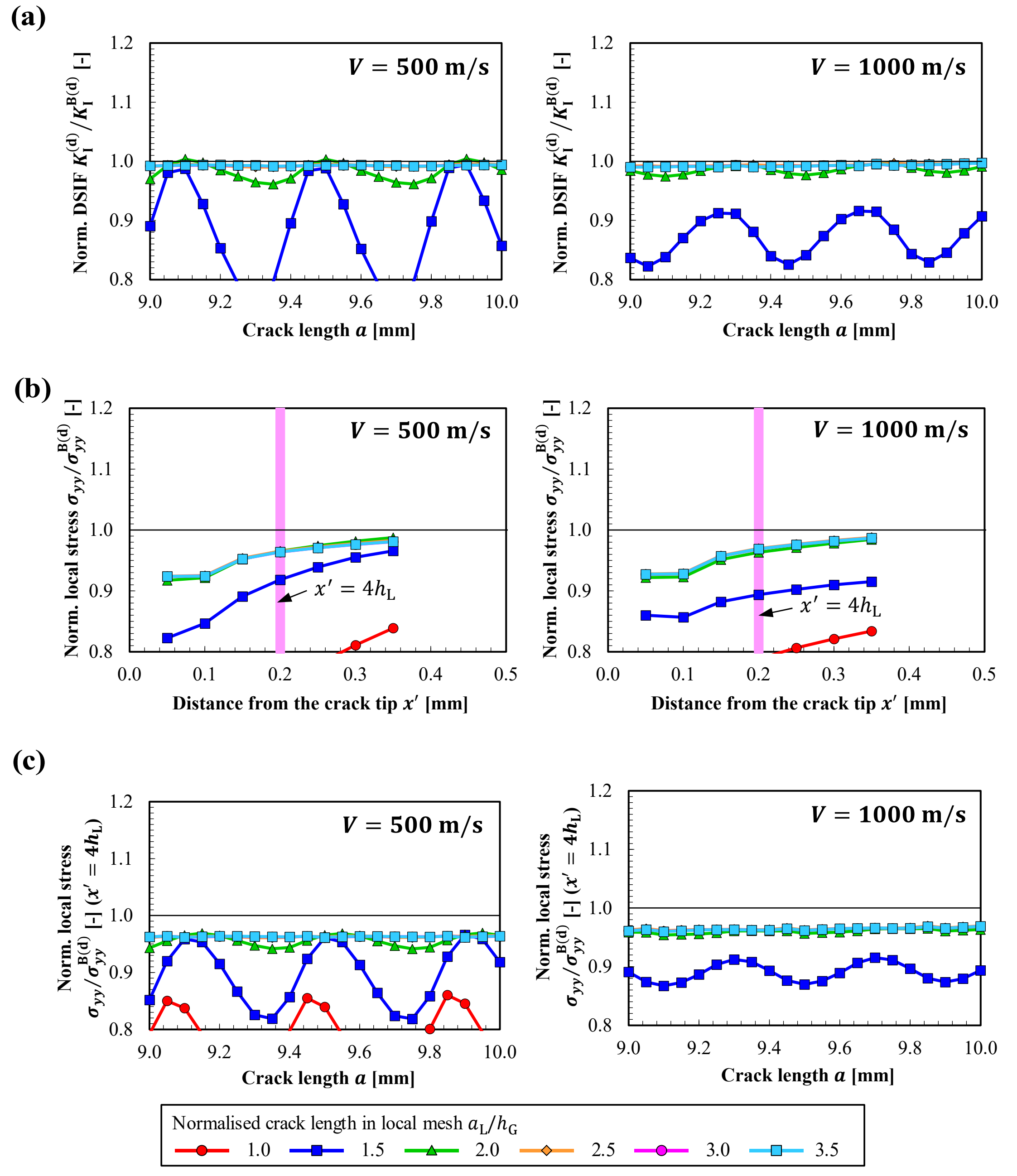

**Fig. 27.** Influence of the crack length in the local domain $a_L$ ($a = 10$ mm, $r_{GL} = 8$, $l_L = 1.2h_G$, $H_L = 1.8h_G$): (a) Histories of dynamic stress intensity factor; (b) local stress distributions in crack propagation direction; (c) histories of local stress at $x' = 0.2$ mm ($= 4h_L$).

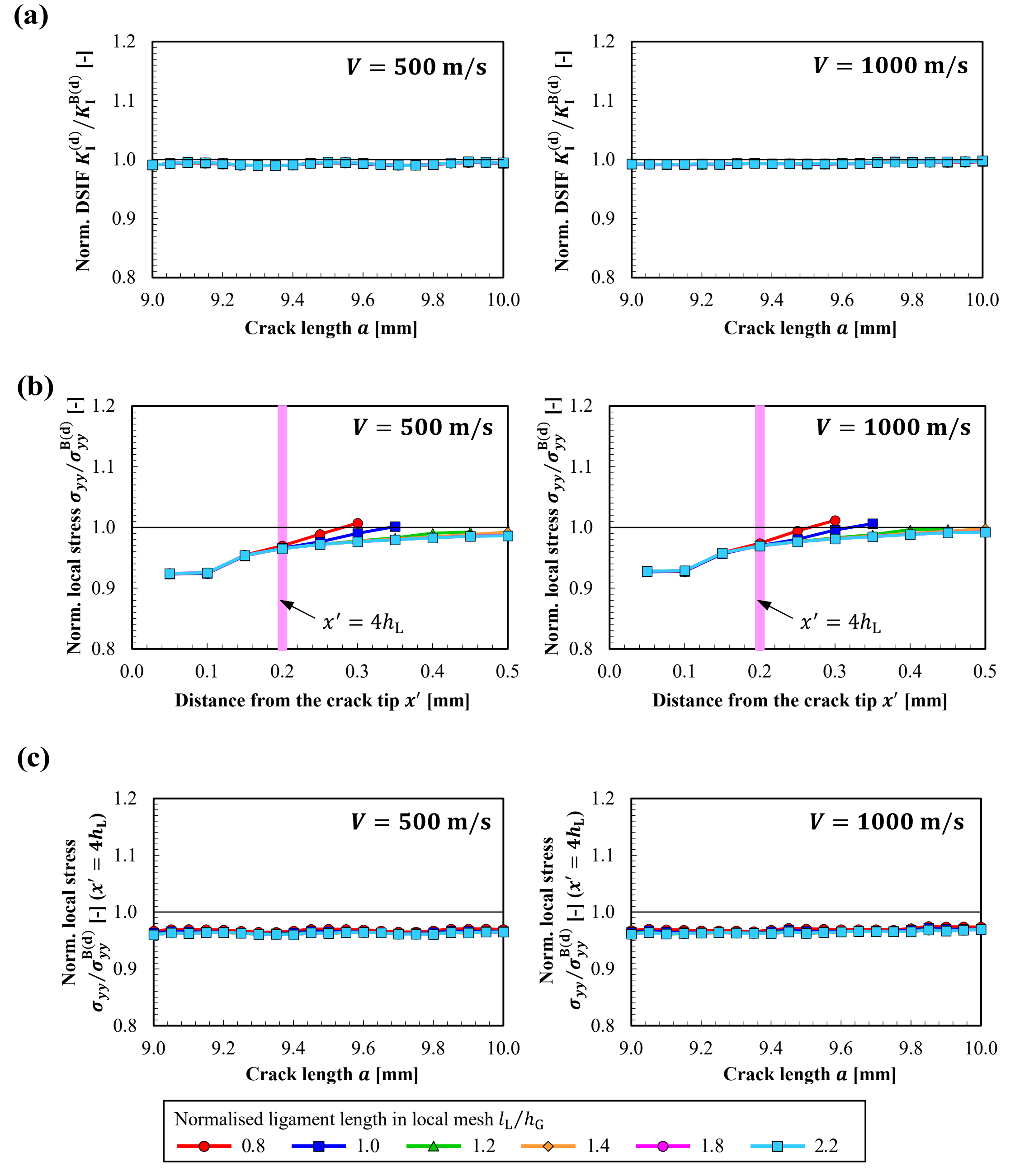

**Fig. 28.** Influence of the ligament length in the local domain $l_L$ ($a = 10$ mm, $r_{GL} = 8$, $a_L = 2.5h_G$, $H_L = 1.8h_G$): (a) Histories of dynamic stress intensity factor; (b) local stress distributions in crack propagation direction; (c) histories of local stress at $x' = 0.2$ mm ($= 4h_L$).

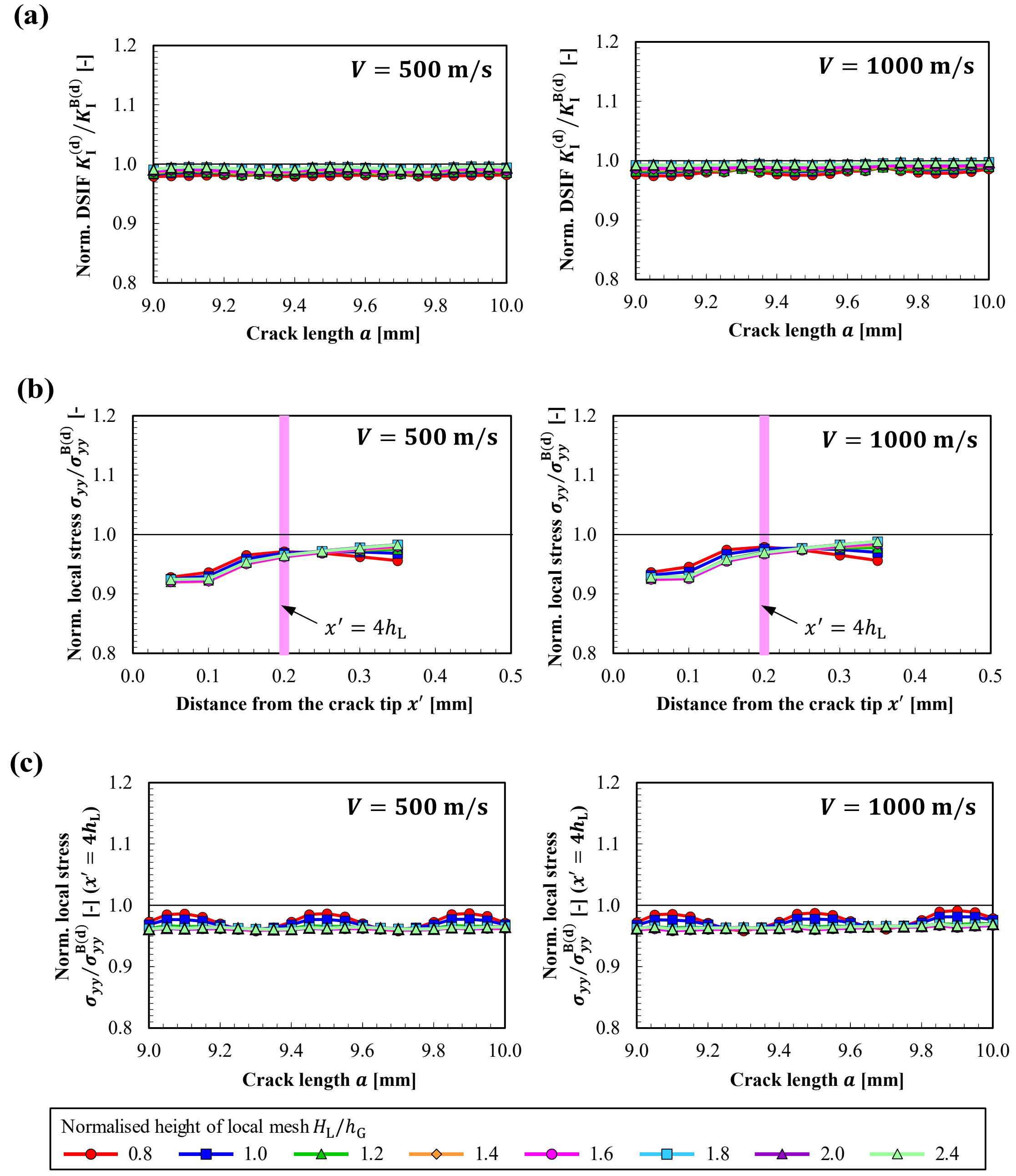

**Fig. 29.** Influence of the height of the local domain $H_L$ ($a = 10$ mm, $r_{GL} = 8$, $a_L = 2.5h_G$, $l_L = 1.2h_G$): (a) Histories of dynamic stress intensity factor; (b) local stress distributions in crack propagation direction; (c) histories of local stress at $x' = 0.2$ mm ($= 4h_L$).

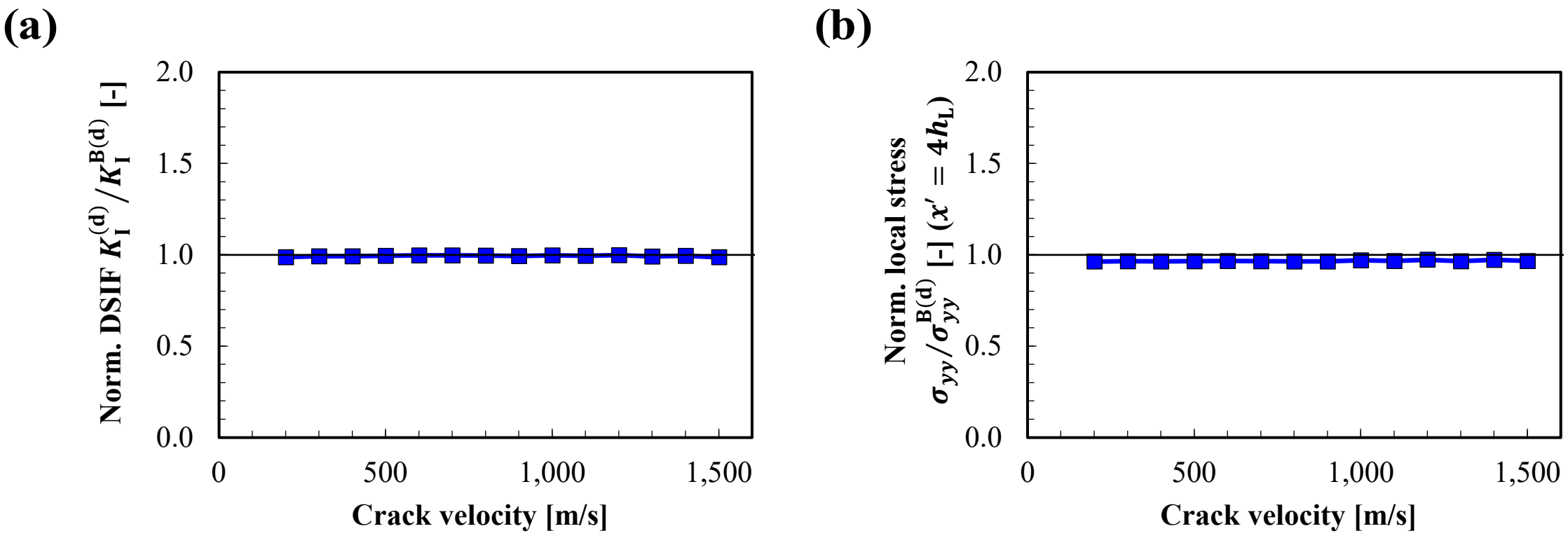

**Fig. 30.** Verification for two-dimensional dynamic crack problem over a wide range of crack velocities: (a) Normalised dynamic stress intensity factor; (b) normalised local stress at $x' = 0.2$ mm $(= 4h_L)$.

To further assess the effectiveness of the proposed strategy, crack propagation analyses based on the hS-IGA strategy are compared with those based on the standard FEM and the conventional s-method. The comparison analyses are conducted for maximum crack lengths $a_{max} = 2.5,\ 6.25,\ 12.5,\ 25$ and $50$ mm, corresponding to 50 to 1000 nodal-force-release steps. In all cases, the same target domain and the same time-dependent displacement boundary conditions are applied, so that the differences among the three methods arise from their discretisation frameworks and integration treatments.

For the hS-IGA strategy, the parameters are set to $r_{GL} = 8$, $a_L = 2.5h_G$, $l_L = 1.2h_G$, and $H_L = 1.8h_G$ according to the conditions identified in the present study. For the conventional s-method, the parameters are set to $r_{GL} = 8$, $a_L = 2.0h_G$, $l_L = 15h_L$, and $H_L = 1.2h_G$, based on the conditions identified in our previous study [12]. Further implementation details of the conventional s-method can be found in our previous study [13]. For the standard FEM, a single finite element mesh is employed over the same target domain, and the displacement histories obtained from the FEA used for boundary-condition generation, $\mathbf{u}^{FE(d)}$, are imposed on the outer boundary of the mesh, as shown in Fig. 31. The mesh is defined in a graded manner, such that the minimum element size near the crack tip, $h_{min}$, is consistent with the local discretisation adopted in the hS-IGA strategy, $h_L$, while the maximum element size is set equal to $h_G$ (see Fig. 31).

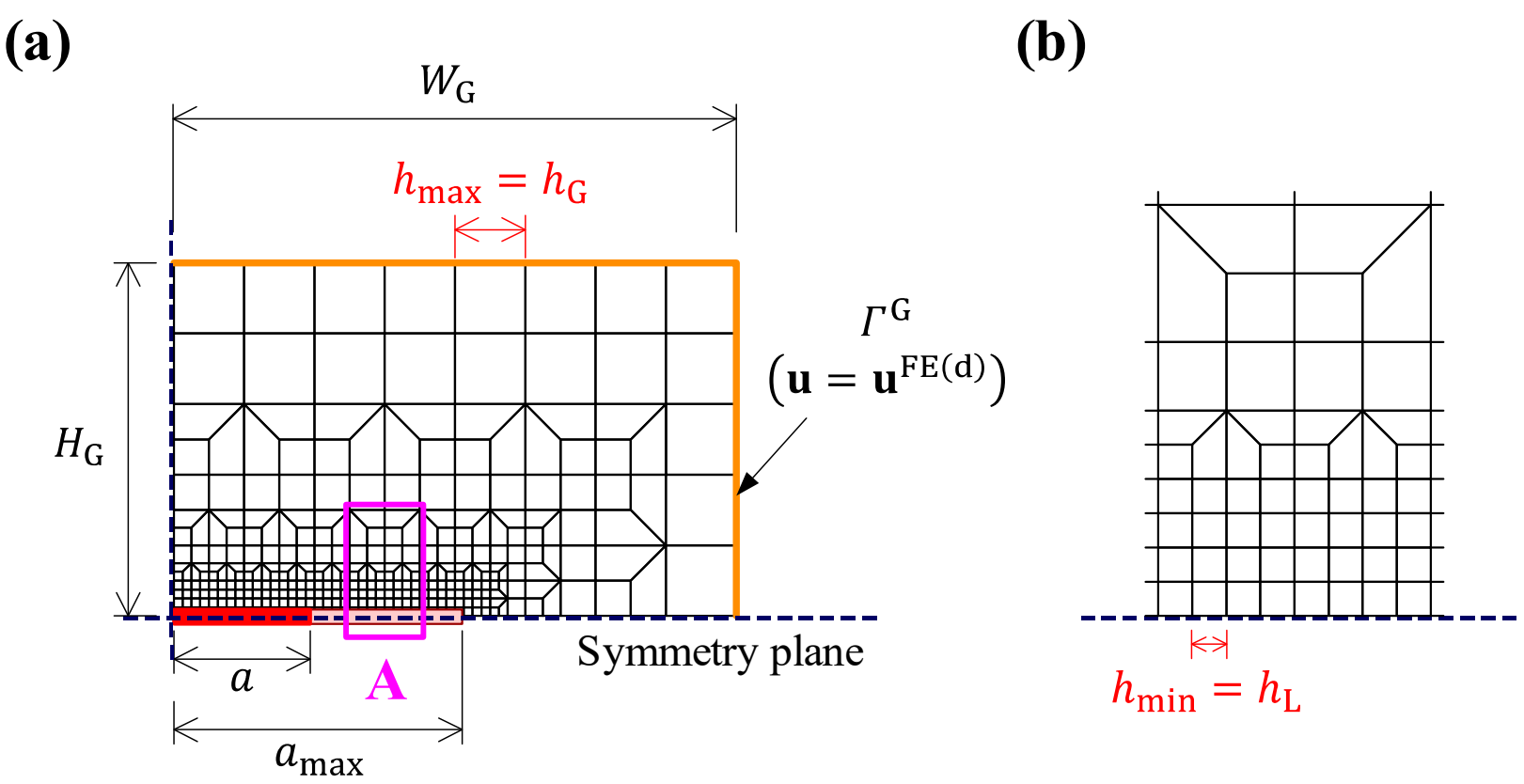

**Fig. 31.** Standard FEM model for comparison in the two-dimensional dynamic crack problem: (a) Mesh and boundary conditions; (b) close-up of A.

Under the above conditions, the accuracies of the proposed hS-IGA strategy, the standard FEM, and the conventional s-method are compared. Fig. 32 summarises the results for the normalised dynamic stress intensity factor and the normalised local stress at the representative crack velocities $V = 500\ \mathrm{m/s}$ and $V = 1000\ \mathrm{m/s}$. The results indicate that all three methods provide good agreement with Broberg's analytical solutions. Overall, the proposed hS-IGA strategy achieves accuracy comparable to that of the standard FEM and the conventional s-method. The maximum deviation of the hS-IGA results from Broberg's analytical solutions remains below 1.4% for the normalised dynamic stress intensity factor and below 3.9% for the normalised local stress.

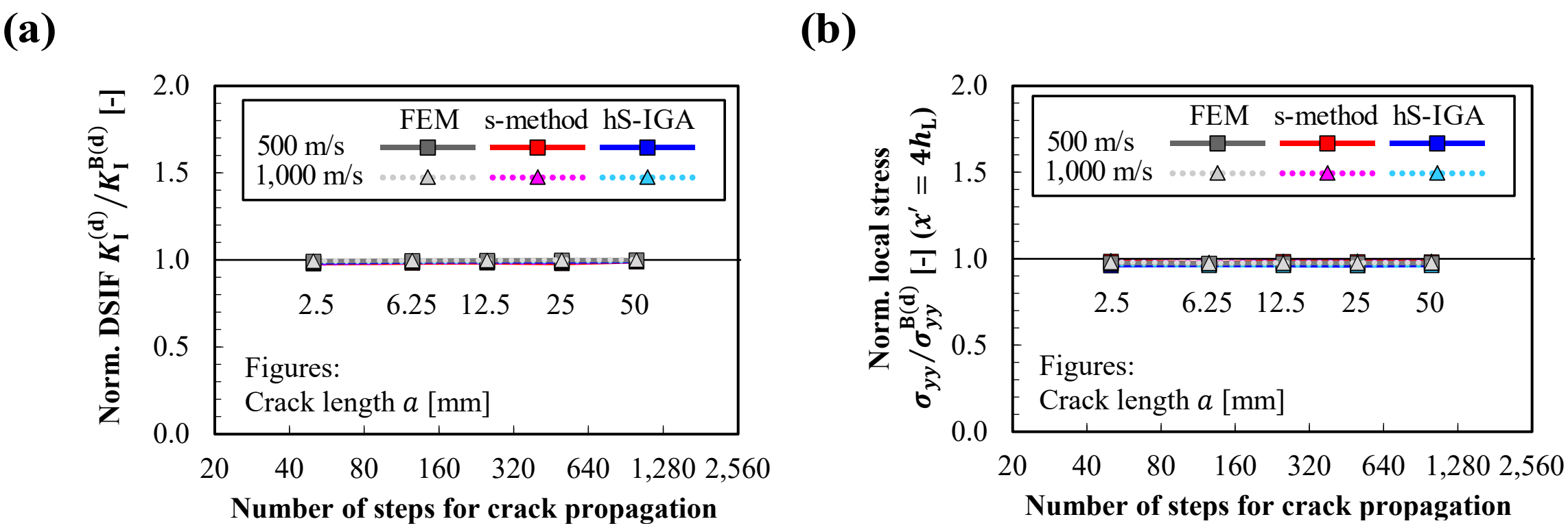

**Fig. 32.** Accuracy comparison between the hS-IGA, the conventional s-method and the standard FEM for two-dimensional dynamic crack propagation analysis: (a) normalised dynamic stress intensity factor; (b) normalised local stress at $x' = 0.2\ \mathrm{mm}\ (= 4h_{\mathrm{L}})$.

To demonstrate the computational efficiency advantage of the proposed hS-IGA strategy, the numerical cost is first examined in terms of the degrees of freedom (DOFs). Fig. 33(a) compares the DOFs required by the three methods to obtain the results shown in Fig. 32. Together with Fig. 32, these results indicate that, for a comparable level of accuracy, the hS-IGA strategy and the conventional s-method require substantially fewer DOFs than the standard FEM, with a reduction of approximately 73% relative to the standard FEM for the maximum crack length considered. In addition, the DOF level of the proposed hS-IGA strategy remains broadly comparable to that of the conventional s-method. These findings indicate that the proposed hS-IGA strategy preserves the principal DOF-related advantage of the s-method while maintaining accuracy comparable to that of the standard FEM.

The principal computational advantage of the proposed hS-IGA strategy becomes more evident when the numerical cost is examined in terms of the number of integration points. Fig. 33(b) compares the total numbers of integration points required by the three methods for the analyses. The figure clearly shows that the number of integration points required by the conventional s-method is far greater than those required by the standard FEM and the proposed hS-IGA strategy. This reflects the main bottleneck of the conventional s-method, namely, the extremely large number of integration points required in the coupling integration. In contrast, the number of integration points required by the proposed hS-IGA strategy is markedly smaller than that required by the conventional s-method, with a reduction of approximately 81% for the maximum crack length considered. These results therefore demonstrate that, for a comparable level of accuracy, the proposed hS-IGA strategy drastically reduces the integration cost relative to the conventional s-method, thereby significantly improving the overall computational efficiency.

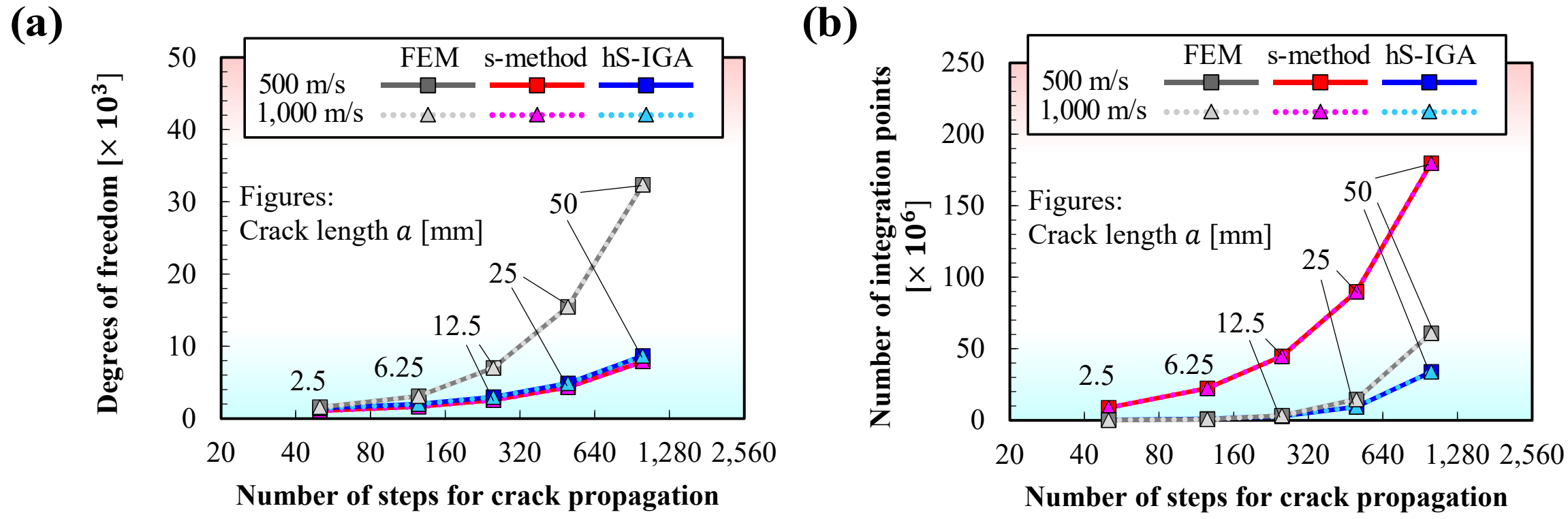

**Fig. 33.** Computational cost comparison between the hS-IGA, the conventional s-method and the standard FEM for two-dimensional dynamic crack propagation analysis: (a) degrees of freedom; and (b) number of integration points required for the analyses.

## 5. Verification and discussion: three-dimensional crack problems

This section further assesses the proposed hS-IGA strategy through three-dimensional crack problems. As in Section 4, both stationary and dynamic crack problems are considered. A circular crack in an infinite three-dimensional solid is adopted as the fundamental benchmark problem, following the conventional three-dimensional s-method framework [13]. The purpose of this section is to examine whether the proposed hS-IGA strategy can reproduce the accuracy of the conventional s-method while retaining the advantage of reduced integration cost in three-dimensional crack-front analyses. Accordingly, the accuracy is assessed through the evaluation of fracture quantities, whereas the computational efficiency is discussed mainly in terms of the number of integration points required for global–local coupling. The same material properties as those in Section 4 are adopted.

### *5.1. Three-dimensional stationary crack problem*

The proposed hS-IGA strategy is verified using a stationary circular crack in an infinite three-dimensional solid. This benchmark provides a fundamental basis for evaluating whether the accuracy and integration-cost advantages identified in the two-dimensional verification are retained in a three-dimensional crack-front problem.

#### *5.1.1. Problem description*

The stationary benchmark considered here is a circular crack in an infinite three-dimensional solid, as shown in Fig. 34. Although this problem is one of the simplest three-dimensional crack problems, it is effective for verifying the fundamental applicability of the proposed strategy for two reasons. First, the global and local elements are superposed at various relative orientations in each analysis, allowing the three-dimensional effects of mesh superposition to be examined. Second, the exact solution of this problem is available [58], enabling quantitative verification. Therefore, this problem setting provides a suitable basis for the fundamental verification of the proposed hS-IGA strategy.

The exact displacement field for this problem was provided by Sneddon [58] as

$$u_r^{\mathrm{ex(s)}} = \frac{a\sigma_\infty}{E}\left(-\nu\bar{r} + \frac{2(1+\nu)}{\pi E}\int_0^\infty (1-2\nu-\bar{y}\eta)\frac{d}{d\eta}\left(\frac{\sin\eta}{\eta}\right)e^{-\bar{y}\eta}J_1(\bar{r}\eta)\,d\eta\right), \tag{57}$$

$$u_\theta^{\mathrm{ex(s)}} = 0, \tag{58}$$

$$u_y^{\mathrm{ex(s)}} = \frac{a\sigma_\infty}{E}\left(\bar{y} - \frac{4(1-\nu^2)}{\pi}\int_0^\infty \left(1+\frac{\bar{y}\eta}{2(1-\nu)}\right)\frac{d}{d\eta}\left(\frac{\sin\eta}{\eta}\right)e^{-\bar{y}\eta}J_0(\bar{r}\eta)\,d\eta\right), \tag{59}$$

where $J_\alpha$ $(\alpha = 0, 1)$ denotes the Bessel function of the first kind. The solutions in Eqs. (57)–(59) are expressed in the cylindrical coordinate system $(r, \theta, y)$. The normalised coordinates $\bar{r}$ and $\bar{y}$, defined with respect to the crack radius $a$, are given by

$$\bar{r} = \frac{r}{a}, \tag{60}$$

$$\bar{y} = \frac{y}{a}. \tag{61}$$

The numerical domain for the target stationary crack problem is shown in Fig. 34. Owing to symmetry, a one-eighth model was employed. Although a numerical model generally represents a finite body, the application of the exact solution on the outer boundaries enables the stationary circular-crack problem in an infinite body to be simulated appropriately [12,13,59,60]. Specifically, Sneddon's solution, given in Eqs. (57)–(59), was imposed as the prescribed displacement on the outer boundaries of the model. In this target problem, the crack radius, applied remote stress, domain width, and domain height were set as $a = 1$, $\sigma_\infty = 1$, $W_\mathrm{G} = 2$, and $H_\mathrm{G} = 1$, respectively.

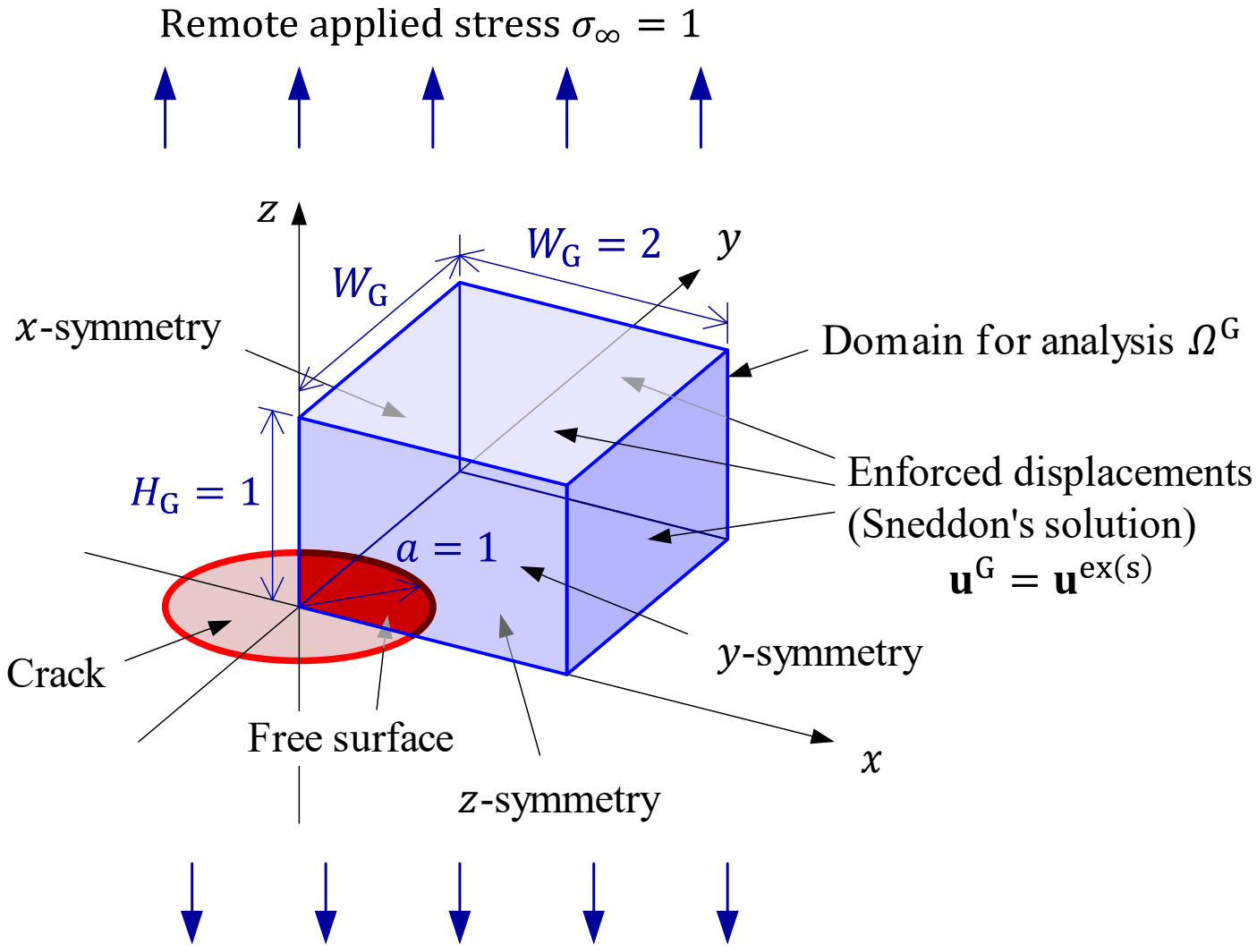

**Fig. 34.** Three-dimensional stationary crack benchmark problem.

The exact solution of the tensile stress distribution ahead of the crack front, $\sigma_{yy}^{\mathrm{ex(s)}}$, which is used to verify the local stress $\sigma_{yy}$, was also provided by Sneddon [58] and is expressed as

$$\sigma_{yy}^{\mathrm{ex(s)}}\Big|_{y=0} = \sigma_\infty \left(\frac{2}{\pi}\left(\frac{1}{\sqrt{\bar{r}^2 - 1}} - arcsin\left(\frac{1}{\bar{r}}\right)\right) + 1\right). \tag{62}$$

Furthermore, the exact solution for the stress intensity factor for the stationary circular crack problem, $K_\mathrm{I}^{\mathrm{ex(s)}}$, is expressed as [61]

$$K_\mathrm{I}^{\mathrm{ex(s)}} = \frac{2}{\pi}\sigma_\infty\sqrt{\pi a}. \tag{63}$$

Because the exact displacement solution is available, the relative $L_2$ error norm of the displacement field, $e_{L_2}$, can be evaluated using the same definition as that for the two-dimensional problem, given in Eq. (54) in Section 4.1.1. As in the two-dimensional stationary crack problem, the optimal convergence slope of $e_{L_2}$ for the standard finite element approximation is $1/2$, owing to the singularity at the crack front [12,13].

*5.1.2. Verification of the three-dimensional stationary crack problem*

The verification is performed using an isotropic local mesh, in which the local element size in the $z'$-direction is equal to that in the $x'y'$-section normal to the crack front; that is, $h_{\mathrm{L}(z')} = h_{\mathrm{L}}$. The employed global and local meshes and their boundary conditions are shown in Fig. 35. The width of the local mesh in the $x'$-direction, corresponding to the crack propagation direction, is defined as $W_{\mathrm{L}} = 0.5$. This width is divided into the crack-surface part ($x' < 0$), with length $a_{\mathrm{L}} = 0.25$, and the ligament part ($x' \geq 0$), with length $l_{\mathrm{L}} = 0.25$. The height of the local mesh is defined as $H_{\mathrm{L}} = 0.25$.

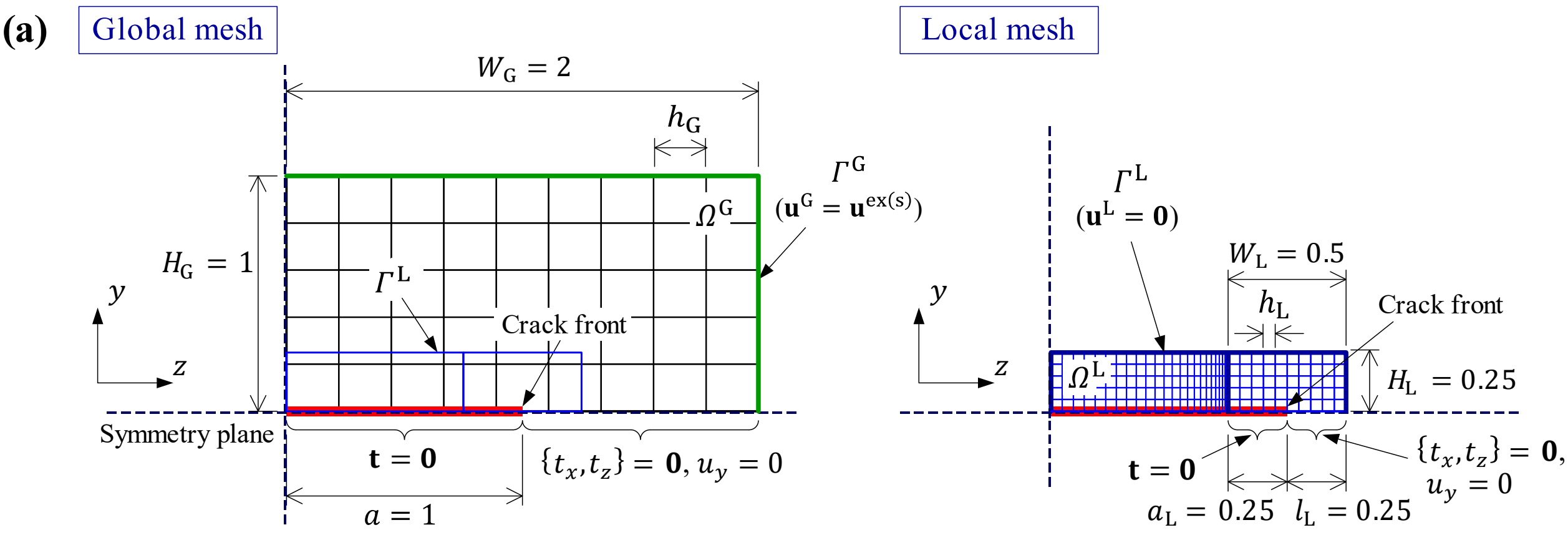

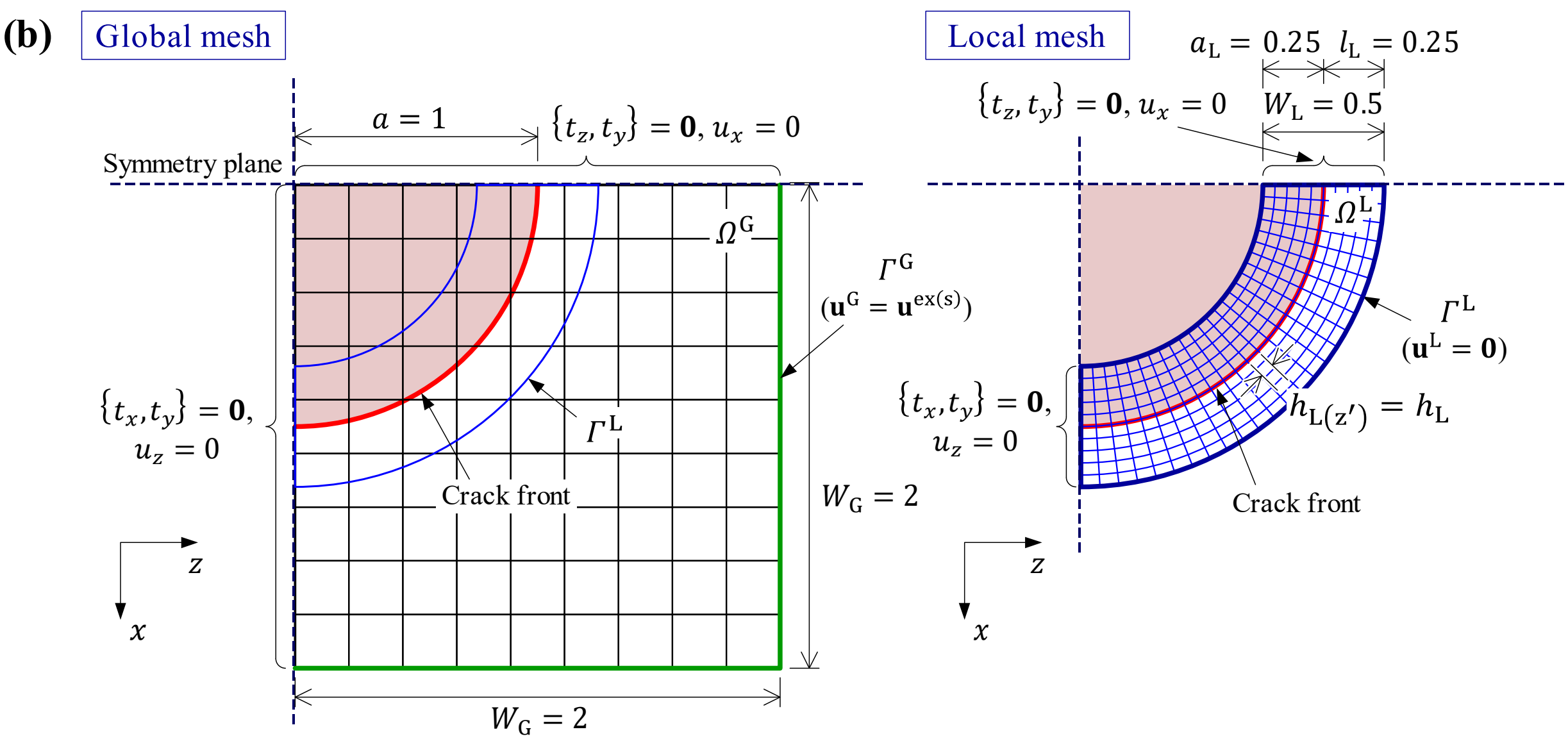

**Fig. 35.** Global and local meshes and their boundary conditions in the three-dimensional stationary crack problem: (a) $yz$ view; (b) $xz$ view.

Convergence studies of the relative $L_2$ error norm, $e_{L_2}$, defined in Eq. (54), were conducted for the global-to-local element-size ratios $r_{\mathrm{GL}} = h_{\mathrm{G}}/h_{\mathrm{L}} = 2$, 4 and 8. Fig. 36(a) summarises the convergence results of $e_{L_2}$ versus the degrees of freedom (DOF). For all examined $r_{\mathrm{GL}}$, the proposed hS-IGA strategy eventually exhibits the optimal convergence behaviour, indicating that the fundamental approximation accuracy is maintained in the three-dimensional stationary crack problem.

The proposed strategy is further compared with the conventional s-method for the representative cases $r_{\mathrm{GL}} = 4$ and 8, as shown in Figs. 36(b1) and 36(b2), respectively. In both cases, once a stable optimal convergence slope is achieved, the hS-IGA results show smaller $e_{L_2}$ values than the conventional s-method at the same DOF. These results demonstrate that the proposed strategy preserves the fundamental approximation capability of the conventional s-method in three-dimensional problems while achieving comparable or better displacement-field accuracy.

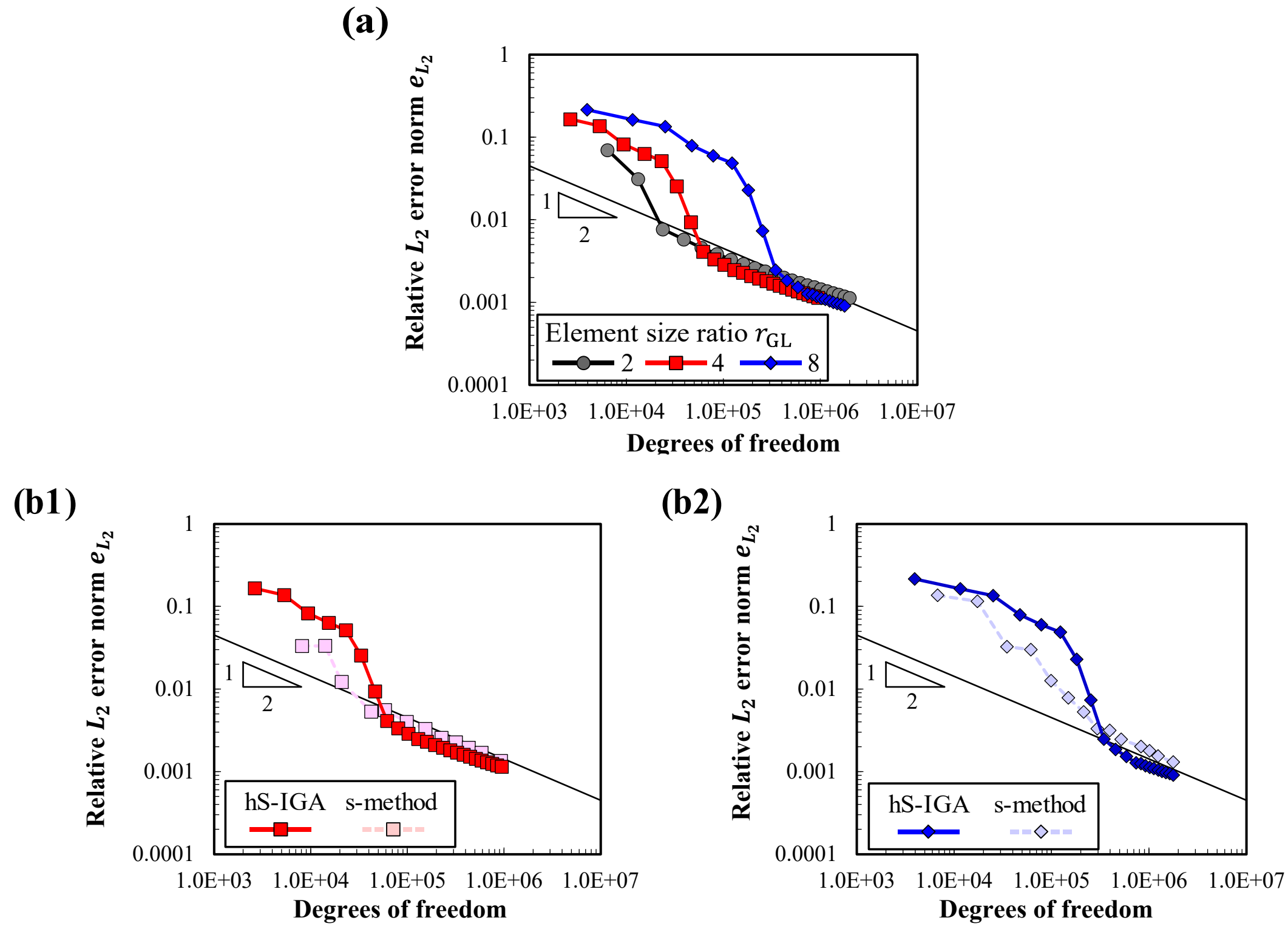

**Fig. 36.** Convergence results of the $L_2$ error norm under fixed $r_{GL}$: (a) proposed hS-IGA strategy for different $r_{GL}$; and comparison with the conventional Lagrange-based s-method for (b1) $r_{GL} = 4$, and (b2) $r_{GL} = 8$.

To clarify the principal computational advantage of the proposed strategy, the total numbers of integration points required by the hS-IGA strategy and the conventional s-method are compared for the representative cases $r_{GL} = 4$ and 8, corresponding to the results shown in Figs. 36(b1) and 36(b2). As in the two-dimensional verification in Section 4.1.2, the relationship between the relative $L_2$ error norm and the number of integration points is plotted on logarithmic scales, as summarised in Fig. 37. For both values of $r_{GL}$, the results of the two methods follow similar trends, whereas the hS-IGA curves are shifted markedly downward relative to those of the conventional s-method. This indicates that, for a comparable level of displacement-field accuracy, the proposed strategy requires substantially fewer integration points.

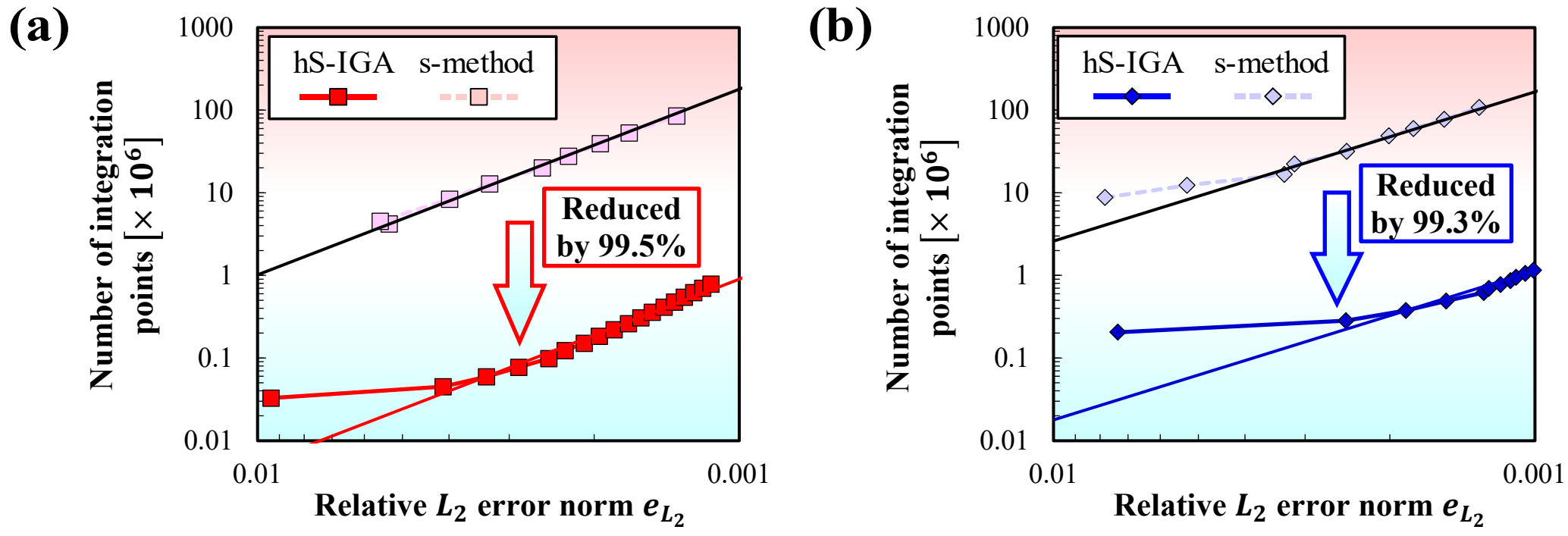

**Fig. 37.** Relationship between the relative $L_2$ error norm and the number of integration points for the hS-IGA and the conventional Lagrange-based s-method under: (a) $r_{GL} = 4$, (b) $r_{GL} = 8$.

The reduction in integration cost is quantified using the same shared-slope regression procedure as that introduced in Section 4.1.2. Based on the fitted relations, the number of integration points required to achieve a matched error level is reduced by approximately 99.5% for $r_{\mathrm{GL}} = 4$ and 99.3% for $r_{\mathrm{GL}} = 8$. This reduction is more pronounced than that observed in the two-dimensional cases, reflecting the fact that subdivision-based coupling integration becomes considerably more expensive in three-dimensional mesh superposition. These results confirm that the proposed hS-IGA strategy effectively removes the principal integration-cost bottleneck of the conventional s-method in the three-dimensional stationary crack problem while preserving comparable accuracy.

Based on the methodology presented in Section 2.3, the static stress intensity factor $K_{\mathrm{I}}^{(\mathrm{s})}$ and local stress $\sigma_{yy}$ ahead of and along the crack front are evaluated. To verify the accuracy of the proposed strategy, the results are compared with those obtained by the conventional s-method. Two representative local mesh resolutions are examined: $h_{\mathrm{L}} = 1/48$ and $1/88$. For both the hS-IGA strategy and the conventional s-method, the corresponding global element sizes are set to $h_{\mathrm{G}} = 2/25$ and $2/45$, respectively.

Representative deformed configurations and local $\sigma_{yy}$ fields for these two mesh cases are shown in Fig. 38. The deformation is magnified by a scale factor of 10, and the $\sigma_{yy}$ fields are displayed only in the local meshes. Although the global and local meshes are generated independently in the s-method framework, the magnified views show smooth deformed configurations and consistent near-front stress distributions around the crack front. These results visually demonstrate that the proposed hS-IGA strategy can accurately represent the local crack-front fields while retaining the independent global–local meshing structure.

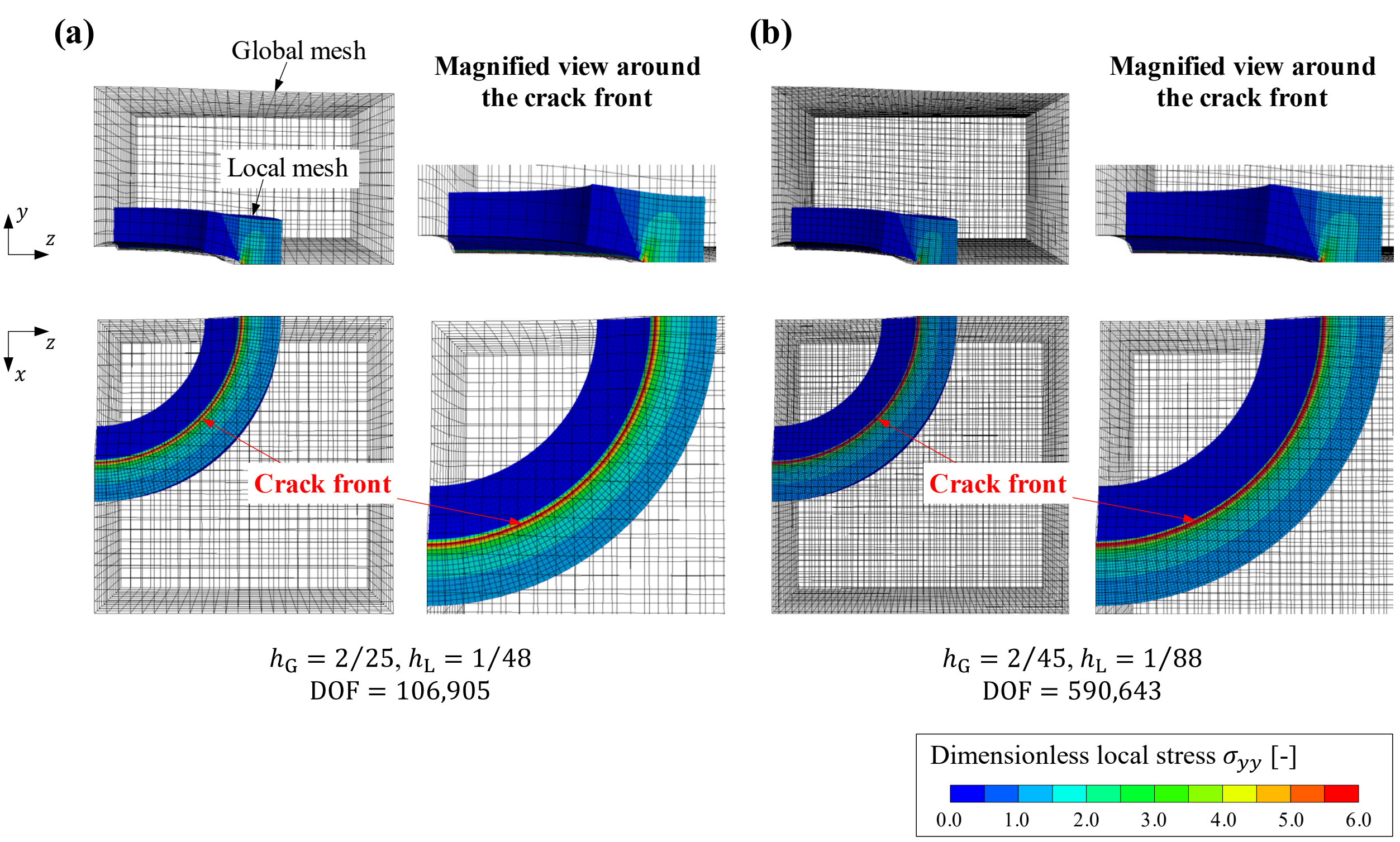

**Fig. 38.** Representative deformed configurations, magnified views around the crack front, and dimensionless $\sigma_{yy}$ fields for the three-dimensional stationary circular-crack problem (scale factor = 10; $\sigma_{yy}$ fields are displayed only in the local meshes): (a) $h_{\mathrm{G}} = 2/25$, $h_{\mathrm{L}} = 1/48$, DOF = 106,905; (b) $h_{\mathrm{G}} = 2/45$, $h_{\mathrm{L}} = 1/88$, DOF = 590,643.

The corresponding quantitative comparisons are presented in Fig. 39, which shows the normalised stress intensity factor, $K_{\mathrm{I}}^{(\mathrm{s})}/K_{\mathrm{I}}^{\mathrm{ex(s)}}$, and the normalised local stress, $\sigma_{yy}/\sigma_{yy}^{\mathrm{ex(s)}}$, obtained by the two methods. Fig. 39(a) presents $K_{\mathrm{I}}^{(\mathrm{s})}/K_{\mathrm{I}}^{\mathrm{ex(s)}}$ along the $z'$-direction, where the circumferential coordinate is defined as $\theta = \tan^{-1}(z/x)$. The results show that the proposed hS-IGA strategy evaluates the stress intensity factor with accuracy comparable to, or slightly higher than, that of the conventional s-method.

Fig. 39(b) presents the average value of $\sigma_{yy}/\sigma_{yy}^{\mathrm{ex(s)}}$ along the $x'$-direction. The error bars indicate the range from the minimum to maximum values along the $z'$-direction. Both methods accurately evaluate the local stress, and errors of less than 2.0% are obtained when the evaluation point satisfies $x' \geq 5h_{\mathrm{L}}$ near the crack front. However, the proposed hS-IGA strategy provides slightly higher accuracy when the evaluation point is close to the crack front. In addition, as in the two-dimensional stationary crack problem discussed in Section 4.1, the conventional s-method shows reduced accuracy when the evaluation point approaches $\Gamma^{\mathrm{GL}}$. This behaviour is attributed to the dependence of the local-stress accuracy on $h_{\mathrm{G}}$, as reported in our previous work [12]. By contrast, the hS-IGA results show no clear $h_{\mathrm{G}}$-dependence in this region.

Fig. 39(c) presents $\sigma_{yy}/\sigma_{yy}^{\mathrm{ex(s)}}$ along the $z'$-direction at $x' = 5h_{\mathrm{L}}$. The results further confirm that the proposed hS-IGA strategy provides more accurate local-stress evaluation along the crack front than the conventional s-method.

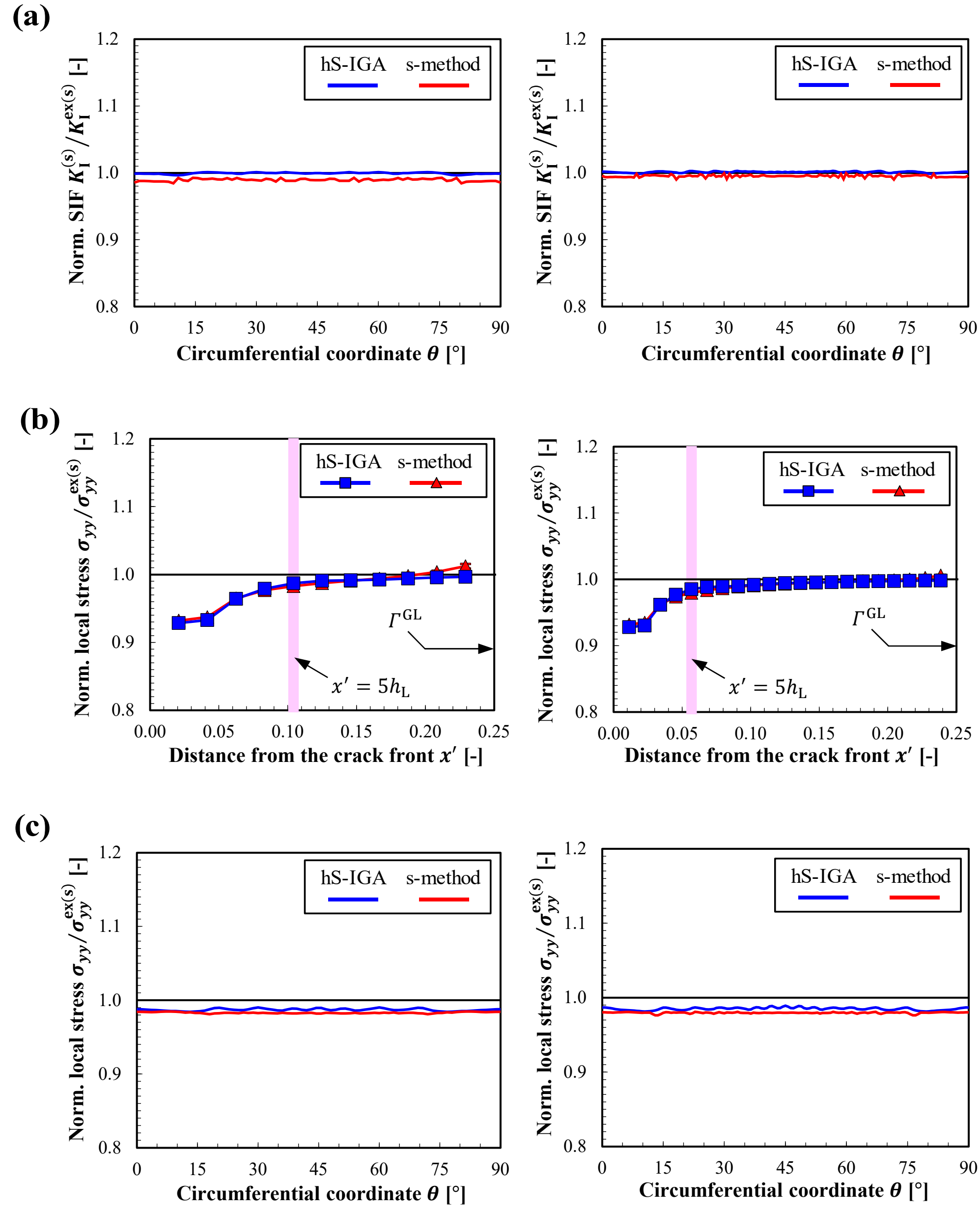

**Fig. 39.** Accuracy of fracture-quantity evaluation in the three-dimensional stationary crack problem: (a) $K_{\mathrm{I}}^{(\mathrm{s})}/K_{\mathrm{I}}^{\mathrm{ex(s)}}$ along the $z'$-direction; (b) $\sigma_{yy}/\sigma_{yy}^{\mathrm{ex(s)}}$ along the $x'$-direction; and (c) $\sigma_{yy}/\sigma_{yy}^{\mathrm{ex(s)}}$ along the $z'$-direction at $x' = 5h_{\mathrm{L}}$.

### 5.2. Three-dimensional dynamic crack problem

The proposed hS-IGA strategy is further verified through a dynamically propagating circular crack in an infinite three-dimensional solid. In this benchmark, the appropriate mesh conditions required for accurate and computationally efficient evaluation of the dynamic stress intensity factor and the local stress are clarified. The results obtained under the appropriate conditions are then compared with those of the conventional s-method to assess whether comparable fracture-quantity accuracy and reduced coupling-integration cost are maintained during three-dimensional crack propagation.

#### 5.2.1. Problem description

The dynamic benchmark considered in this section is a circular crack propagating dynamically in an infinite three-dimensional solid at a prescribed constant velocity $V$, as shown in Fig. 40. Although this problem is one of the simplest three-dimensional dynamic crack propagation problems, it is suitable for verifying the fundamental applicability of the proposed strategy because various relative configurations between the global and local elements appear during crack propagation.

The target problem is shown schematically in Fig. 40. As in the stationary crack problem described in Section 5.1, a one-eighth symmetric model is employed. The width and height of the global domain are defined as 8 mm and 4 mm, respectively. The circular crack propagates dynamically with its radius increasing from 0 mm, corresponding to the initial uncracked state, to $a_{\max} = 4$ mm.

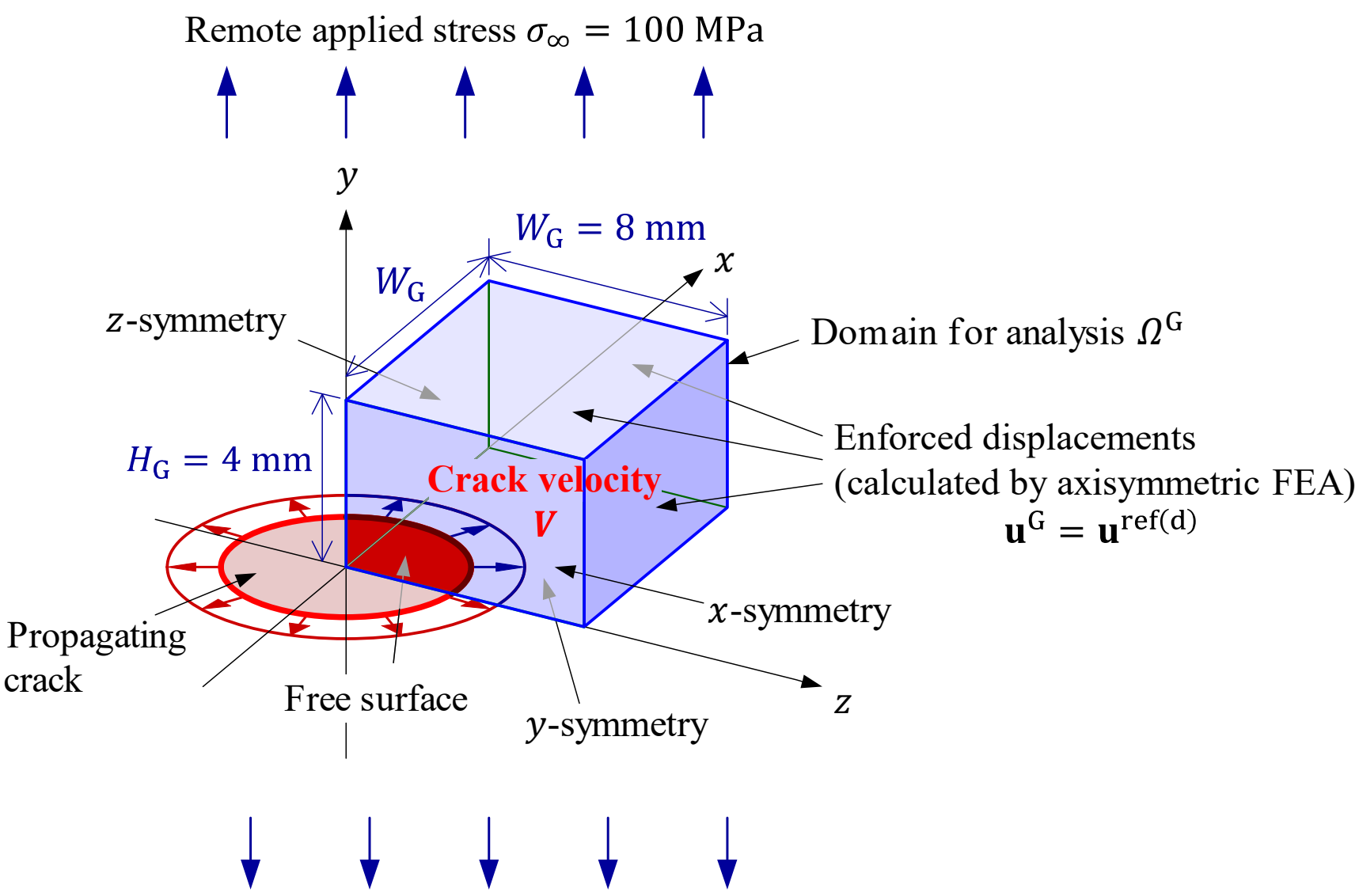

**Fig. 40.** Three-dimensional dynamic crack benchmark problem.

As in the two-dimensional dynamic problem discussed in Section 4.2.1, the exact solutions of the displacement and stress fields for the dynamically propagating circular-crack problem have not yet been clarified. Therefore, a reference-analysis procedure similar to that in Section 4.2.1 is employed. Specifically, two-dimensional axisymmetric finite element analyses (2D-A-FEA) are conducted with sufficiently fine meshes in the vicinity of the crack surface to obtain accurate reference solutions. The detailed procedure for obtaining the reference solution is provided in our previous study [13]. The reference displacement histories obtained by 2D-A-FEA, $\mathbf{u}^{\mathrm{ref(d)}}$, are imposed as time-dependent Dirichlet boundary conditions on the outer boundary of the target domain, $\Gamma^{\mathrm{G}}$, as shown in Fig. 40. In addition, the dynamic stress intensity factor $K_{\mathrm{I}}^{\mathrm{ref(d)}}$ and the local stress $\sigma_{yy}^{\mathrm{ref(d)}}$ obtained from 2D-A-FEA are used as reference values for the following verification.

*5.2.2. Verification of the three-dimensional dynamic crack problem*

The global and local meshes, together with the associated boundary conditions used in the verification analyses, are shown in Fig. 41. As described in Section 5.2.1, the time-dependent Dirichlet boundary conditions imposed on the outer boundary of the target domain, $\Gamma^{\mathrm{G}}$, are prescribed from the displacement histories obtained by the 2D-A-FEA reference analysis, $\mathbf{u}^{\mathrm{ref(d)}}$. The global mesh is defined over the target domain with width $W_{\mathrm{G}} = 8$ mm and height $H_{\mathrm{G}} = 4$ mm, while the local mesh is superposed in the vicinity of the propagating crack front. The local element size in the $x'$-direction is fixed at $h_{\mathrm{L}} = 0.04$ mm, so that the local stress is evaluated at $x' = 5h_{\mathrm{L}} =$ 0.2mm, which satisfies the accuracy condition identified in the stationary verification and corresponds to the characteristic distance adopted in previous studies on brittle crack propagation in steels [10,27,50,51]. The local element size in the $y'$-direction is set to $h_{\mathrm{L}(y')} = h_{\mathrm{L}}$. For the crack-front direction, a constant angular interval is used throughout the analysis. This interval is determined from the final crack radius $a_{\max} = 4$ mm, so that the local element size in the $z'$-direction at the crack front becomes $h_{\mathrm{L}(z')}\big|_{x'=0} = 2h_{\mathrm{L}}$, following the mesh condition adopted in previous studies[17,18]. In the present analysis, this gives an angular interval of $90/79°$, as shown in Fig. 41.

With the analysis framework defined above, the local-domain and mesh conditions required for accurate and computationally efficient evaluation of the near-front fracture quantities are identified for the three-dimensional dynamic crack problem. Following the same strategy as in the two-dimensional dynamic verification, four parameters are examined: (a) the global-to-local element-size ratio, $r_{\mathrm{GL}} = h_{\mathrm{G}}/h_{\mathrm{L}}$; (b) the crack length in the local domain, $a_{\mathrm{L}}$; (c) the ligament length in the local domain, $l_{\mathrm{L}}$; and (d) the height of the local domain, $H_{\mathrm{L}}$. Their effects on the dynamic stress intensity factor, $K_{\mathrm{I}}^{(\mathrm{d})}$, and the local stress, $\sigma_{yy}$, are assessed at the maximum crack length, $a_{\max} = 4$ mm, which corresponds to $100h_{\mathrm{L}}$. Accordingly, crack propagation from $a = 0$ to $a = a_{\max}$ is simulated using 100 nodal-force-release steps.

Systematic analyses are performed for crack velocities of $V = 500\,\mathrm{m/s}$ and $V = 1000\,\mathrm{m/s}$. The resulting accuracy requirements are expressed in a unified form for both $K_{\mathrm{I}}^{(\mathrm{d})}$ and $\sigma_{yy}$ as $r_{\mathrm{GL}} \geq 4$, $a_{\mathrm{L}} \geq 2.5h_{\mathrm{G}}$, $l_{\mathrm{L}} \geq \sqrt{2}h_{\mathrm{G}}$, and $H_{\mathrm{L}} \geq 1.8h_{\mathrm{G}}$. Under these conditions, the deviations from the corresponding FEA reference solutions remain within 1.3% for $K_{\mathrm{I}}^{(\mathrm{d})}$ and 2.5% for $\sigma_{yy}$.

Representative deformed configurations and local $\sigma_{yy}$fields at selected propagation stages under the mesh conditions satisfying the above accuracy requirements are shown in Fig. 42. This representative case corresponds to $V = 500\,\mathrm{m/s}$, $r_{\mathrm{GL}} = 8$, $a_{\mathrm{L}} = 2.5h_{\mathrm{G}}$, $l_{\mathrm{L}} = \sqrt{2}h_{\mathrm{G}}$, and $H_{\mathrm{L}} = 1.8h_{\mathrm{G}}$. The configurations are shown with a deformation scale factor of 400, and the $\sigma_{yy}$ fields are displayed only in the local meshes to highlight the near-front stress field. The magnified views around the crack front further visualise the local mesh deformation and the evolution of the near-front stress distribution during crack propagation.

Since the parameter effects are examined using the same procedure as in the two-dimensional dynamic verification, only representative results are presented below to focus on the three-dimensional verification. Unless otherwise stated, the parameter values satisfying the above criteria are used for all parameters except the one being varied.

*(a) Global-to-local element-size ratio $r_{GL}$ (Fig. 43)*

Instabilities in the evaluated $K_{\mathrm{I}}^{(\mathrm{d})}$ and $\sigma_{yy}$ are observed when $r_{\mathrm{GL}} < 4$, suggesting that $r_{\mathrm{GL}} \geq 4$ is required for stable evaluation. This indicates that, as in the two-dimensional dynamic problem, the accuracies of these near-front quantities are governed mainly by the local resolution rather than by further refinement of the global mesh. It should be noted that, if the ligament-length condition $l_{\mathrm{L}} = \sqrt{2}h_{\mathrm{G}}$, identified below, is applied directly, then for $r_{\mathrm{GL}} = 3$ the resulting ligament length becomes smaller than $5h_{\mathrm{L}}$, making the prescribed local-stress evaluation at $x' = 5h_{\mathrm{L}}$ impossible. Therefore, for this case, $l_{\mathrm{L}}$ is instead set to $6h_{\mathrm{L}}$.

*(b) Crack length in the local domain $a_L$ (Fig. 44)*

The crack length in the local domain, $a_{\mathrm{L}}$, has a strong influence on the evaluation accuracy. The results indicate that the accuracies of $K_{\mathrm{I}}^{(\mathrm{d})}$ and $\sigma_{yy}$ are governed by the ratio $a_{\mathrm{L}}/h_{\mathrm{G}}$, and remain stable when $a_{\mathrm{L}} \geq 2.5h_{\mathrm{G}}$.

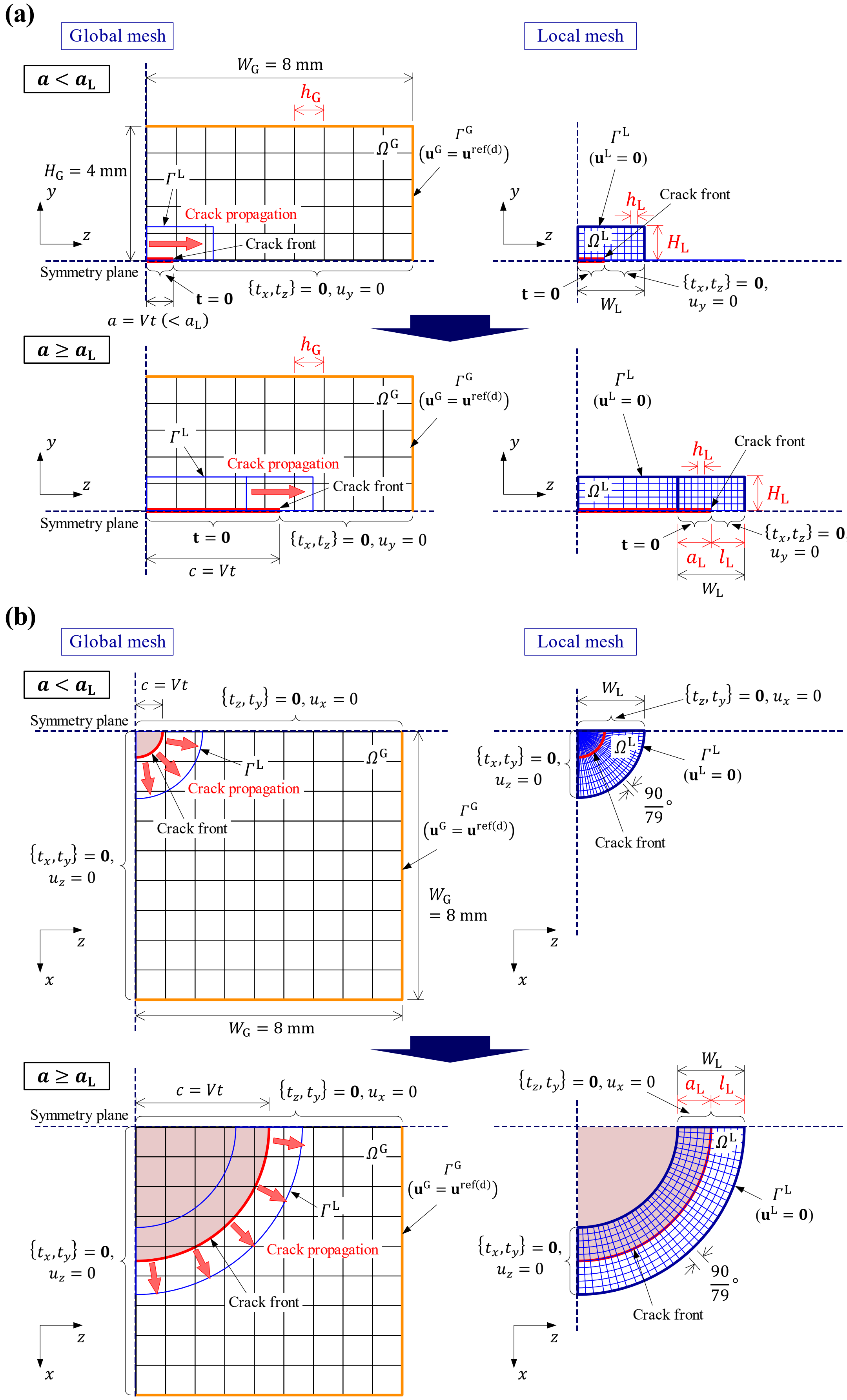

**Fig. 41.** Global and local meshes and their boundary conditions in the three-dimensional dynamic crack propagation problem: (a) $yz$ view; (b) $xz$ view.

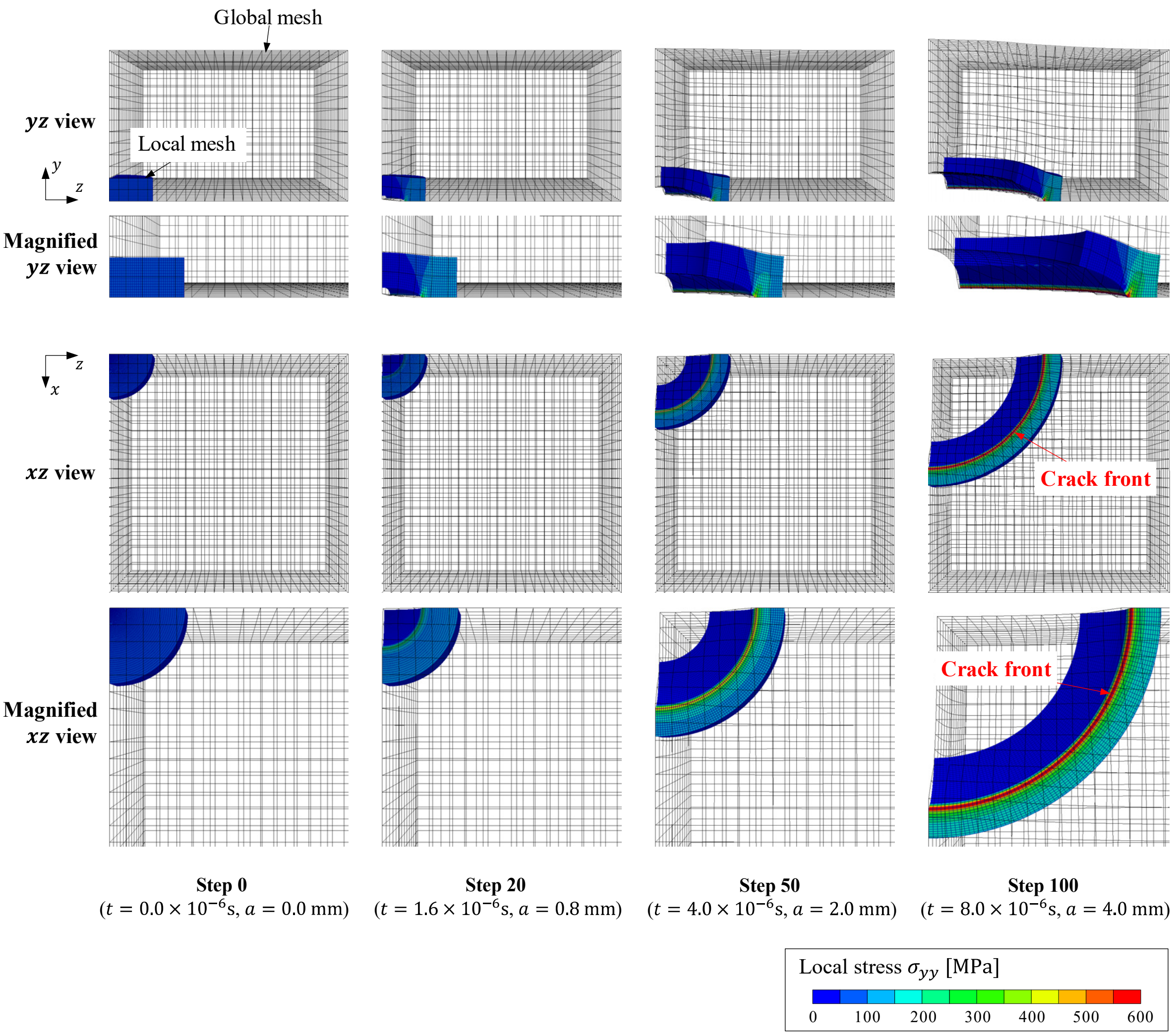

**Fig. 42.** Representative deformed configurations, magnified crack-front views, and $\sigma_{yy}$ fields for the three-dimensional dynamic crack propagation analysis at $V = 500$ m/s ($r_{\mathrm{GL}} = 8$, $a_{\mathrm{L}} = 2.5h_{\mathrm{G}}$, $l_{\mathrm{L}} = \sqrt{2}h_{\mathrm{G}}$, and $H_{\mathrm{L}} = 1.8h_{\mathrm{G}}$; scale factor = 400; $\sigma_{yy}$ fields are displayed only in the local meshes).

*(c) Ligament length in the local domain $l_L$ (Fig. 45)*

The ligament length also affects the stability of the evaluation. As in the two-dimensional dynamic problem, the accuracy deterioration near $\Gamma^{\mathrm{GL}}$, which is observed in the conventional s-method, is not significant in the proposed hS-IGA strategy. However, unlike in the two-dimensional case, the three-dimensional problem requires consideration of the geometric relationship between the local domain and the global elements not only in the $x'$-direction but also in the crack-front direction. Instabilities appear when $l_{\mathrm{L}}$ is smaller than the diagonal length of a global element in the $x'z'$-plane. Therefore, the appropriate condition is identified as $l_{\mathrm{L}} \geq \sqrt{2}h_{\mathrm{G}}$.

*(d) Height of the local domain $H_L$ (Fig. 46)*

The results indicate that the appropriate condition for ensuring accurate evaluation of both $K_{\mathrm{I}}^{(\mathrm{d})}$ and $\sigma_{yy}$ is $H_{\mathrm{L}} \geq 1.8h_{\mathrm{G}}$.

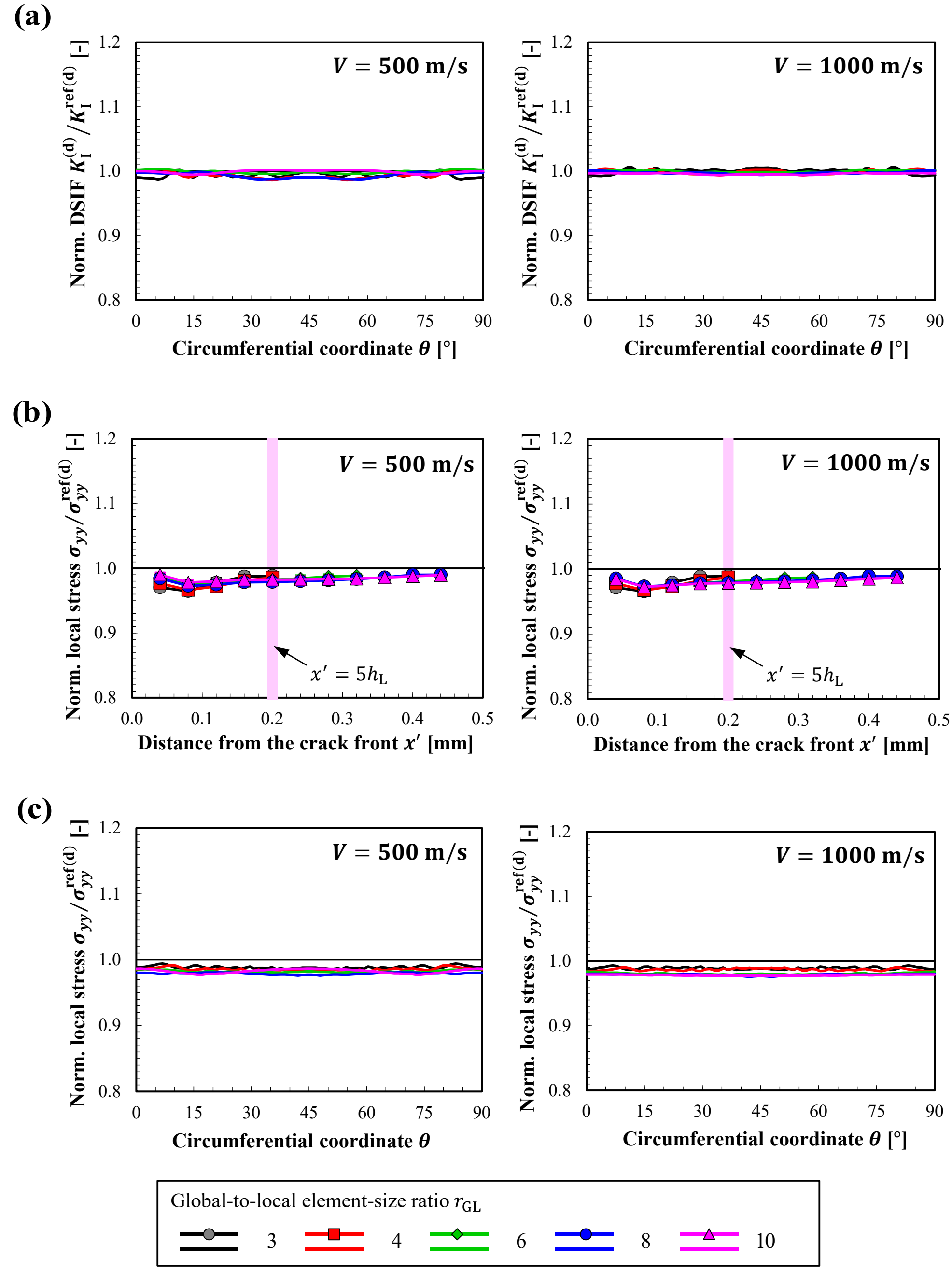

**Fig. 43.** Influence of the global-to-local element-size ratio $r_{GL}$ ($a$ = 4 mm, $a_L = 2.5h_G$, $l_L = \sqrt{2}h_G$, $H_L = 1.8h_G$): (a) Dynamic stress intensity factor distributions along the crack front; (b) local stress distributions in crack propagation direction; (c) local stress distributions at $x' = 0.2$ mm ($= 5h_L$) along the crack front.

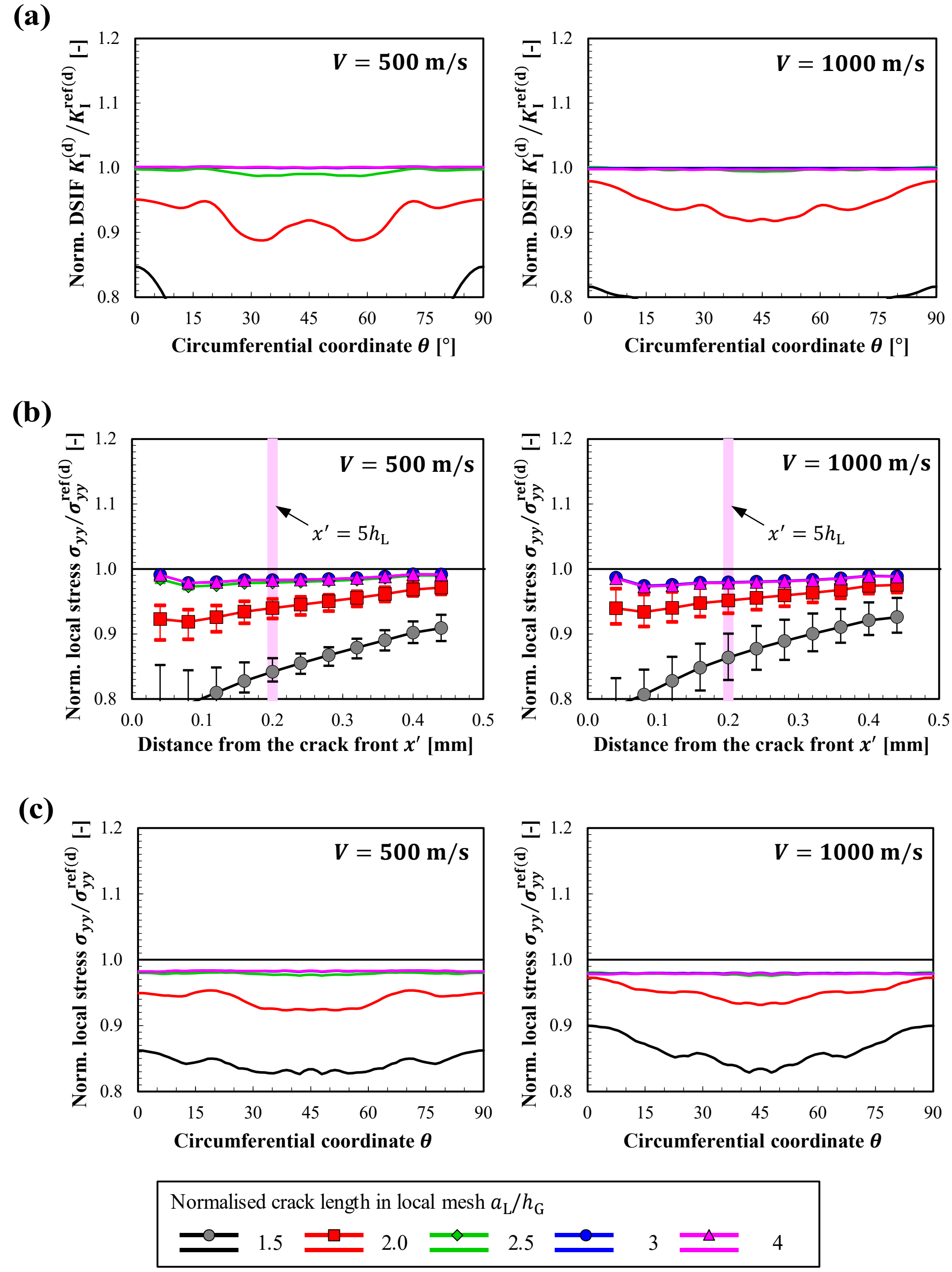

**Fig. 44.** Influence of the crack length in the local domain $a_L$ ($a$ = 4 mm, $r_{GL}$ = 8, $l_L = \sqrt{2}h_G$, $H_L = 1.8h_G$): (a) Dynamic stress intensity factor distributions along the crack front; (b) local stress distributions in crack propagation direction; (c) local stress distributions at $x'$ = 0.2 mm (= $5h_L$) along the crack front.

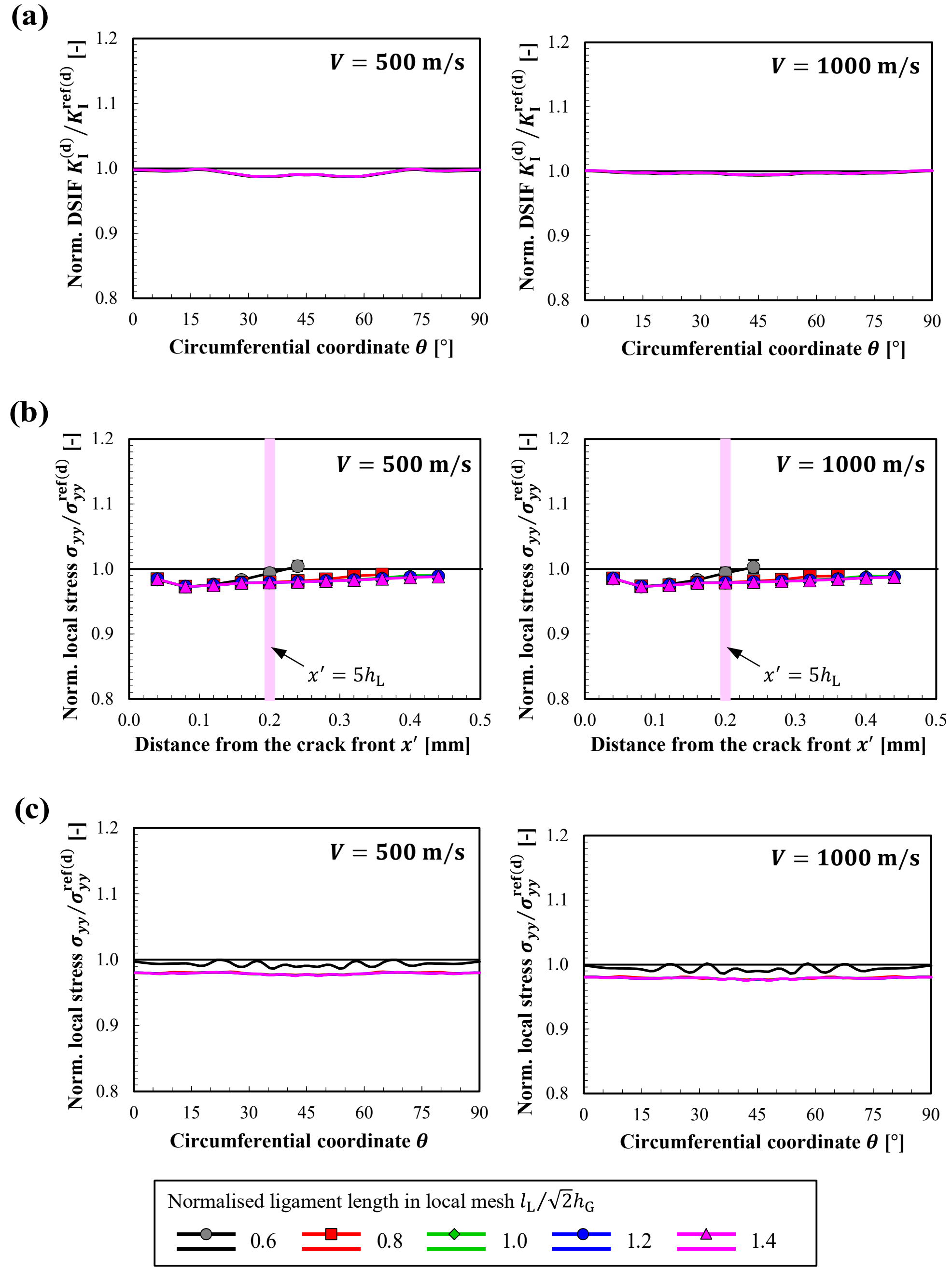

**Fig. 45.** Influence of the ligament length in the local domain $l_L$ ($a = 4$ mm, $r_{GL} = 8$, $a_L = 2.5h_G$, $H_L = 1.8h_G$): (a) Dynamic stress intensity factor distributions along the crack front; (b) local stress distributions in crack propagation direction; (c) local stress distributions at $x' = 0.2$ mm ($= 5h_L$) along the crack front.

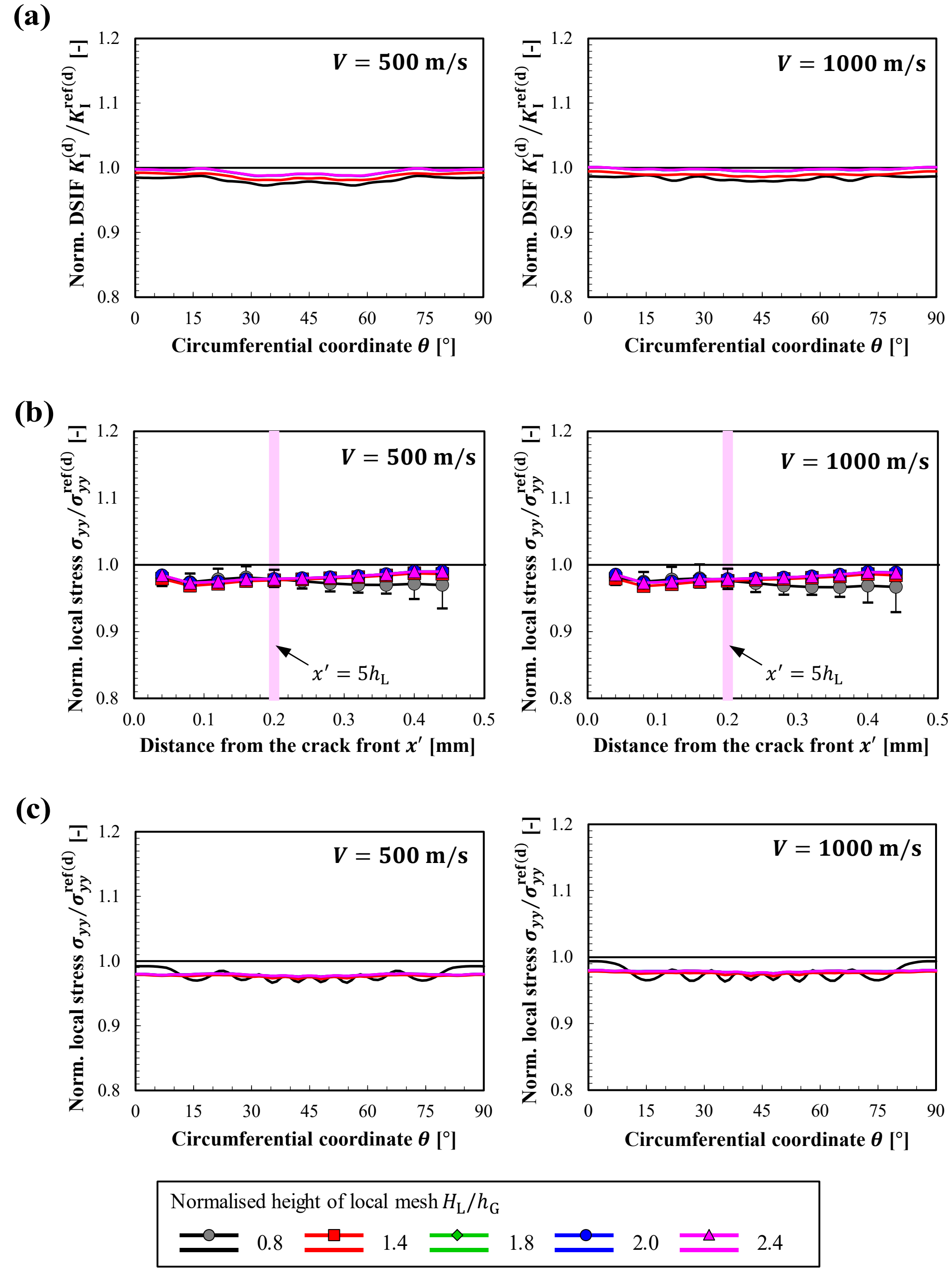

**Fig. 46.** Influence of the height of the local domain $H_L$ ($a$ = 4 mm, $r_{GL}$ = 8, $a_L = 2.5h_G$, $l_L = \sqrt{2}h_G$): (a) Dynamic stress intensity factor distributions along the crack front; (b) local stress distributions in crack propagation direction; (c) local stress distributions at $x' = 0.2$ mm $(= 5h_L)$ along the crack front.

As in the two-dimensional dynamic problem, the stability of the proposed hS-IGA strategy is examined over a wide range of crack velocities. The parameters are set to $r_{\mathrm{GL}} = 8$, $a_{\mathrm{L}} = 2.5h_{\mathrm{G}}$, $l_{\mathrm{L}} = \sqrt{2}h_{\mathrm{G}}$, and $H_{\mathrm{L}} = 1.8h_{\mathrm{G}}$, satisfying the conditions identified above. The accuracies of $K_{\mathrm{I}}^{(\mathrm{d})}$ and $\sigma_{yy}$ are then evaluated for crack velocities ranging from $V = 200\ \mathrm{m/s}$ to $1500\ \mathrm{m/s}$ in increments of $100\ \mathrm{m/s}$, with the maximum crack length fixed at $a_{\max} = 4$ mm. As summarised in Fig. 47, the proposed hS-IGA strategy agrees well with the FEA reference solutions over the entire velocity range. The maximum deviations are less than 0.8% for the normalised dynamic stress intensity factor and less than 2.3% for the normalised local stress. In addition, the maximum differences from the corresponding mean values along the crack front are only 0.9% for $K_{\mathrm{I}}^{(\mathrm{d})}$ and 0.7% for $\sigma_{yy}$, confirming the stability of the proposed strategy in the crack-front direction.

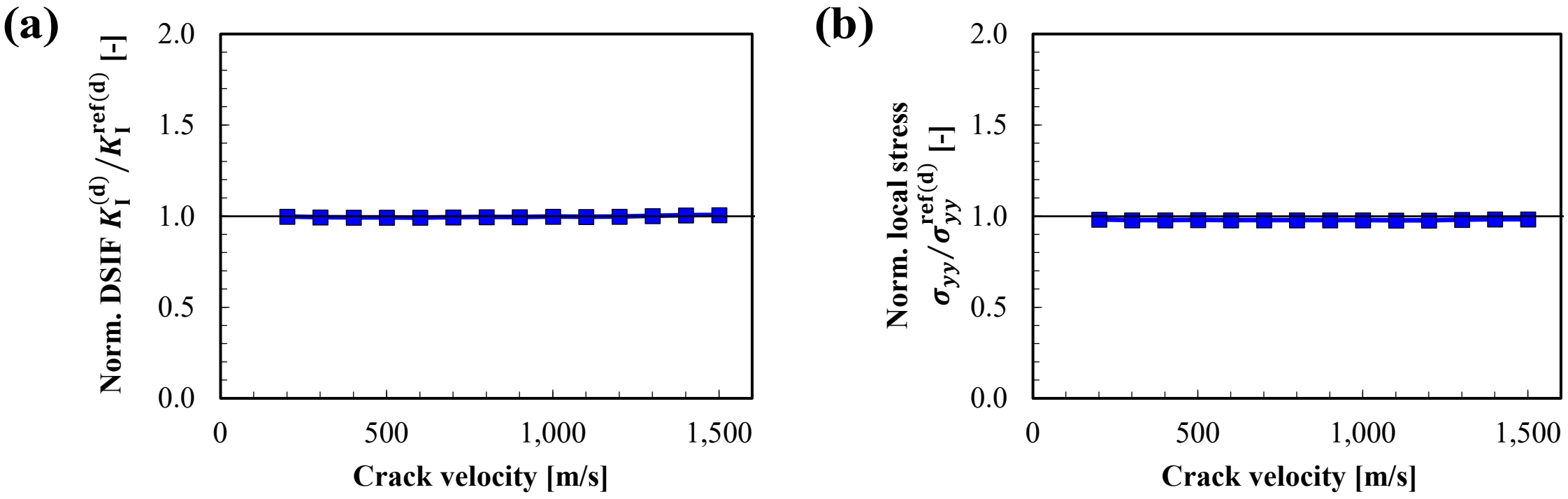

**Fig. 47.** Verification for three-dimensional dynamic crack problem over a wide range of crack velocities: (a) Normalised dynamic stress intensity factor; (b) normalised local stress at $x' = 0.2$ mm $(= 5h_{\mathrm{L}})$.

To further assess the effectiveness of the proposed strategy in the three-dimensional dynamic crack problem, the results obtained by the hS-IGA strategy are compared with those obtained by the conventional s-method. In the present three-dimensional analysis, the target domain size is fixed, and the circular crack propagates up to $a_{\max} = 4.0$ mm. Therefore, the comparison is performed at several representative propagation stages: $a = 0.8$, 1.6, 2.4, 3.2, and 4.0 mm, corresponding to 20, 40, 60, 80, and 100 nodal-force-release steps, respectively. In all cases, the same target domain and the same time-dependent displacement boundary conditions are applied, so that the differences among the compared models arise only from their approximation and coupling-integration frameworks.

For the hS-IGA strategy, the parameters are set to $r_{\mathrm{GL}} = 8$, $a_{\mathrm{L}} = 2.5h_{\mathrm{G}}$, $l_{\mathrm{L}} = \sqrt{2}h_{\mathrm{G}}$, and $H_{\mathrm{L}} = 1.8h_{\mathrm{G}}$ according to the conditions identified in the present study. For the conventional s-method, two global mesh densities are considered based on the conditions identified in our previous study [13]: a coarse-global-mesh model with $r_{\mathrm{GL}} = 8$, and a fine-global-mesh model with $r_{\mathrm{GL}} = 6$. In both conventional s-method models, the local mesh conditions are set to $a_{\mathrm{L}} = 20h_{\mathrm{L}}$, $l_{\mathrm{L}} = 15h_{\mathrm{L}}$, and $H_{\mathrm{L}} = 10h_{\mathrm{L}}$.

Under these conditions, the accuracies of the proposed hS-IGA strategy and the two conventional s-method models are compared. Fig. 48 summarises the results for the normalised dynamic stress intensity factor and the normalised local stress at the representative crack velocities $V = 500\ \mathrm{m/s}$ and $V = 1000\ \mathrm{m/s}$. The results show that all three models agree well with the FEA reference solutions throughout the crack propagation process. The proposed hS-IGA strategy achieves accuracy comparable to that of both conventional s-method models, with maximum deviations from the reference solutions remaining below 1.0% for the normalised dynamic stress intensity factor and below 2.4% for the normalised local stress.

Following the accuracy comparison in Fig. 48, the computational efficiency of the three models is examined in terms of both the number of degrees of freedom (DOFs) and the number of integration points. The coarse- and fine-global-mesh conventional s-method models require $\mathrm{DOF}_{\text{s-method}} = 82{,}288$ and $161{,}190$, respectively, whereas the proposed hS-IGA strategy requires $\mathrm{DOF}_{\text{hS-IGA}} = 106{,}350$. Thus, the DOF level of the proposed strategy lies between those of the two conventional s-method models. The differences in DOFs mainly originate from the global discretisation: the fine conventional s-method model increases the number of DOFs by refining the global mesh, whereas the hS-IGA strategy introduces additional global control points owing to the support and arrangement of the B-spline basis functions. In contrast, the local crack-front resolution, which is essential for evaluating the near-front fracture quantities, is kept at a comparable level among the three models.

As shown in Fig. 48, despite these differences in global discretisation and total DOFs, the three models provide comparable accuracy for both $K_{\mathrm{I}}^{(\mathrm{d})}$ and $\sigma_{yy}$. This result is consistent with the observations in Sections 4.1.2 and 4.2.2, where the accuracies of the near-tip/front quantities were found to be governed mainly by the local mesh resolution rather than by further refinement of the global mesh. Therefore, within the present global–local framework and parameter range, the accuracy of the proposed hS-IGA strategy should not be attributed simply to an increase in DOFs. Instead, once the local mesh provides sufficient resolution around the crack front, the main computational issue is the cost of global–local coupling integration.

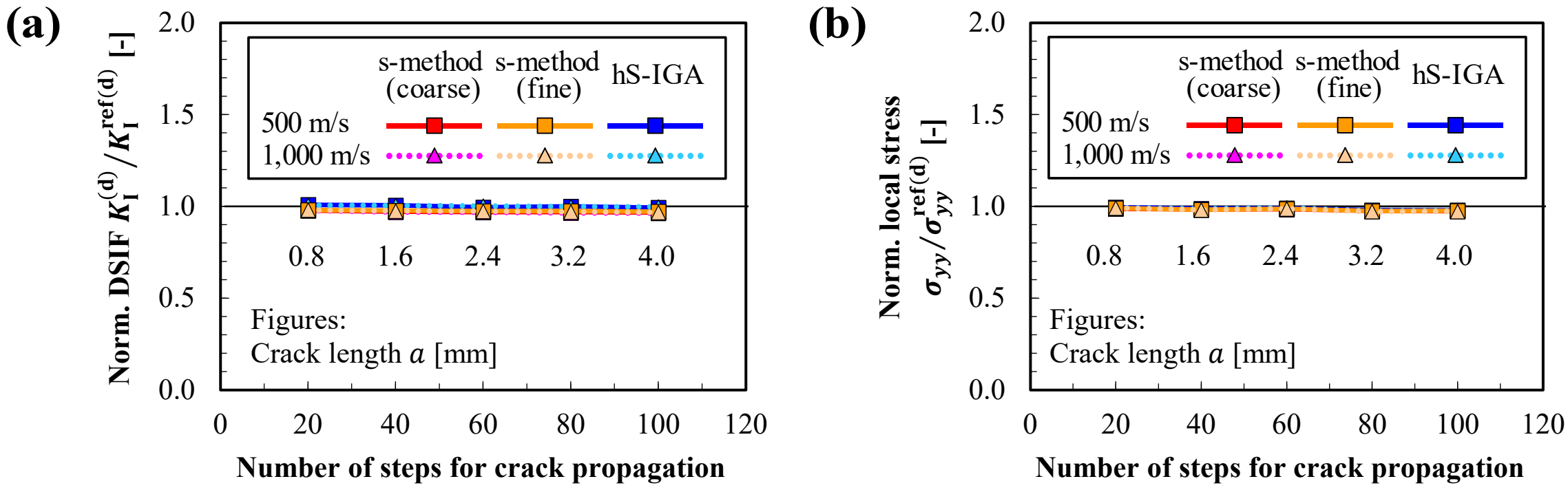

**Fig. 48.** Accuracy comparison between the hS-IGA and the conventional s-method for three-dimensional dynamic crack propagation analysis at representative propagation stages: (a) normalised dynamic stress intensity factor; (b) normalised local stress at $x' = 0.2$ mm $(= 5h_{\mathrm{L}})$.

This difference in integration cost is clarified in Fig. 49, which compares the cumulative number of integration points required by the proposed hS-IGA strategy and the two conventional s-method models at the representative propagation stages considered in Fig. 48. Although all three models achieve comparable fracture-quantity accuracy, the proposed hS-IGA strategy requires substantially fewer integration points throughout the crack propagation analysis. At the maximum crack length considered, $a_{\max} = 4.0$ mm, the cumulative number of integration points is reduced by approximately 95.6% relative to the coarse-global-mesh conventional s-method model and 96.6% relative to the fine-global-mesh conventional s-method model. This reduction is even more pronounced than that observed in the two-dimensional dynamic problem, indicating that the computational advantage of the proposed hS-IGA strategy becomes more significant in three-dimensional crack propagation analysis. These results confirm that the proposed strategy effectively alleviates the principal coupling-integration bottleneck of the conventional s-method while maintaining comparable accuracy.

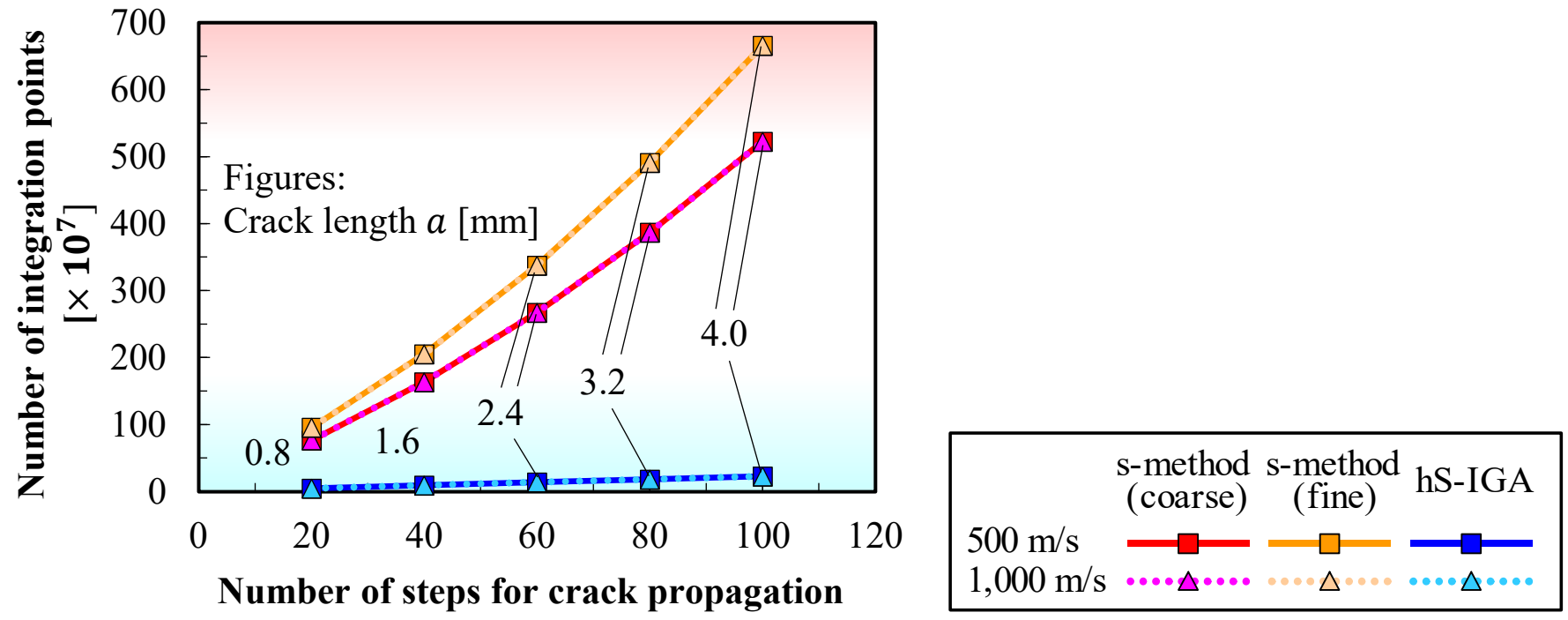

**Fig. 49.** Computational efficiency comparison for the three-dimensional dynamic crack problem: cumulative number of integration points required by the proposed hS-IGA strategy and the conventional s-method at representative propagation stages.

## 6. Conclusions

In this study, a hybrid s-version of isogeometric analysis (hS-IGA) strategy was proposed for the accurate and computationally efficient evaluation of near-tip/front fracture quantities in dynamic crack propagation analysis. The proposed strategy retains the global–local superposition framework of the conventional s-method, while introducing B-spline basis functions only into the global discretisation and preserving a standard Lagrange discretisation in the local crack domain.

The significance of this hybrid global/local discretisation lies in its ability to address the main computational bottleneck of the conventional Lagrange-based s-method without sacrificing the advantages of the local crack representation. In the conventional s-method, the limited inter-element continuity of the global Lagrange approximation causes discontinuities in the global–local coupling integrands, which necessitates recursive subdivision for accurate coupling integration. By contrast, the use of B-spline basis functions in the global mesh removes these discontinuities within local elements and enables accurate coupling integration by standard Gauss quadrature. At the same time, retaining Lagrange elements in the local mesh allows the dynamically propagating crack tip/front to be represented naturally and preserves the conventional procedures for nodal force release, local mesh update, and post-processing of the DSIF and local stress.

The proposed strategy was first verified through two-dimensional stationary and dynamically propagating straight-crack problems. The results showed that the hS-IGA strategy accurately reproduced the near-tip displacement, stress, DSIF, and local-stress fields, with accuracy comparable to that of the standard FEM and the conventional s-method. The two-dimensional benchmarks also confirmed that the proposed strategy retains the degree-of-freedom advantage of the s-method over the standard FEM, while effectively reducing the mesh-interface sensitivity observed in the conventional s-method. In the dynamic crack propagation problem, the total number of integration points was reduced by approximately 81% relative to the conventional s-method.

The applicability of the proposed strategy was further verified through three-dimensional stationary and dynamically propagating circular-crack problems. The hS-IGA strategy accurately evaluated the displacement field, stress intensity factor, DSIF, and local stress along the crack front, with accuracy comparable to that of the conventional three-dimensional s-method. More importantly, the computational advantage became more pronounced in the three-dimensional crack-front analyses. In the dynamically propagating circular-crack problem, the cumulative number of integration points was reduced by approximately 95.6% at the maximum crack length considered, compared with the conventional s-method model having the same global-to-local element-size ratio.

These results demonstrate that the proposed hS-IGA strategy provides an accurate and efficient global–local framework for dynamic crack propagation analysis requiring reliable near-tip/front fracture-quantity evaluation. Although the present study has been restricted to planar linear-elastic crack problems, the same concept is expected to be promising for more advanced applications, particularly three-dimensional elastoplastic fracture analyses, in which the computational burden associated with repeated coupling integration becomes even more significant.

## CRediT authorship contribution statement

**Tianyu He**: Writing – original draft, Visualization, Validation, Software, Methodology, Investigation, Funding acquisition, Formal analysis, Data curation. **Kosei Kurosaki**: Validation, Visualization. **Naoki Morita**: Software, Methodology. **Naoto Mitsume**: Methodology, Conceptualization. **Kazuki Shibanuma**: Writing – review & editing, Supervision, Project administration, Funding acquisition, Conceptualization.

## Funding

This study was supported by JSPS KAKENHI [grant numbers 22H00242 and 25KJ1181].

## Declaration of conflict of interest

The authors declare that they have no known competing financial interests or personal relationships that could have appeared to influence the work reported in this paper.

## Appendix A. Basic formulations of the conventional s-method

In the s-method, the displacement $\mathbf{u}(\mathbf{x})$ and acceleration $\ddot{\mathbf{u}}(\mathbf{x})$ at the physical coordinate $\mathbf{x}$ are defined as

$$\mathbf{u}(\mathbf{x}) = \begin{cases} \mathbf{u}^{\mathrm{G}}(\mathbf{x}) & \text{in } \Omega^{\mathrm{G}} \setminus \Omega^{\mathrm{L}} \\ \mathbf{u}^{\mathrm{G}}(\mathbf{x}) + \mathbf{u}^{\mathrm{L}}(\mathbf{x}) & \text{in } \Omega^{\mathrm{L}} \end{cases}, \tag{A1}$$

$$\ddot{\mathbf{u}}(\mathbf{x}) = \begin{cases} \ddot{\mathbf{u}}^{\mathrm{G}}(\mathbf{x}) & \text{in } \Omega^{\mathrm{G}} \setminus \Omega^{\mathrm{L}} \\ \ddot{\mathbf{u}}^{\mathrm{G}}(\mathbf{x}) + \ddot{\mathbf{u}}^{\mathrm{L}}(\mathbf{x}) & \text{in } \Omega^{\mathrm{L}} \end{cases}, \tag{A2}$$

where $\mathbf{u}^{\mathrm{G}}(\mathbf{x})$, $\mathbf{u}^{\mathrm{L}}(\mathbf{x})$ and $\ddot{\mathbf{u}}^{\mathrm{G}}(\mathbf{x})$, $\ddot{\mathbf{u}}^{\mathrm{L}}(\mathbf{x})$ denote the global and local contributions of displacement and acceleration, respectively. Thus, within the local domain $\Omega^{\mathrm{L}}$, both fields are represented as superpositions of global and local components.

Continuity of the displacement field on the local-domain boundary, $\Gamma^{\mathrm{GL}}$, is enforced through the Dirichlet condition

$$\mathbf{u}^{\mathrm{L}}(\mathbf{x}) = \mathbf{0} \quad \text{on } \Gamma^{\mathrm{GL}}. \tag{A3}$$

Accordingly, the corresponding velocity and acceleration on $\Gamma^{\mathrm{GL}}$ also vanish.

As mentioned in Section 2.1, we use the term "conventional (Lagrange-based) s-method" to denote the widely adopted implementation in which both the global and local domains are discretised using Lagrange basis functions [12,13,17–19,28–37]. Under this setting, the respective components of the displacement and acceleration in Eqs. (A1) and (A2) are approximated using Lagrange basis functions as:

$$\mathbf{u}^{\mathrm{G}}(\mathbf{x}) = \boldsymbol{\phi}^{\mathrm{G}}(\mathbf{x})\mathbf{d}^{\mathrm{G}}, \tag{A4}$$

$$\mathbf{u}^{\mathrm{L}}(\mathbf{x}) = \boldsymbol{\phi}^{\mathrm{L}}(\mathbf{x})\mathbf{d}^{\mathrm{L}}, \tag{A5}$$

$$\ddot{\mathbf{u}}^{\mathrm{G}}(\mathbf{x}) = \boldsymbol{\phi}^{\mathrm{G}}(\mathbf{x})\ddot{\mathbf{d}}^{\mathrm{G}}, \tag{A6}$$

$$\ddot{\mathbf{u}}^{\mathrm{L}}(\mathbf{x}) = \boldsymbol{\phi}^{\mathrm{L}}(\mathbf{x})\ddot{\mathbf{d}}^{\mathrm{L}}, \tag{A7}$$

where $\boldsymbol{\phi}^{\mathrm{G}}(\mathbf{x})$ and $\boldsymbol{\phi}^{\mathrm{L}}(\mathbf{x})$ denote the Lagrange basis functions corresponding to the global and local meshes, respectively. $\mathbf{d}^{\mathrm{G}}$, $\mathbf{d}^{\mathrm{L}}$, $\ddot{\mathbf{d}}^{\mathrm{G}}$, and $\ddot{\mathbf{d}}^{\mathrm{L}}$ are vectors of the nodal degrees of freedom.

The discretised weak form of the equation of motion is expressed as

$$\mathbf{M}\ddot{\mathbf{d}} + \mathbf{C}\dot{\mathbf{d}} + \mathbf{K}\mathbf{d} = \mathbf{f}, \tag{A8}$$

where $\ddot{\mathbf{d}}$, $\dot{\mathbf{d}}$, and $\mathbf{d}$ are the assembled vectors of the nodal degrees of freedom corresponding to the acceleration, velocity, and displacement, respectively, as

$$\ddot{\mathbf{d}} = \begin{Bmatrix} \ddot{\mathbf{d}}^{\mathrm{G}} \\ \ddot{\mathbf{d}}^{\mathrm{L}} \end{Bmatrix}, \tag{A9}$$

$$\dot{\mathbf{d}} = \begin{Bmatrix} \dot{\mathbf{d}}^{\mathrm{G}} \\ \dot{\mathbf{d}}^{\mathrm{L}} \end{Bmatrix}, \tag{A10}$$

$$\mathbf{d} = \begin{Bmatrix} \mathbf{d}^{\mathrm{G}} \\ \mathbf{d}^{\mathrm{L}} \end{Bmatrix}. \tag{A11}$$

The matrix $\mathbf{K}$ in Eq. (A8) is the assembled stiffness matrix, expressed as

$$\mathbf{K} = \begin{bmatrix} \mathbf{K}^{\mathrm{G}} & \mathbf{K}^{\mathrm{GL}} \\ \mathbf{K}^{\mathrm{LG}} & \mathbf{K}^{\mathrm{L}} \end{bmatrix}, \tag{A12}$$

where $\mathbf{K}^{\mathrm{G}}$, $\mathbf{K}^{\mathrm{L}}$, $\mathbf{K}^{\mathrm{GL}}$, and $\mathbf{K}^{\mathrm{LG}}$are partial stiffness matrices defined as follows:

$$\mathbf{K}^{\mathrm{G}} = \int_{\Omega^{\mathrm{G}}} \mathbf{B}_{\phi}^{\mathrm{G}}(\mathbf{x})^{\mathrm{T}} \mathbf{D} \mathbf{B}_{\phi}^{\mathrm{G}}(\mathbf{x}) d\Omega, \tag{A13}$$

$$\mathbf{K}^{\mathrm{L}} = \int_{\Omega^{\mathrm{L}}} \mathbf{B}_{\phi}^{\mathrm{L}}(\mathbf{x})^{\mathrm{T}} \mathbf{D} \mathbf{B}_{\phi}^{\mathrm{L}}(\mathbf{x}) d\Omega, \tag{A14}$$

$$\mathbf{K}^{\mathrm{GL}} = \int_{\Omega^{\mathrm{L}}} \mathbf{B}_{\phi}^{\mathrm{G}}(\mathbf{x})^{\mathrm{T}} \mathbf{D} \mathbf{B}_{\phi}^{\mathrm{L}}(\mathbf{x}) d\Omega, \tag{A15}$$

$$\mathbf{K}^{\mathrm{LG}} = \int_{\Omega^{\mathrm{L}}} \mathbf{B}_{\phi}^{\mathrm{L}}(\mathbf{x})^{\mathrm{T}} \mathbf{D} \mathbf{B}_{\phi}^{\mathrm{G}}(\mathbf{x}) d\Omega \ \left(= \mathbf{K}^{\mathrm{GL}^{\mathrm{T}}}\right), \tag{A16}$$

where $\mathbf{B}_{\phi}^{\mathrm{G}}$ and $\mathbf{B}_{\phi}^{\mathrm{L}}$ are the strain–displacement matrices formed from the derivatives of $\boldsymbol{\phi}^{\mathrm{G}}$ and $\boldsymbol{\phi}^{\mathrm{L}}$ with respect to the physical coordinates.

The matrix $\mathbf{M}$ in Eq. (A8) represents the assembled mass matrix, expressed as

$$\mathbf{M} = \begin{bmatrix} \mathbf{M}^{\mathrm{G}} & \mathbf{M}^{\mathrm{GL}} \\ \mathbf{M}^{\mathrm{LG}} & \mathbf{M}^{\mathrm{L}} \end{bmatrix}, \tag{A17}$$

where $\mathbf{M}^{\mathrm{G}}$, $\mathbf{M}^{\mathrm{L}}$, $\mathbf{M}^{\mathrm{GL}}$, and $\mathbf{M}^{\mathrm{LG}}$ are the partial mass matrices defined as follows:

$$\mathbf{M}^{\mathrm{G}} = \int_{\Omega^{\mathrm{G}}} \rho \boldsymbol{\phi}^{\mathrm{G}}(\mathbf{x})^{\mathrm{T}} \boldsymbol{\phi}^{\mathrm{G}}(\mathbf{x}) d\Omega, \tag{A18}$$

$$\mathbf{M}^{\mathrm{L}} = \int_{\Omega^{\mathrm{L}}} \rho \boldsymbol{\phi}^{\mathrm{L}}(\mathbf{x})^{\mathrm{T}} \boldsymbol{\phi}^{\mathrm{L}}(\mathbf{x}) d\Omega, \tag{A19}$$

$$\mathbf{M}^{\mathrm{GL}} = \int_{\Omega^{\mathrm{L}}} \rho \boldsymbol{\phi}^{\mathrm{G}}(\mathbf{x})^{\mathrm{T}} \boldsymbol{\phi}^{\mathrm{L}}(\mathbf{x}) d\Omega, \tag{A20}$$

$$\mathbf{M}^{\mathrm{LG}} = \int_{\Omega^{\mathrm{L}}} \rho \boldsymbol{\phi}^{\mathrm{L}}(\mathbf{x})^{\mathrm{T}} \boldsymbol{\phi}^{\mathrm{G}}(\mathbf{x}) d\Omega \ \left(= \mathbf{M}^{\mathrm{GL}^{\mathrm{T}}}\right). \tag{A21}$$

The matrix $\mathbf{C}$ in Eq. (A8) is a damping matrix. In the present study, Rayleigh damping is adopted to improve the numerical stability of the dynamic crack propagation analysis. In the Rayleigh damping method, $\mathbf{C}$ is defined as the linear combination of $\mathbf{M}$ and $\mathbf{K}$, expressed as

$$\mathbf{C} = \alpha_{\mathrm{R}} \mathbf{M} + \beta_{\mathrm{R}} \mathbf{K}, \tag{A22}$$

where $\alpha_{\mathrm{R}}$ and $\beta_{\mathrm{R}}$ are non-negative constants representing the degrees of damping [62]. Yanagimoto et al. [38] reported that, in dynamic crack propagation analyses based on the nodal force release technique adopted in the present study (see Section 2.2), Rayleigh damping with appropriately chosen parameters can effectively suppress numerical oscillations, and that such stabilisation is essential for reliable application phase analyses. The recommended Rayleigh coefficients $\alpha_R$ and $\beta_R$ in Eq. (A22) are given by

$$\alpha_{\mathrm{R}} = 0, \tag{A23}$$

$$\beta_{\mathrm{R}} = 2.57 \times h\ [\mathrm{m}] \times \sqrt{\frac{\rho\ [\mathrm{kg/m^3}]}{E\ [\mathrm{MPa}]}}, \tag{A24}$$

where $\rho$ and $E$ are the density and Young's modulus, respectively, and $h$ denotes the characteristic element size in the vicinity of the crack tip/front. In the present study, we set $h = h_{\mathrm{L}}$, where $h_{\mathrm{L}}$ is the local element size.

The vector $\mathbf{f}$ in Eq. (A8) is the assembled force vector expressed as

$$\mathbf{f} = \begin{Bmatrix} \mathbf{f}^{\mathrm{G}} \\ \mathbf{f}^{\mathrm{L}} \end{Bmatrix}, \tag{A25}$$

where $\mathbf{f}^{\mathrm{G}}$ and $\mathbf{f}^{\mathrm{L}}$ are partial nodal force vectors corresponding to the global and local nodes, respectively, calculated as follows:

$$\mathbf{f}^{\mathrm{G}} = \int_{\Omega^{\mathrm{G}}} \boldsymbol{\Phi}^{\mathrm{G}}(\mathbf{x})^{\mathrm{T}} \bar{\mathbf{b}}(\mathbf{x}) d\Omega + \int_{\Gamma_t} \boldsymbol{\Phi}^{\mathrm{G}}(\mathbf{x})^{\mathrm{T}} \bar{\mathbf{t}}(\mathbf{x}) d\Gamma, \tag{A26}$$

$$\mathbf{f}^{\mathrm{L}} = \int_{\Omega^{\mathrm{L}}} \boldsymbol{\Phi}^{\mathrm{L}}(\mathbf{x})^{\mathrm{T}} \bar{\mathbf{b}}(\mathbf{x}) d\Omega + \int_{\Gamma_t \cap \Omega^{\mathrm{L}}} \boldsymbol{\Phi}^{\mathrm{L}}(\mathbf{x})^{\mathrm{T}} \bar{\mathbf{t}}(\mathbf{x}) d\Gamma, \tag{A27}$$

where $\bar{\mathbf{b}}$ and $\bar{\mathbf{t}}$ are the body force and traction vectors, respectively. Body forces $\bar{\mathbf{b}}$ are assumed to be negligible in the present study.

The well-known implicit Hilber-Hughes-Taylor (HHT) method [63] is employed for time integration. Utilising this method, the equation of motion in Eq. (A8) is rewritten as

$$\mathbf{M}\ddot{\mathbf{d}}_{t+\Delta t} + \mathbf{C}\{(1+\alpha_{\mathrm{HHT}})\dot{\mathbf{d}}_{t+\Delta t} - \alpha_{\mathrm{HHT}}\dot{\mathbf{d}}_t\} + \mathbf{K}\{(1+\alpha_{\mathrm{HHT}})\mathbf{d}_{t+\Delta t} - \alpha_{\mathrm{HHT}}\mathbf{d}_t\} = \mathbf{f}\{t + (1+\alpha_{\mathrm{HHT}})\Delta t\}, \tag{A28}$$

and the kinematic relationships among displacement, velocity, and acceleration are given by

$$\mathbf{d}_{t+\Delta t} = \mathbf{d}_t + \Delta t \dot{\mathbf{d}}_t + \Delta t^2 \left(\frac{1}{2} - \beta_{\mathrm{HHT}}\right) \ddot{\mathbf{d}}_t + \Delta t^2 \beta_{\mathrm{HHT}} \ddot{\mathbf{d}}_{t+\Delta t}, \tag{A29}$$

$$\dot{\mathbf{d}}_{t+\Delta t} = \dot{\mathbf{d}}_t + \Delta t (1 - \gamma_{\mathrm{HHT}}) \ddot{\mathbf{d}}_t + \Delta t \gamma_{\mathrm{HHT}} \ddot{\mathbf{d}}_{t+\Delta t}, \tag{A30}$$

where $\beta_{\mathrm{HHT}}$ and $\gamma_{\mathrm{HHT}}$ are dimensionless parameters defined as:

$$\beta_{\mathrm{HHT}} = \frac{(1-\alpha_{\mathrm{HHT}})^2}{4}, \tag{A31}$$

$$\gamma_{\mathrm{HHT}} = \frac{1}{2} - \alpha_{\mathrm{HHT}}. \tag{A32}$$

Here, $\alpha_{\mathrm{HHT}}$ is also a dimensionless parameter and is set to $\alpha_{\mathrm{HHT}} = -0.02$ based on preliminary evaluation. Substituting Eqs. (A29) and (A30) into Eq. (A28), $\ddot{\mathbf{d}}$, $\dot{\mathbf{d}}$, and $\mathbf{d}$ at $t + \Delta t$ can be solved using the values at $t$.

## Appendix B. Broberg's analytical solutions for the local stress and dynamic stress intensity factor

The analytical solution of the local stress field ahead of a dynamically propagating crack tip, $\sigma_{yy}^{\mathrm{B(d)}}$, formulated by Broberg [48], is expressed as

$$\begin{aligned} \frac{\sigma_{yy}^{\mathrm{B(d)}} - \sigma_{yy}^{\infty}}{\sigma_{yy}^{\infty}} = -\frac{1}{\beta^2 g_1(\beta)} \Bigg\{ & \beta^2[4k^4 + (1-4k^2)\beta^2]\boldsymbol{F}(\kappa_1, q_1) - [8k^4 - 4k^2(1+k^2)\beta^2 + \beta^4]\boldsymbol{E}(\kappa_1, q_1) \\ & + [8k^4 - 4k^2(1+k^2)\beta^2 + \beta^4 - 4k^4\beta^2(1-\beta^2)\xi^2] \frac{\sqrt{1 - \frac{1}{\xi^2}}}{\sqrt{1-\beta^2\xi^2}} \Bigg\} \\ & + H_0\left(\xi - \frac{1}{k}\right) \frac{4k^2(1-\beta^2)}{\beta^2 g_1(\beta)} \left\{ \beta^2 \boldsymbol{F}(\kappa_2, q_2) - 2k^2 \boldsymbol{E}(\kappa_2, q_2) + k^2(2-\beta^2\xi^2) \frac{\sqrt{1 - \frac{1}{k^2\xi^2}}}{\sqrt{1-\beta^2\xi^2}} \right\}, \end{aligned} \tag{B1}$$

$$\beta = \frac{V}{V_{\mathrm{p}}}, \tag{B2}$$

$$\xi = \frac{V_{\mathrm{p}} a}{x' + a}, \tag{B3}$$

$$k = \sqrt{\frac{1-2\nu}{2\,(1-\nu)}}, \tag{B4}$$

$$V_{\mathrm{p}} = \sqrt{\frac{1-\nu}{(1+\nu)(1-2\nu)}\frac{E}{\rho}}, \tag{B5}$$

$$\kappa_1 = \sin^{-1}\frac{\sqrt{1-1/\xi^2}}{\sqrt{1-\beta^2}}, \tag{B6}$$

$$\kappa_2 = \sin^{-1}\frac{\sqrt{k^2-1/\xi^2}}{\sqrt{k^2-\beta^2}}, \tag{B7}$$

$$q_1 = \sqrt{1-\beta^2}, \tag{B8}$$

$$q_2 = \sqrt{1-\beta^2/k^2}, \tag{B9}$$

$$H_0(s) = \begin{cases} 0 & s \le 0 \\ 1 & s > 0 \end{cases}, \tag{B10}$$

$$\begin{aligned} g_1(\beta) = {} & -4(1-\beta^2)k^2\,\overline{\boldsymbol{K}}\left(\sqrt{1-\frac{\beta^2}{k^2}}\right) + \left(4k^4+\beta^2(1-4k^2)\right)\overline{\boldsymbol{K}}\left(\sqrt{1-\beta^2}\right) + \frac{8(1-\beta^2)k^4}{\beta^2}\overline{\boldsymbol{E}}\left(\sqrt{1-\frac{\beta^2}{k^2}}\right) \\ & - \frac{(\beta^4+8k^4-4\beta^2(k^2+1)k^2)}{\beta^2}\overline{\boldsymbol{E}}\left(\sqrt{1-\beta^2}\right). \end{aligned} \tag{B11}$$

Here, $\sigma_{yy}^{\infty}$ is the remotely applied stress, $V$ is the crack velocity, $a$ is the crack length, $x'$ is the distance from the crack tip, $E$ is the Young's modulus, $\nu$ is the Poisson's ratio, $\rho$ is the density, $V_{\mathrm{p}}$ denotes the P-wave velocity, $\boldsymbol{F}$ and $\boldsymbol{E}$ denote the elliptic integrals of the first and second kinds, respectively, whereas $\overline{\boldsymbol{K}}$ and $\overline{\boldsymbol{E}}$ denote the complete elliptic integrals of the first and second kinds, respectively.

The corresponding analytical solution of the dynamic stress intensity factor, $K_{\mathrm{I}}^{\mathrm{B(d)}}$, is given as

$$R(\beta) = 4k^3\sqrt{1-\beta^2}\sqrt{k^2-\beta^2} - (2k^2-\beta^2)^2, \tag{B12}$$

$$K_{\mathrm{I}}^{\mathrm{B(d)}} = \frac{\sigma_{yy}^{\infty}\sqrt{1-\beta^2}\,R(\beta)}{\beta^2\,g_1(\beta)}\sqrt{\pi a}, \tag{B13}$$

where $a$ is the crack length, and the remaining quantities are defined above.